\documentclass[article]{aa} 

\usepackage{graphicx}
\usepackage{txfonts}
\usepackage{natbib}
\usepackage{upgreek}
\usepackage{subcaption}
\usepackage{url}
\usepackage[rightcaption]{sidecap}
\usepackage{multirow}

\bibpunct{(}{)}{;}{a}{}{,}
\begin{document}
\title{Dynamical parallax, physical parameters and evolutionary status of the components of the bright eclipsing binary $\alpha$~Draconis
	\thanks{Based on observations made with the Mercator Telescope, operated on the island of La Palma by the Flemish Community, at the Spanish Observatorio del Roque de los Muchachos of the Instituto de Astrofísica de Canarias.}
	}

\author{K.~Pavlovski\inst{1}
\and
C.~A.~Hummel\inst{2}
\and
A.~Tkachenko\inst{3}
\and
A.~Dervi\c{s}o\u{g}lu\inst{4,5}
\and
C.~Kayhan\inst{4}
\and
R.~T.~Zavala\inst{6}
\and
D.~J.~Hutter\inst{7}
\and 
C.~Tycner\inst{7}
\and
T.~\c{S}ahin\inst{8}
\and
J.~Audenaert\inst{3}
\and
R.~Baeyens\inst{3}
\and
J.~Bodensteiner\inst{3,2}
\and
D.~M.~Bowman\inst{3}
\and
S.~Gebruers\inst{3}
\and
N.~E.~Jannsen\inst{3}
\and
J.~S.~G.~Mombarg\inst{3}
}

\institute{Department of Physics, Faculty of Science, University of Zagreb, 
10\,000 Zagreb, Croatia \\ 
\email{pavlovski@phy.hr}
\and
European Southern Observatory,
Karl-Schwarzschild-Str.~2, 85748 Garching, Germany
\and
Institute of Astronomy, KU Leuven, Celestijnenlaan 200D, 3001 Leuven, Belgium
\and
Department of Astronomy  Space Sciences, Science Faculty, Erciyes 
University, 38030 Melikgazi, Kayseri, Turkey
\and
Astronomy and Space Sciences Observatory and Research Center, 
Erciyes University, 38281 Talas, Kayseri, Turkey
\and
U.S. Naval Observatory, Flagstaff Station,
10391 W. Naval Obs. Rd., Flagstaff, AZ 86001, USA
\and
Central Michigan University, Department of Physics, 
Mt.\ Pleasant, MI 48859, USA
\and
Akdeniz University, Faculty of Science, Department of Space Sciences and Technologies, 07058, Antalya, Turkey}



\abstract
{}
{
Altough both components of the bright eclipsing binary $\alpha$~Dra having been resolved using long baseline interferometry and the secondary component shown to contribute some 15\% of the total flux, a spectroscopic 
detection of the companion star was so far unsuccessful.
We aim for a firm spectroscopic detection of the secondary component of $\alpha$~Dra using state-of-the-art spectroscopic analysis methods for very-high quality spectroscopic observations. This will allow the determination of  fundamental and atmospheric properties of the components in the system with high precision and accuracy.
}
{
To achieve our goals, we use a combined data set from interferometry with the Navy Precision Optical Interferometer (NPOI), photometry with the TESS space observatory, and high-resolution spectroscopy with the {\sc HERMES} fibre-fed spectrograph at the La Palma observatory. We use the method of spectral disentangling to search for the contribution of a companion star in the observed composite {\sc HERMES} spectra,
to separate the spectral contributions of both components, and to determine orbital elements of the $\alpha$~Dra system. TESS light curves are analysed in an iterative fashion with spectroscopic inference of stellar atmospheric parameters to determine fundamental stellar properties and their uncertainties. Finally, NPOI interferometric measurements are used for determination of the orbital parameters of the system and angular diameters of both binary components.
}
{
We report the first firm spectroscopic detection of the secondary component in $\alpha$~Dra and deliver disentangled spectra of both binary components. The components’ masses and radii are inferred with high precision and accuracy, and are $M_{\rm A} = 3.186\pm0.044$~M$_\odot$, $R_{\rm A} = 4.932\pm0.036$~R$_\odot$, and $M_{\rm B} = 2.431\pm0.019$~M$_\odot$, $R_{\rm B} = 2.326\pm0.052$ R$_\odot$ for the primary and secondary
components, respectively. Combined astrometric and spectroscopic analysis yields the semi-major axis of the system, which is ultimately used to derive the dynamical parallax of $\pi = 11.48\pm0.13$~mas, and the distance $d = 87.07\pm1.03$~pc to the $\alpha$~Dra system.
Evolutionary analysis of both binary components  with {\sc mesa} stellar structure and evolution models  suggests the primary is an evolved post-TAMS A-type star, while the companion is a main-sequence A-type star with a convective core mass of $M_{cc} = 0.337 \pm 0.011$~M$_\odot$. Positions of both binary components in
the Kiel- and HR-diagrams suggest a value of the convective core overshooting parameter $f_{\rm ov}$ well below 0.010~$H_{\rm p}$, and we infer the age of the system to be 310$\pm$25~Myr.
}
{
The inferred near-core mixing properties of both components do not support a dependence of the convective core overshooting on the stellar mass. At the same time, the $\alpha$~Dra system provides extra support to hypothesise that the mass discrepancy in eclipsing spectroscopic double-lined binaries is associated with inferior atmospheric modelling of intermediate- and high-mass stars, and less so with the predictive power
of stellar structure and evolution models as to the amount of near-core mixing and mass of the convective core.
}


      \titlerunning{Bright eclipsing binary $\alpha$ Draconis: dynamical parallax, physical properties, and evolutionary status}
   \authorrunning{Pavlovski et al.}

   \maketitle
%
\section{Introduction}
\label{sec:intro}

It is a well-known fact that a considerable fraction of stars are born in binary or higher-order multiple systems. In the case of hot massive OB-stars, binarity or higher multiplicity is found to have a strong effect on their evolution as over 50\% of those stars are found in binaries or higher-order multiple systems whose components will interact one or multiple times in the course of their lifetime \citep[e.g.][]{Sana2012,Sana2013}. Moreover, observational evidence was recently presented that the binary fraction among massive stars is independent of their environment. As such, \citet{Almeida2017} demonstrate that O-type stars in the 30~Dor region of the Large Magellanic Cloud (LMC) show the observed binary fraction of some 55-60\% that is similar to the existing Galactic samples of massive stars. Furthermore, \citet{Bodensteiner2021} and \citet{Banyard2021} report similar binary fraction of 50-55\% among SMC and Galactic B-type stars, respectively. At the same time, \citet{Luo2021} report a somewhat lower binary fraction of some 40\% among Galactic OB-stars from LAMOST low-resolution multi-epoch spectroscopy. Finally, \citet{Cohen2020} report the binary fraction among solar-type stars to be about 42\% suggesting that binarity is a key player in the evolution and fate of stars, independently of their initial mass.

Given the large binary fraction among solar-type \citep{Duchene_Kraus_2013}, and more massive stars \citep{Sana_2014}, binarity can become a serious obstacle for interpretation of astrophysical results when it is not properly accounted for. For example, detailed chemical composition analysis of stars may reveal complex and spurious abundance patterns when a faint and spectroscopically undetected companion is not accounted for in the spectroscopic analysis \citep{Scholler_2010}. Furthermore, observationally undetected stellar pairs can lead to the appearance of multiple parallel main sequences in the Hertzsprung-Russell (HR) diagrams from photometric surveys \citep{Kroupa_1993, Elson_1998}. Moreover, ignorance of the orbital motion of the two stars in a binary system also leads to a large bias in the astrometric measurement of the stellar parallax \citep{Makarov_Kaplan_2005, Frankowski_2007, Kervella_2019}. Finally, classification of a star as an X-ray source or a flaring object can be erroneous when a possible companion star remains undetected \citep{Schroder_Schmitt_2007, Stelzer_2011}.

On the other hand, when the subject of targeted research, spectroscopic double-lined or eclipsing binaries, if spatially resolved, become a principal source of the fundamental stellar quantities such as masses and radii of stars. With the advent of observational instruments and techniques, such as high-precision and high-resolution spectroscopy, high-angular resolution interferometry, and micro-magnitude precision and high duty cycle space-based photometry, stellar masses and radii are more routinely measured than ever before, with accuracy better than 3\% \citep{Torres_2010, Serenelli_2021}. High-precision and accuracy measurements of fundamental and atmospheric properties of stars, on the one hand, put theory of stellar structure and evolution (SSE) at a challenge, and offer a powerful tool for the improvement of the SSE theory, on the other hand. Over the last two decades, spectroscopic double-lined eclipsing binaries have been extensively exploited to study the effects of the stabilization of thermal convection by magnetic fields and of atomic diffusion in low-mass stars \citep[e.g.,][]{Feiden2012,Feiden2013,Feiden2014,Torres_2014,Higl_Weiss_2017}, probe the levels of near-core mixing (typically in the form of convective core overshooting) in intermediate- and high-mass stars \citep[e.g.,][]{Pols_1997,Guinan2000,Claret_Torres_2016,Claret2017,Claret2018,Claret2019,Martinet_2021}, investigate chemical element transport from accurate measurements of surface abundances in massive stars \citep{Pavlovski_2018}, and measure convective core masses in intermediate and high-mass BA-type stars \citep[e.g.,][]{Johnston2019b,Tkachenko_2020}. 

Nowadays, binary stars are often found to be synergistic with intrinsically variable stars, where the variability is typically caused by stellar pulsations, rotational modulation of surface inhomogeneities, and activity of cool companion stars. Hence, combining complementary research approaches (e.g., asteroseismology and binarity) and/or observational techniques (e.g., spectroscopy, astrometry, and photometry) allows us to attack the problem of uncertainties in SSE models from various angles, where SB2 EBs play a crucial role.

In this work, we study $\alpha$~Dra (also known as Thuban), a bright ($V$\,=\,3.7 mag) binary system that was first reported as such by \citet{Harper_1907}. Following the recent detection of eclipses in the photometric data recorded with the Transiting Exoplanet Survey Satellite \citep[TESS,][]{Ricker2014,Ricker2015} by \citet{Bedding_2019} and the possible detection of a faint companion star in high-resolution spectra by Hey et al. (2021, in review), we acquired a new set of high-resolution optical spectra, complementing already existing high-angular resolution interferometric data to enable a high-precision study of the system. In Sect.~\ref{sec:alphaDra}, we provide a short overview of the previous studies of the $\alpha$~Dra system. Results of the analysis of Mark~III and NPOI interferometric, {\sc HERMES} high-resolution optical spectroscopic, and TESS space-based photometric data are presented in Sects.~\ref{sec:interf}, \ref{sec:spectroscopy}, and \ref{sect:tess}, respectively. We assess the evolutionary status of both binary components in Sect.~\ref{sec:SSE}, while the dynamical parallax measurement is provided in Sect.~\ref{sec:parallax} along with discussion on all the so far available distance estimates to the system. We present a general discussion of the obtained results in Sect.~\ref{sec:dis} and close the paper with conclusions and an outline of future work provided in Sect.~\ref{sec:conclusions}.

\section{An overview of $\alpha$~Dra}\label{sec:alphaDra}

Since the first determination of the orbital parameters by \citet{Harper_1907},
$\alpha$~Dra (11 Dra, HD~123299) was frequently observed spectroscopically
as a SB1 system. A high degree of consistency is observed for orbital solutions reported by different research groups, yet some of the reported differences are statistically significant as they are larger than the quoted uncertainties. For example, \citet{Elst_Nelles_1983} report the RV semi-amplitude of $K = 49.8\pm0.3$~km\,s$^{-1}$ for the visible component, while more recent studies by \citet{Kallinger_2004} and \citet{Bischoff_2017} find $K = 48.488\pm0.080$~km\,s$^{-1}$ and $K = 47.48\pm0.21$~km\,s$^{-1}$, respectively. Since $\alpha$~Dra was not recognised as a SB2 system so far, all previous inferences of RV semi-amplitude are likely to be affected by the broad spectral features of an (at that time) undetected companion star. No appreciable changes in the longitude of periastron $\omega$ indicative of an apsidal motion are reported in the literature; the system is found to reside in a highly eccentric orbit, $e = 0.426\pm0.004$ \citep{Bischoff_2017}. The star was also first resolved with the Navy Precision Optical Interferometer (NPOI), and an estimate of the brightness difference of $\Delta m = 1.83\pm0.07$ mag at $\lambda$\,7000 {\AA} was derived by \citet{Hutter_2016}.

\citet{Adelman_2001} performed a detailed abundance analysis of $\alpha$~Dra based on DAO spectrograms and classified the system as a $\lambda$~Bootis star owing to the detected metal deficiencies. However, more recently, $\alpha$~Dra has been withdrawn from a list of $\lambda$~Boo candidate stars by \citet{Murphy_2015}. A historical disagreement as to the true value of the projected rotational velocity of the star was finally resolved thanks to the advent of electronic detectors and high spectral resolution of the instruments. In doing so, \citet{Gray_2014} measured $v \sin i$ of 26.2~km\,s$^{-1}$ from a high-resolution and high signal-to-noise spectrum using the Fourier transform technique. The measurement is in a good agreement with the findings by \citet{Royer_2002b} and \citet{Shorlin_2002} who report $v \sin i$ of $25\pm2$~km\,s$^{-1}$ and $28\pm2$~km\,s$^{-1}$, respectively.

A dedicated spectroscopic observational campaign was organised by \citet{Kallinger_2004} to search for possible intrinsic variability of the star through a study of line profile variations. Their data set consists of 140 echelle spectra secured during 10 nights in January 2003 at the Observatoire de Haute-Provence (OHP) and 1150 time resolved high-resolution spectra acquired during 45 nights in 2003-2004 at the T\"{u}ringer Landessternwarte Tautenburg. \citet{Kallinger_2004} report the detection of periodic RV variations at frequency of 26.5~d$^{-1}$ and with variable, orbital phase-dependent amplitude. The authors conclude that the observed periodic RV variability is caused by the interaction between tides in the system and a pulsation driving mechanism intrinsic to the star.

In the HR diagram $\alpha$~Dra is positioned  in between the instability strips of $\delta$~Sct and Slowly Pulsating B-type (SPB) stars. This fact triggered a speculation that $\alpha$~Dra might belong to a loosely defined class of ``Maia variable stars'', named after the B8-type prototype Maia star in the Pleiades \citep{Struve_1955}. Although the variability of Maia itself was disproved by \citet{Struve_1957}, the speculation about a new class of intrinsically variable stars was reintroduced by \citet{Mowlavi2013} who reported the detection of 36 stars in NGC~3766 that were found to reside in between the $\delta$~Sct and SPB instability strips and whose variability was attributed to stellar pulsations. Moreover, a recent study of the seven brightest B-type stars in the Pleiades by \citet{White_2017} concludes that six of those objects are SPB stars while Maia itself exhibits photometric variability with a period of about 10 days that is caused by rotational modulation of a large chemical spot. Absence of rapid line profile variations is also reported by \citet{Monier_2021} based on the analysis of FEROS and NARVAL high-resolution spectra. Finally, \citet{Balona_2015} presented a study of a large sample of B-type stars in the $Kepler$ field and classified all stars with high-frequency variability that are cooler than 20\,000~K as ``Maia variables''. It is worth noting that the study by \citet{Balona_2015} does not provide a solid proof of the existence of ``Maia variables'', instead it once again reintroduced the speculation about an unknown class of variables among B-type stars.

\citet{Bedding_2019} investigated two sectors of TESS space-based photometric data of $\alpha$~Dra and report the detection of primary and secondary eclipses separated by some 38.5 d. The authors also inspect the TESS light curve for signatures of intrinsic variability and find no evidence for it at the level of 10 parts per million. This result is in contradiction with the findings by \citet{Kallinger_2004} who reported spectroscopic variability on the timescales of 1~hr. More recently, Hey et al. (2021, in review) performed a study of $\alpha$~Dra based on the combined TESS photometry and newly obtained high-resolution SONG spectroscopy. The authors find a hint for a faint secondary component in the strongest absorption lines in the archival SOPHIE spectra of the system, yet spectral disentangling proved difficult owing to the low flux contribution from and a rapid rotation ($v \sin i \sim 200$~km\,s$^{-1}$) of a companion star. Based on the results of a combined light- and RV-curve analysis, and using MIST isochrones, Hey et al. (2021, in review) constrain $T_{\rm eff}$ and the mass of the primary to be 9975$\pm$125~K and 3.7$\pm$0.1~M$_{\odot}$, respectively. From the measured inclination and mass function of the system, the authors report a minimum mass of the secondary component to be 2.5$\pm$0.1~M$_{\odot}$. In addition, their chemical composition analysis of the high signal-to-noise ratio co-added SOPHIE spectrum reveals a complex abundance pattern for the primary component, and the authors confirm the claim of \citet{Bedding_2019} on the absence of intrinsic variability in the TESS photometric data.

\section{High-angular resolution interferometry}
\label{sec:interf}

\subsection{Observations, data reduction, and calibration}
\label{subsec:obs}

$\alpha$ Dra was observed both with the Mark III interferometer\footnote{Decommissioned in 1992.} \citep{Shao_1988} and the NPOI \citep{Armstrong_1998}.  The observation log is given in Table~\ref{tab:log} with information about the baseline lengths and calibrators used.  The Mark III data, recorded in narrow-band channels centered at 500 nm, 550 nm, and 800 nm, were reduced and the visibilities calibrated as described by \citet{Mozurkewich_2003}. Even though baselines of different lengths could be configured,  the Mark III operated only a single baseline during a night, so that no closure phases were obtained. 

The reduction of early NPOI data using a 3-way combiner was described by \citet{Hummel_1998}, and for the 6-way combiner (data from 2002) by \citet{Hummel_2003}. The data consist of visibilities and closure phases.

The calibrators were taken from a list of single stars maintained at NPOI with diameters estimated from $V$ and $(V-K)$ using the surface brightness relation published by \citet{Mozurkewich_2003} and \citet{van_Belle_2009}.  Values for $E(B-V)$ were derived from comparing the observed colours to theoretical colors as a function of spectral type by Schmidt-Kaler in \citet{1982lbg6.conf.....A}. Values for the extinction derived from $E(B-V)$ were compared to estimates based on maps by \citet{Drimmel_2003}, and used to correct $V$ if they agreed within 0.5 magnitudes.  Even though the surface brightness
relationship based on $(V-K)$ colours is to first order independent of reddening, we included this small correction. The minimum (squared) visibility amplitudes corresponding to the diameter estimates are given in Table~\ref{tab:cal} for any NPOI observation performed and show that the calibrator stars are mostly unresolved or only weakly resolved.

\begin{table}
\small
\caption{List of NPOI calibrators. $V^2_{\rm min}$ is minimum (squared) visibility on the longest baselines used.}
\label{tab:cal}
\begin{tabular}{llccccc}
\hline
\hline
HD& Sp. &$V$&$V-K$&$E(B-V)$&$\theta_{V-K}$&$V^2_{\rm min}$ \\   
  & type   & [mag]  & [mag]    & [mag]  & [mas]&     \\
\hline
FK5\,423&A2V& 3.34& 0.26&-0.06&0.86&0.88\\
FK5\,447&A0V& 2.44& 0.01& 0.02&1.11&0.81\\
FK5\,456&A3Vv & 3.31& 0.21& 0.00&0.84&0.90\\
FK5\,472&B6IIIp& 3.87& 0.05& 0.02&0.59&0.89\\
HR\,5062&A5V& 4.01& 0.87& 0.01&0.67&0.95\\
FK5\,509&B3V& 1.86&-0.41& 0.01&1.11&0.81\\
HR\,5329&A8IV& 4.54& 0.44&-0.05&0.56&0.79\\
FK5\,622&O9.5V& 2.60&-0.08& 0.29&0.00&1.00\\
FK5\,668&A0V& 3.75& 0.13& 0.06&0.65&0.93\\
FK5\,677&B5Ib& 3.97&-0.03& 0.12&0.50&0.97\\
FK5\,913&A1Vn& 4.36& 0.10& 0.01&0.48&0.99\\
\hline
\end{tabular}

\end{table}

\begin{figure}
\centering
\resizebox{\hsize}{!}{\includegraphics{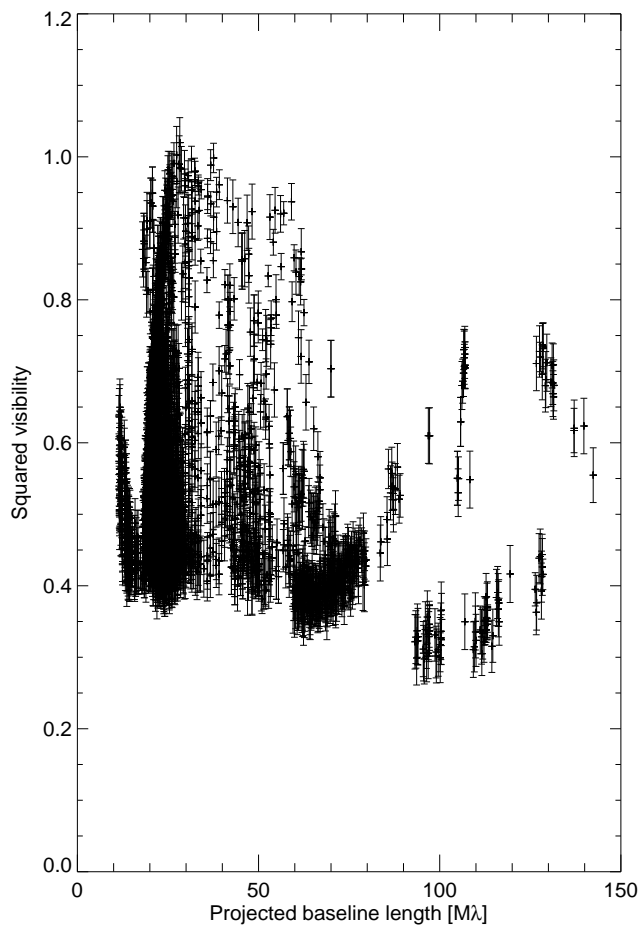}}
\caption{
	The measured (squared) visibility amplitudes plotted against $uv$ radius. Only data with errors less than 4\% are shown for clarity. The decreasing upper envelope of the visibility amplitudes indicates that the primary component is well resolved, while the weaker trend of the lower envelope indicates a smaller diameter for the secondary. The strong variation of the amplitude in between is due to the interference pattern caused by the binarity of the source structure. 
}
\label{fig:uvr} 
\end{figure} 

\subsection{Orbital analysis}
\label{subsec:orb}

\label{subsec:astro}

\begin{figure}
\centering
\includegraphics[width=9.cm]{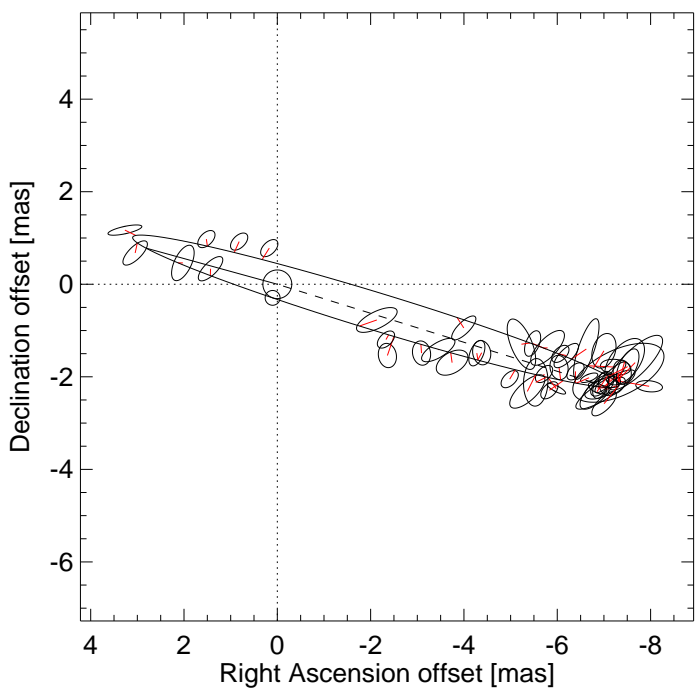}
\caption{
	The apparent orbit of $\alpha$ Draconis. The solid straight line indicates the periastron, and the dashed line the ascending node. The two small circles at/near the origin indicate the interferometric sizes of the primary and secondary components, respectively, at inferior conjunction.
}
\label{fig:apporb} 
\end{figure}

Modelling of the (squared) visibility amplitudes (Fig.~\ref{fig:uvr}) and closure phases was performed with OYSTER\footnote{\url{http://www.eso.org/~chummel/oyster/oyster.html}}. Each night's data were fit using a binary model parameterized with the angular diameters of the two stars, their magnitude difference, and their relative position. The stellar parameters were not free to vary in this step, but were constrained with the orbital elements replacing the relative positions in a fit directly to all visibility data. To better constrain the orbital period, we added radial velocity measurements from historical sources \citep{Pearce_1957, Elst_Nelles_1983,Adelman_2001,Bischoff_2017}. The smaller diameter of the secondary, which is not well constrained in this fit, was estimated in a second fit by adopting the effective temperatures and surface gravities given in Table~\ref{tab:abspar} and the diameter of the primary. The astrometric results are summarized in Table~\ref{tab:astro_results}, and the apparent orbit of resolved binary system of $\alpha$ Dra is shown in Fig.~\ref{fig:apporb}. The astrometric error ellipses were derived from fitting the $4\sigma$ contour of the minimum of the $\chi^2$-surface around the secondary's position, normalized to unity.

\begin{table}
\caption{Orbital elements and component parameters for $\alpha$ Dra. Ascending node is for epoch J2000.0, and the periastron angle is that of the primary. }
\centering
\begin{tabular}{lcr@{\,$\pm$\,}l}
\hline\hline
Parameter	& Unit		& Value		& Error\\
\hline
Period $P$                    & [d]  & $51.417350$ 	& $0.00034$     \\
Periastron epoch $T_{\rm pp}$ & [JD]  & $2453498.8$ 	& $0.1$     \\
Eccentricity $e$              &		 & $0.43$    	& $0.01$    \\ 
Periastron angle $\omega$     &[deg] & $22.4$    	& $0.5$     \\
Ascending node  $\Omega$      &[deg] & $252.6$   	& $0.4$     \\
Inclination $i$               &[deg] & $85.4$ 	    & $0.5$     \\
Semi-major axis $a''$         &[mas] & $5.52$ 	    & $0.06$    \\
\\
Magnitude difference $\Delta R$      & [mag]	& $1.79$	& $0.02$ \\
Primary's diameter $D_{\mathrm A}$   & [mas]	& $0.62$	& $0.05$ \\
Secondary's diameter $D_{\mathrm B}$ & [mas]    & $0.28$	& $0.05$ \\
\hline
\end{tabular}
\label{tab:astro_orbit}
\end{table}

\section{High-resolution spectroscopy}\label{sec:spectroscopy}

\subsection{New {\sc HERMES} high-resolution spectroscopy}

New high-resolution echelle spectra were obtained from 
10 February to 1 April 2021 at the 1.2-m Mercator telescope at the 
Observatorio del Roque de los Muchachos, La Palma, Canary Islands, Spain.
The spectra were secured with the {\sc HERMES} fibre-fed  high-resolution ($R = 85\,000$) spectrograph \citep{Raskin_2011}. The {\sc HERMES} spectra cover the 
entire optical spectral range from 3770 to 9000 {\AA} in 55 spectral 
orders. Depending on zenith distance, exposure times ranged between 40~sec and 120~sec for a typical signal-to-noise (S/N) ratio of over 200.  In total 50 spectra were collected, of which 4 were discarded for an inferior S/N.

The reduction of the raw observations was performed with a dedicated {\sc HERMES} data reduction pipeline. The individual reduction steps include background and bias subtraction, 
flat fielding, wavelength calibration using a ThArNe lamp, and echelle order merging. Reduced spectra are resampled in constant velocity-bins preserving the size of the detector pixels. Continuum normalisation is done manually by fitting a spline function to a carefully selected set of pseudo-continuum points.

\begin{figure}[]
\centering
\includegraphics[width=8.5cm]{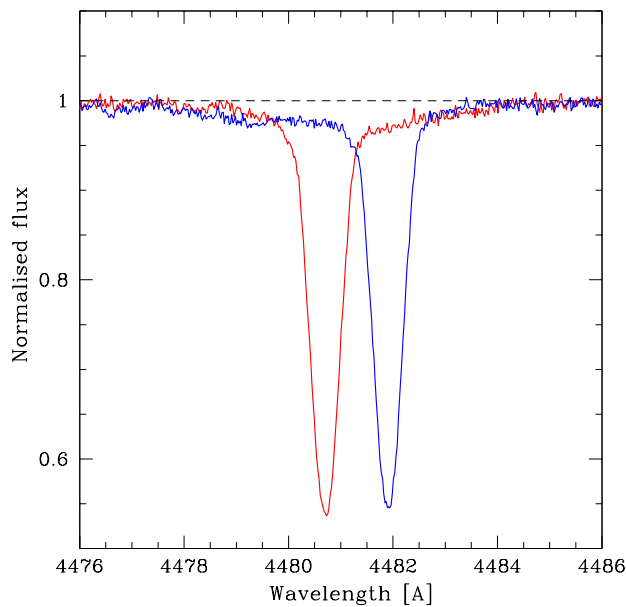} 
\caption{Two observed composite {\sc HERMES} spectra taken at quadrature in the region of the \ion{Mg}{ii}~4481~\AA\ spectral line. A broad and shallow spectral feature of the secondary component is clearly seen at about 1\% level of the continuum. A fiducial pseudo-continuum is indicated with the dashed line for reference. The spectra were obtained on BJD 2\,459\,286.64516 (red line), and BJD 2\,459\,305.62250 (blue line), and correspond to the phases of the orbital cycle 0.486 and 0.856, respectively.}
\label{fig:mag}
\end{figure}

\subsection{Spectroscopic detection of the secondary component}

\begin{figure*}[t!]
\includegraphics[width=6cm]{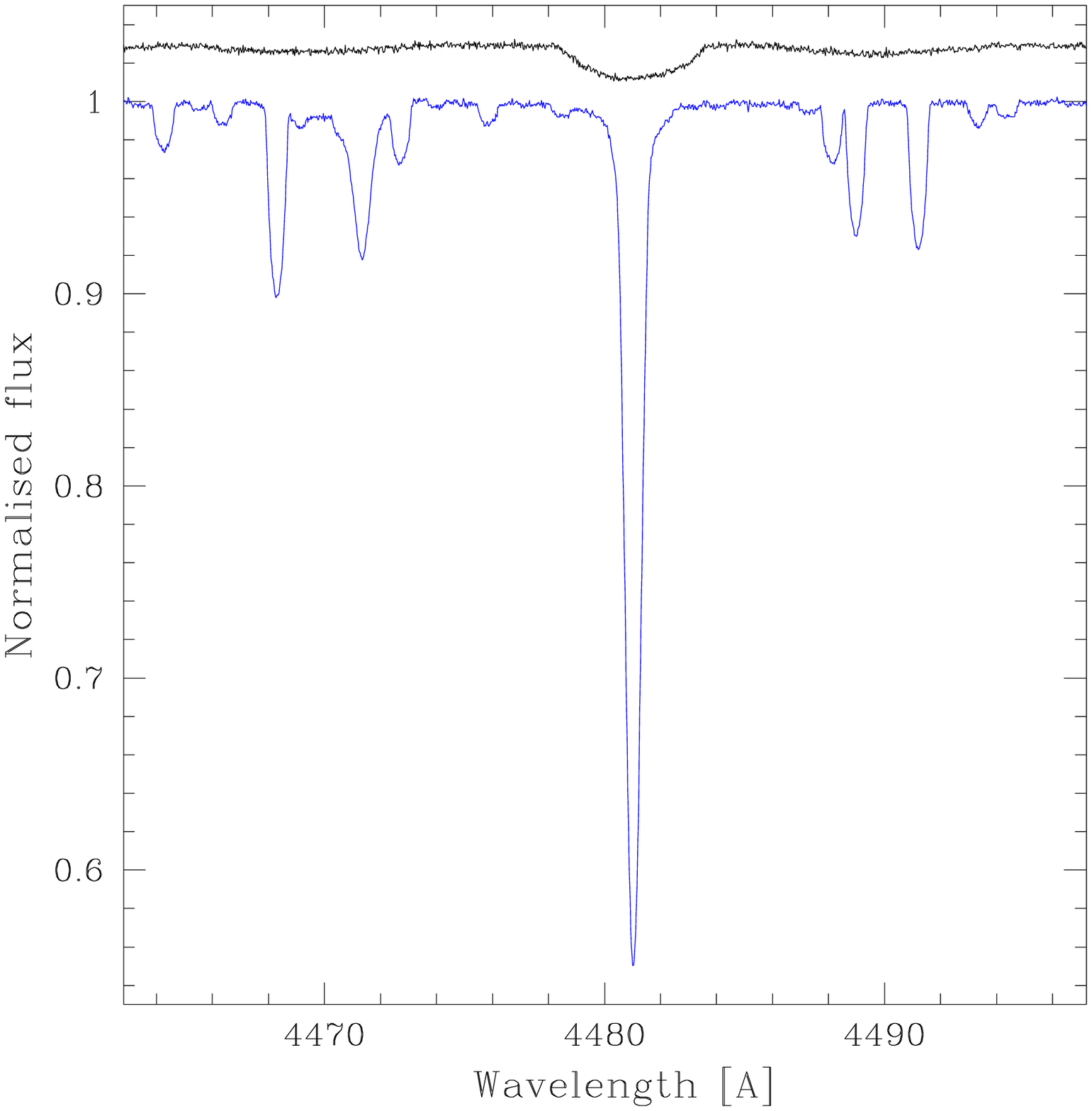}
\includegraphics[width=6cm]{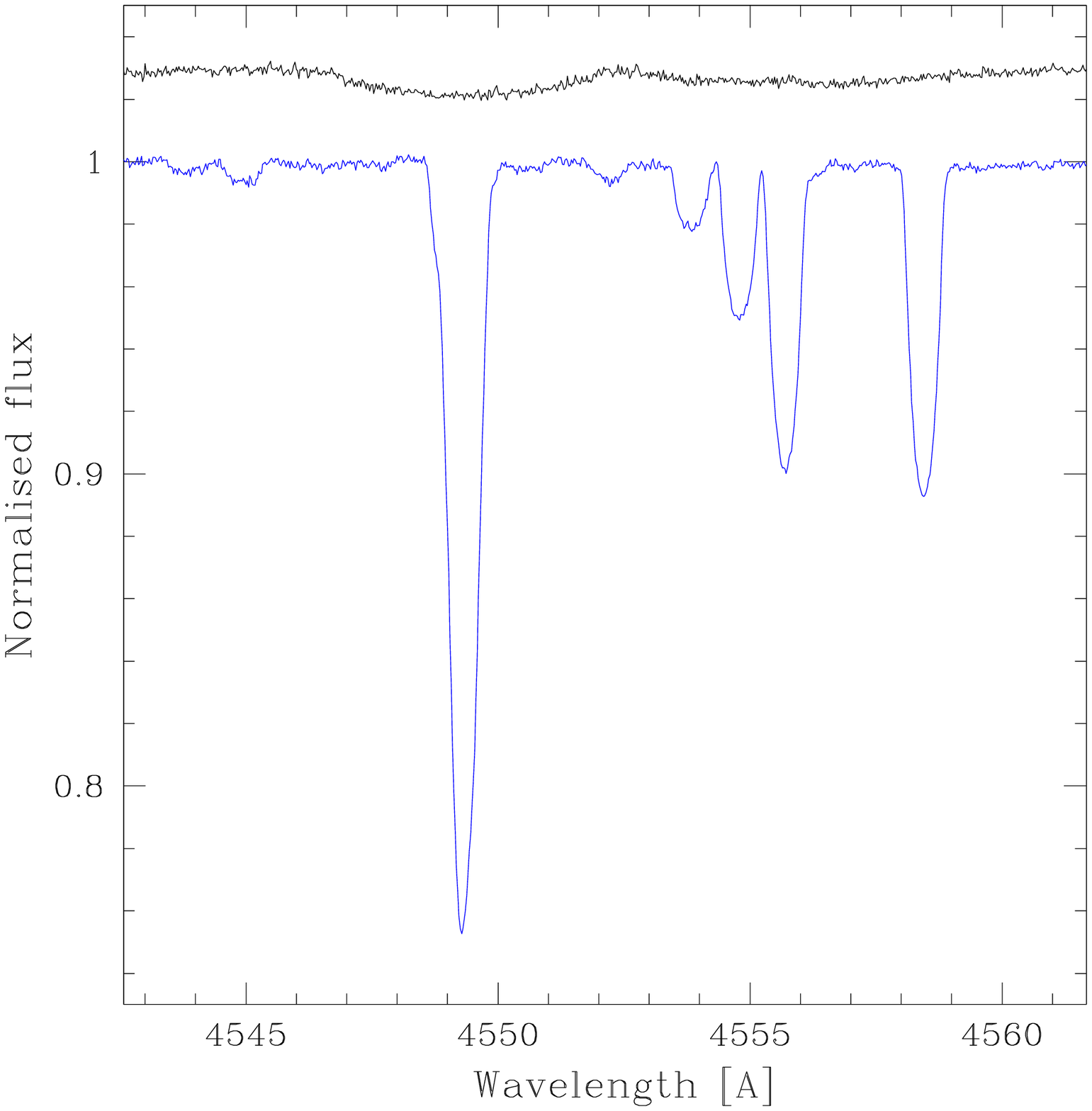}  \includegraphics[width=6cm]{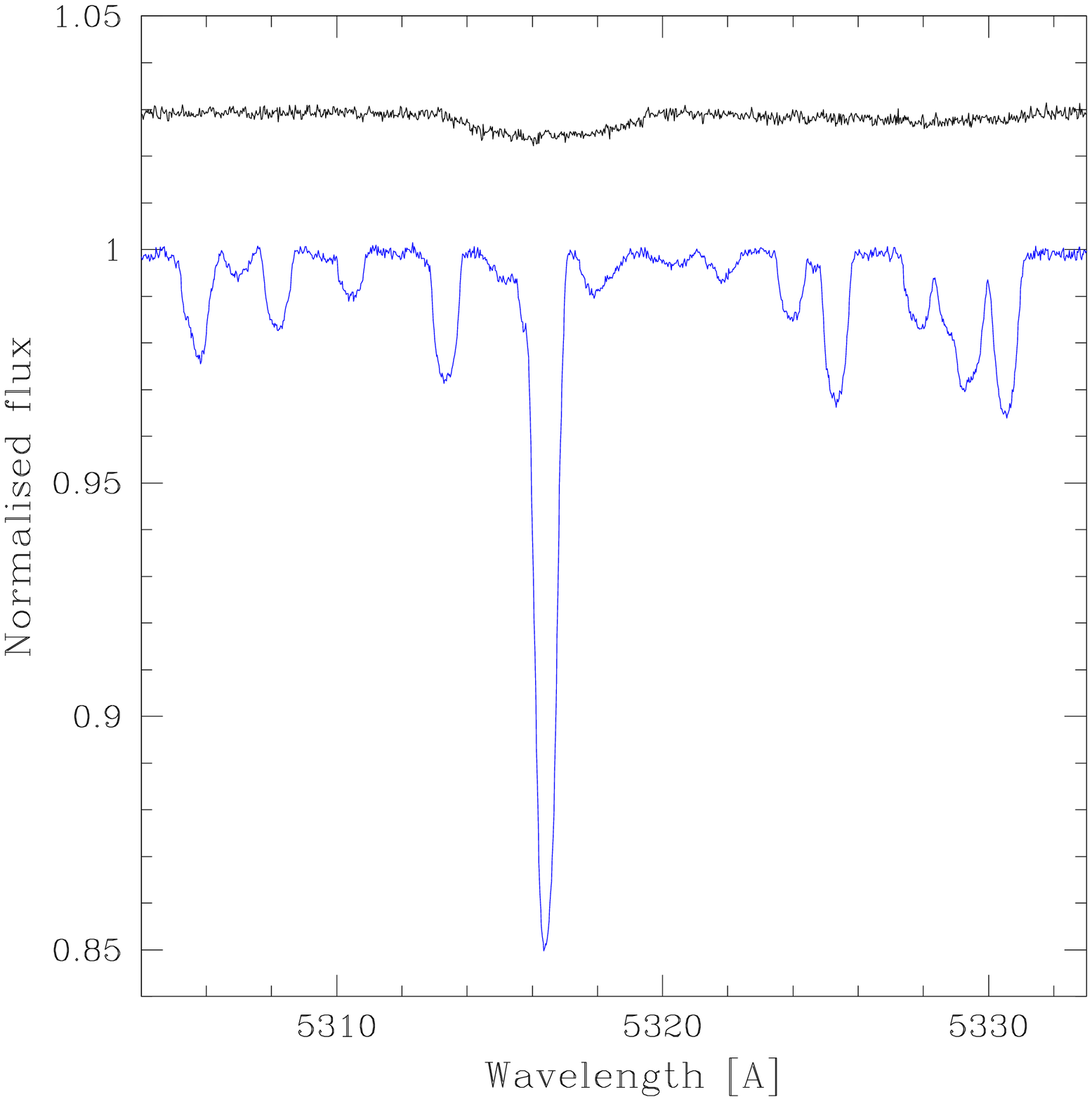} 
\caption{
Small portions of the disentangled spectra of the primary (blue line) and secondary (black line) components of the $\alpha$~Dra system. From left to right: \ion{Mg}{ii}~4481~\AA\ doublet, the \ion{Fe}{ii}~4549.5/\ion{Ti}{ii}~4549.7~\AA\ spectral blend, and the \ion{Fe}{ii}~5316.6/\ion{Fe}{ii}~5316.8~\AA\ spectral blend. The secondary's disentangled spectrum is arbitrary shifted for clarity.}
\label{fig:metals}
\end{figure*}

As discussed in detail in Sect.~\ref{sec:alphaDra}, $\alpha$~Dra has so far been classified as a SB1 system, and only recently Hey et al. (2021, in review) found a hint for a companion star in a high S/N archival SOPHIE spectrum. According to the interferometric measurements, the magnitude difference between the two stars is $\Delta m = 1.81\pm0.02$ mag, such that a companion star contributes some 15\% to the total light. Therefore, we suspect that the lack of detection of signatures of the secondary component in the observed composite spectra of the system so far is probably associated with its rapid rotation, just as it is also hypothesised by Hey et al. (2021, in review). Therefore, our {\sc HERMES} spectroscopic campaign was specifically tuned to a detection of broad and shallow lines of a companion star, hence the requirement for high S/N and full orbital phase coverage. A visual inspection of the normalised {\sc HERMES} spectra confirms the suspicion: a broad and shallow ``depression'' in the continuum is clearly visible around prominent metal lines in the composite spectra of the system. This is illustrated in Fig.~\ref{fig:mag} where two spectra taken at quadrature are shown in the region of \ion{Mg}{ii}~4481~\AA\ spectral line.

Spectral disentangling is an efficient method for extracting the individual components' spectra while simultaneously optimising the orbital elements of a binary (or higher-order multiple) system \citep{Simon_Sturm_1994}. A time-series of the observed spectra, preferably of high-resolution and uniformly distributed with the orbital phase is required for the task. The method has proven to be very efficient in revealing even very faint companions, contributing as little as 1-2 percent to the total light of the system \citep[e.g.,][]{Mayer_2013, Torres_2014, Kolbas_2015, Themessl_2018}.

We employ spectral disentangling in the Fourier space as proposed by \citet{Hadrava_1995} and implemented in the {\sc FDBinary} software package \citep{Ilijic_2004}. A complete procedure was worked-out in \citet{Hensberge_2000}, and elaborated and refined in detail in \citet{Pavlovski_2018}. We apply the method to our dataset of {\sc HERMES} high-resolution spectra, starting with prominent metal lines such as those of the \ion{Mg}{ii}~4481~\AA\ doublet. This procedure leads us to unambiguously detect the spectral contribution of the secondary component; Fig.~\ref{fig:metals} shows a portion of the disentangled spectra of both binary components in the regions of the \ion{Mg}{ii}~4481~\AA\ doublet (left panel), the spectral line blends of \ion{Fe}{ii}~4549.5/\ion{Ti}{ii}~4549.7~\AA\ (middle panel) and \ion{Fe}{ii}~5316.6/\ion{Fe}{ii}~5316.8~\AA\ (right panel). Spectral lines of the secondary component are shallow, with depths at the level of 1\% of the continuum, owing to its high rotational velocity. We note that the detection of such a faint contribution is only possible when several conditions are simultaneously met: (i) a high resolving power of the instrument, (ii) a high S/N of the observed composite spectra, and (iii) optimally extracted and normalised spectra. The latter aspect is particularly important for the successful reconstruction of the individual spectra of (faint) binary components. Furthermore, a proper selection of spectral segments for Fourier-based disentangling is essential, where particular attention should be paid to placing the edges of selected spectral intervals in the (pseudo-)continuum regions that are free of spectral lines. Violation of the latter condition leads to large-amplitude continuum undulations in the disentangled spectra, which affects the line depths.

Interferometric and photometric analyses of the $\alpha$~Dra system presented in Sects.~\ref{sec:interf} and \ref{sect:tess}, respectively, unambiguously identify the star as a detached binary system. Both components are found to be well-bounded within their Roche lobes, and there is barely any light variation detected outside the partial eclipses in the TESS light curve. In the cases when changes in the light ratio between the two components are absent and/or eclipse spectra are missing, an ambiguity occurs in the spectral disentangling: Since the zeroth-order term in the Fourier expansion is indeterminable, the line depths in the disentangled spectra are not uniquely determined \citep{Pavlovski_Hensberge_2005}. In that case, the best approach is to perform spectral disentangling in the so-called ``pure separation'' mode \citep{Pavlovski_Hensberge_2010}, of which the net result is the individual components' separated spectra that are diluted according to the light ratio of the two stars. Two options are available for further analysis of the separated spectra: (i) they are re-normalised to their individual continua using the light ratio inferred from the light curve analysis; or (ii) the light ratio can be used as a free parameter in spectroscopic analysis along with atmospheric parameters of the star.

An important aspect of spectral disentangling is a S/N gain in the resulting disentangled spectra. Spectral disentangling is acting as co-addition of the observed spectra, so the expected gain is proportional to $\sqrt{N}$, where $N$ is the number of observed composite spectra. The gain in S/N is distributed to both disentangled spectra proportional to the fractional light contribution of a component to the total light of the system. For $\alpha$~Dra, we obtain S/N of $\sim$1470 and $\sim$260 for the primary and secondary component, respectively.
 
Final separation of the components' spectra of $\alpha$~Dra from our newly obtained set of {\sc HERMES} spectra was performed with the orbital parameters determined in Sect.~\ref{subsect:specorb}. The disentangled spectra cover a wavelength range from 4000~\AA\ to 5650~\AA, including the prominent H$\beta$, H$\gamma$, and H$\delta$ Balmer lines. These spectra are used for our detailed spectrum analysis in Sect.~\ref{subsec:atmos}.

\begin{table}
\caption{Orbital parameters of the $\alpha$~Dra system obtained with the method of spectral disentangling from 46 high-resolution, high S/N {\sc HERMES} spectra. The orbital period was fixed to the value determined from the observed primary minima (cf. Sect.~\ref{sect:tess}). The uncertainties in the orbital parameters were calculated with the bootstrapping method from 10\,000 samples.}
    \centering
    \begin{tabular}{lcc}
    \hline\hline
    Parameter   & Unit  &  Value  \\
    \hline
 $P$  & [d]  &   51.41891 (fix) \\
 $T_{\rm pp}$ & [d] & 2\,451\,441.804$\pm$0.014 \\
 $e$          &     &  0.4229$\pm$0.0012  \\
 $\omega$     & [deg] &  21.28$\pm0.13$    \\
 $K_{\rm A}$  & [km\,s$^{-1}$] &  48.512$\pm$0.054  \\
 $K_{\rm B}$  & [km\,s$^{-1}$] &  63.58$\pm$0.41  \\
 $q$          &              &  0.7642$\pm$0.0064  \\
 $M_{\rm A}\,\sin^3 i$ & $[M_{\odot}]$ & $3.167\pm0.044$  \\   
 $M_{\rm B}\,\sin^3 i$ & $[M_{\odot}]$ & $2.416\pm0.019$\\  
 $A \sin i$   & [R$_{\odot}$] & $103.19\pm0.38$   \\
 \hline
    \end{tabular}
    \label{tab:specorb}
\end{table}

\begin{figure}
\centering
\includegraphics[width=8.6cm]{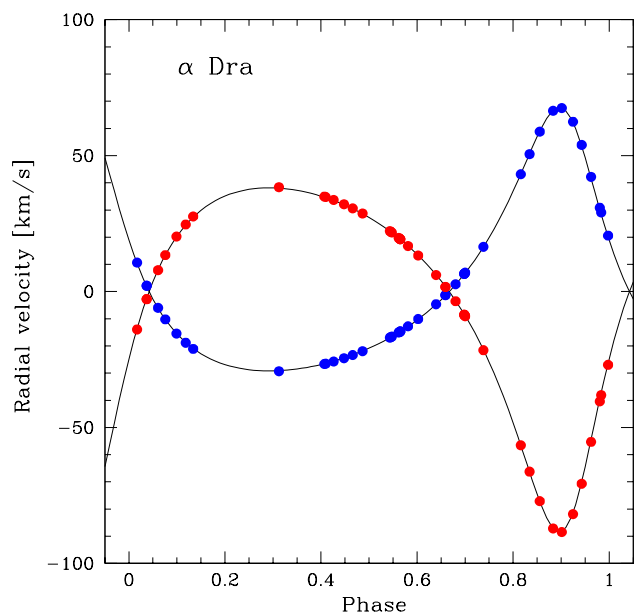}
\caption{RVs computed from the final orbital solution; blue and red dots refer to the primary and secondary component, respectively. The obtained spectroscopic orbital solution is indicated with the black solid line.}
\label{fig:specorb} 
\end{figure} 

\subsection{Spectroscopic orbit from spectral disentangling}
\label{subsect:specorb}

In the spectral disentangling method, orbital elements are optimised based on a time-series of the observed spectra in a self-consistent manner, along with the reconstruction of the individual components' spectra \citep{Simon_Sturm_1994, Hadrava_1995}. In this way, the determination of RVs from the individual exposures is not needed. Spectra of both components of the $\alpha$~Dra system are dominated by strong and broad hydrogen lines. Those are not the best diagnostic lines for determination of orbital parameters of the system, irrespective of whether the parameters are inferred from one of the RV determination methods or from the method of spectral disentangling. The main reasons why Balmer lines in A-type stars are not suitable for the task are: (i) owing to their strengths and widths, hydrogen lines are typically  covered by two echelle orders which make them subject to systematic uncertainties in the order merging and normalisation to the local continuum; and (ii) hydrogen lines in the individual components' spectra are rarely sufficiently well separated in velocity space leading to a more uncertain orbital solution. In the case of the $\alpha$~Dra system, both of the aforementioned obstacles are present, hence we do not use the Balmer lines in the determination of the orbital parameters of the system. 

To ensure that a sufficient number of metal lines are considered for the determination of orbital parameters of the system, and to make the best use of the high quality of the extracted and normalised {\sc HERMES} spectra, we opted for the spectral disentangling in large spectral segments of about 350~\AA, from 4400~\AA\ to 4750~\AA, and about 500~\AA, from 5000~\AA\ to 5500~\AA. Our final solution is given in Table~\ref{tab:specorb}; RVs computed from the finally adopted spectroscopic orbital parameters are shown in Fig.~\ref{fig:specorb} and are representative of the obtained orbital phase coverage. 

\begin{figure*}
\centering
\includegraphics[width=17cm]{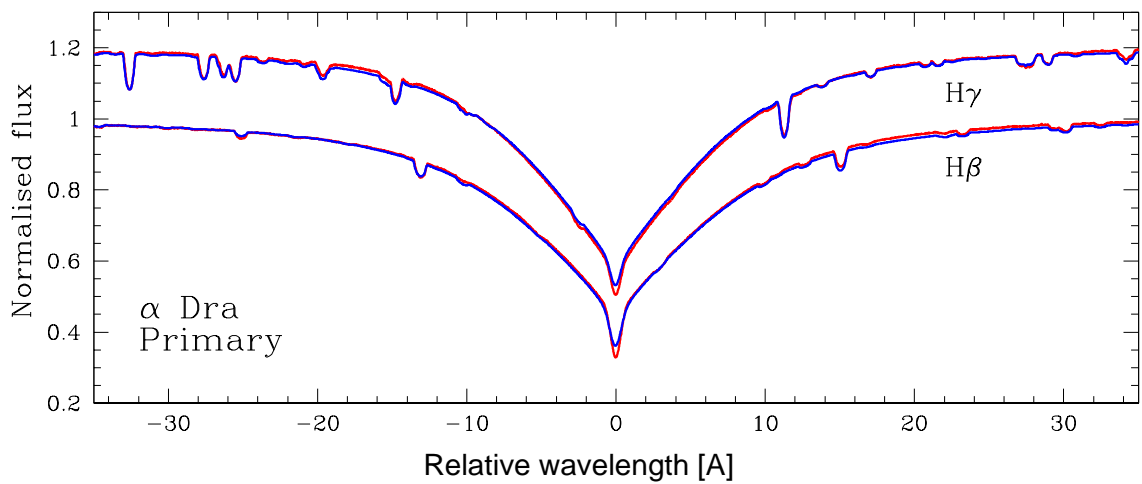}
\caption{Comparison between the best fit synthetic (blue lines) and the observed disentangled (red lines) spectra of the primary component in the regions of H$\gamma$ (upper profile) and H$\beta$ (lower profile) Balmer lines. Both profiles are shown relative to their central wavelengths; the corresponding best fit atmospheric parameters are listed in Table~\ref{tab:gssp_params} (Run 2b). The spectra centered on H$\gamma$ are arbitrary shifted for better visibility.}
\label{fig:gsspBalmerlinesfit} 
\end{figure*}

\subsection{Atmospheric parameters}
\label{subsec:atmos}

\begin{table*}[t!]
\caption{Atmospheric parameters of both components of the $\alpha$~Dra system as inferred from their disentangled spectra with the {\sc GSSP\_single} module. The asterisk symbol ($^*$) refers to the parameters whose values are fixed to those inferred from the TESS light curve. The finally accepted set of the parameters is highlighted in boldface.}
\label{tab:gssp_params}
\centering
\begin{tabular}{lr@{\,$\pm$\,}lr@{\,$\pm$\,}lr@{\,$\pm$\,}l}
\hline\hline
\multirow{2}{*}{Parameter}				& \multicolumn{6}{c}{Value$\pm$Error}\\
 & \multicolumn{2}{c}{Run 1} & \multicolumn{2}{c}{Run 2a} & \multicolumn{2}{c}{Run 2b}\\
\hline
& \multicolumn{6}{c}{{\bf Primary}}\\
$T_{\rm eff}$ (K)			& 10190 	& 125 & 10205 & 100 & {\bf 10225} 	& {\bf 100}\\
$\log\,g$ (dex)			& 3.52 	& 0.09 & 3.5511$^*$ & 0.0081 & {\bf 3.5511$^*$}  	& {\bf 0.0081} \\
${\rm [M/H]}$ (dex) & -0.03 	& 0.05 & 0.01 	& 0.05 & {\bf 0.01} 	& {\bf 0.05}\\
$\xi$ (km\,s$^{-1}$)				& 1.06 	& 0.21 & 1.26 	& 0.30 & {\bf 1.34} 	& {\bf 0.30}\\
$v\,\sin\,i$ (km\,s$^{-1}$)	& 25.4 	& 0.7 & 25.4 	& 0.9 & {\bf 25.4} 	& {\bf 0.9}\\
light factor			& 0.848 	& 0.012 & 0.83 	& 0.01 & {\bf 0.825$^*$} 	& {\bf 0.010} \\
& \multicolumn{6}{c}{{\bf Secondary}}\\
$T_{\rm eff}$ (K)			& \multicolumn{2}{c}{-----} & 10150 	& 300 & {\bf 10165} & {\bf 250}\\
$\log\,g$ (dex)			& \multicolumn{2}{c}{-----} & 4.095$^*$ 	& 0.023 & {\bf 4.095$^*$} & {\bf 0.023}\\
${\rm [M/H]}$ (dex) & \multicolumn{2}{c}{-----} & -0.12& 0.15 & {\bf -0.01} & {\bf 0.09}\\
$\xi$ (km\,s$^{-1}$)				& \multicolumn{2}{c}{-----} & 0.1 & 0.9 & {\bf 0.75} & {\bf 0.75}\\
$v\,\sin\,i$ (km\,s$^{-1}$)	& \multicolumn{2}{c}{-----} & 168 & 10 & {\bf 168} & {\bf 11}\\
light factor			& \multicolumn{2}{c}{-----} & 0.20 & 0.02 & {\bf 0.175$^*$} & {\bf 0.011} \\
\hline
\end{tabular}
\end{table*}

\begin{figure*}[h!]
\centering
\includegraphics[width=8.6cm]{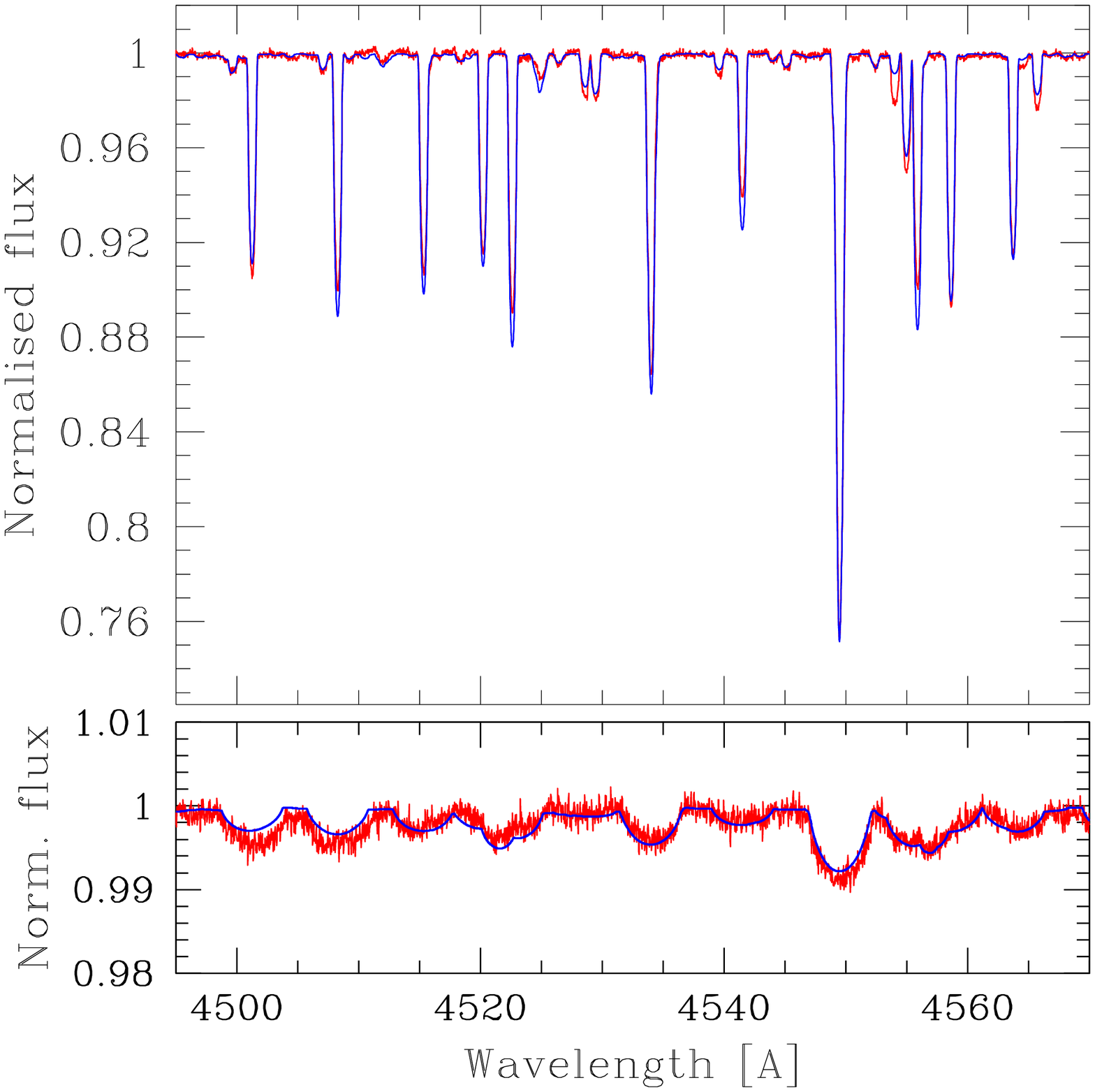} \includegraphics[width=8.6cm]{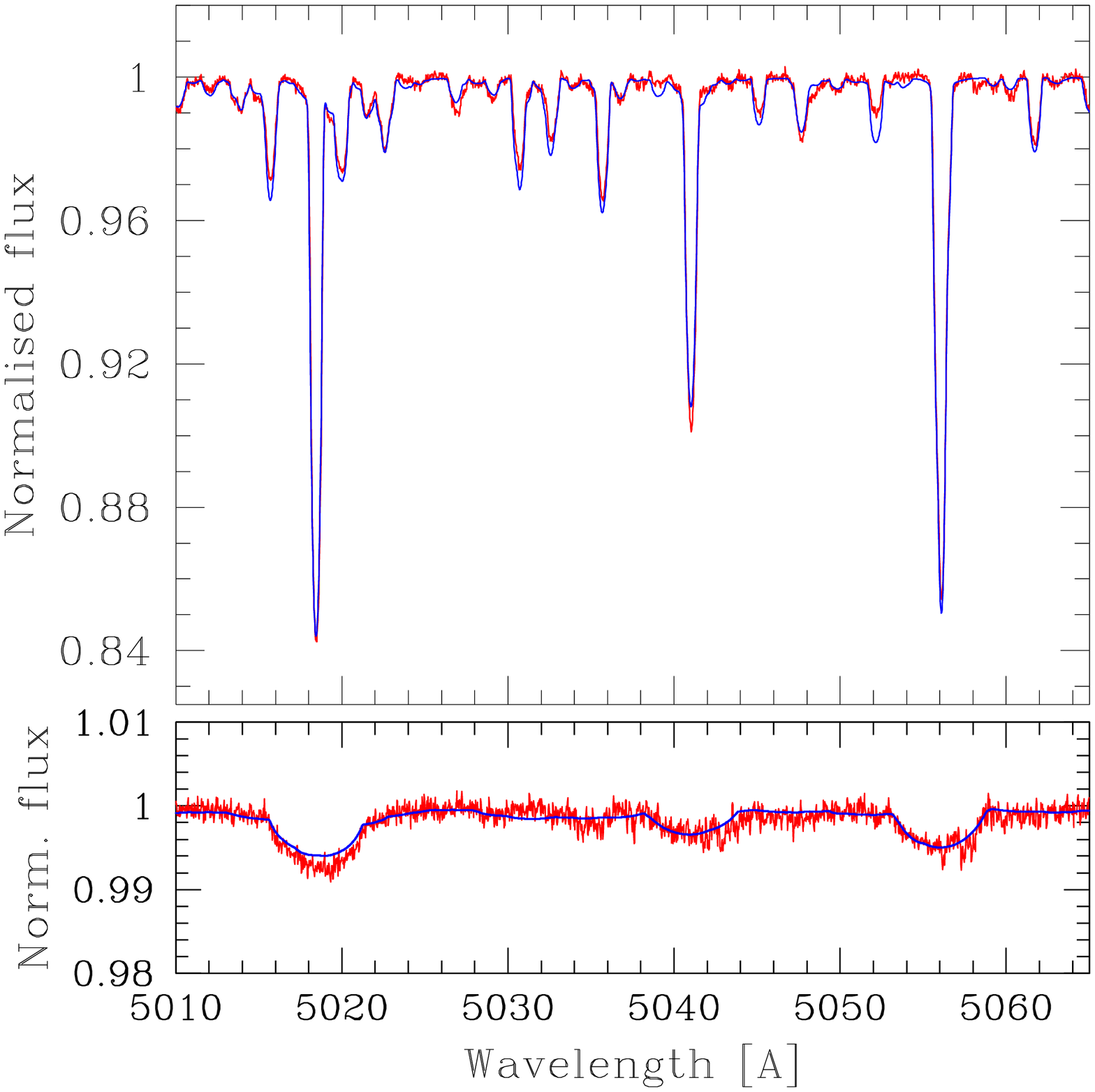} 
\caption{Quality of the fit (blue) to the disentangled spectra (red) of the primary (top panels) and secondary (bottom panels) in two metal line regions: $\lambda\lambda$ 4495 to 4570 {\AA} (left column) and $\lambda\lambda$ 5010 to 5065 {\AA} (right column). We note the difference in the Y-axis between the top and bottom panels owing to the small relative light contribution and rapid rotation of the secondary component.
}
\label{fig:gsspfitlines}
\end{figure*}

For the analysis of the disentangled spectra, we employ the Grid Search in Stellar Parameters \citep[GSSP;][]{Tkachenko2015} software package. The method is based on the comparison of a grid of synthetic spectra to the observed spectrum of the star and the quality of the fit is judged on the $\chi^2$ merit function. {\sc GSSP} allows for simultaneous optimisation of five atmospheric and line-broadening parameters, namely effective temperature $T_{\rm eff}$, surface gravity $\log\,g$, metallicity [M/H], microturbulent velocity $\xi$, and projected rotational velocity $v\,\sin\,i$ of the star. Optionally, the effect of light dilution on the disentangled spectrum can also be taken into account, by either assuming that the light dilution factor is wavelength independent \citep[as implemented in the {\sc GSSP\_single} module;][their Eq. (1)]{Tkachenko2015}, or by replacing it with the ratio of the radii of the two stars where the wavelength dependence of the light dilution effect is introduced via the continuum ratio of the binary components \citep[as implemented in the {\sc GSSP\_binary} module,][their Eq. (4)]{Tkachenko2015}. The latter approach is justified in the cases where binary components have significantly different effective temperatures, otherwise the assumption of the wavelength-independent light dilution factor is robust. Given that our spectroscopic and photometric analyses of the $\alpha$~Dra system  both show that the binary components have nearly identical effective temperatures, we limit ourselves to the analysis with the {\sc GSSP\_single} module where the light dilution factor is assumed to be independent of wavelength. 

The {\sc GSSP} package employs a grid of atmosphere models pre-computed with the {\sc LLmodels} program of \citet{Shulyak2004} \citep[for the summary of the grid properties, see the Table~1 of][]{Tkachenko2015}, while theoretical spectra are computed ``on-the-fly'' with the {\sc SynthV} \citep{Tsymbal1996} spectrum synthesis code. The $\chi^2$ merit function is computed for each pair of the observed-theoretical spectra in the grid, and 1$\sigma$ uncertainties are computed from $\chi^2$ statistics by projecting all $\chi^2$ values on the parameter in question. That way, we take into account possible correlations between the free parameters in our model.

The spectroscopic analysis is performed iteratively with the analysis of the TESS light curve (see Sect.~\ref{sect:tess}). We start with the analysis of the disentangled spectrum of the primary component only, and optimise the following six parameters: $T_{\rm eff}$, $\log\,g$, [M/H], $\xi$, $v\,\sin\,i$, and the light dilution factor. The results of this analysis are summarised in Table~\ref{tab:gssp_params} (column designated as ``Run 1'') and suggest the primary component is  a somewhat evolved A-type star with a close-to-solar metallicity. A light curve solution can then be obtained with the $T_{\rm eff}$ value of the primary component fixed to 10\,190~K, in turn delivering photometric values of the surface gravity and light dilution factor for both binary components.

In the next step of the spectroscopic analysis, we analyse the disentangled spectra of both binary components by fixing the $\log\,g$ values for both stars to their photometric values while keeping all other parameters free. It is important to note that we exclude broad Balmer lines from the analysis of the disentangled spectrum of the faint secondary component as those are found to be non-negligibly distorted in the disentangling process. This issue is not present in the disentangled spectrum of the primary component thanks to its dominance in the observed composite spectra of the $\alpha$~Dra system. The results of the aforementioned spectroscopic analysis are reported in Table~\ref{tab:gssp_params} (column designated as ``Run 2a''). One can see that the parameters of the primary component barely change compared to the previous iteration (columns ``Run 1'' vs. ``Run 2a''). Furthermore, we find a good agreement between the spectroscopic and photometric $T_{\rm eff}$ values for the secondary component. 

On the other hand, we obtain a slightly larger value for the light dilution factor of the secondary as compared to its photometric value, and the secondary appears to be slightly metal deficient compared to the primary component. Because we fit a region of the secondary's spectrum free of hydrogen lines, there is a significant degeneracy between the light dilution factor and the metallicity parameter of the star. Therefore, in the final iteration of the spectroscopic analysis, we decided to fix the light factors of both binary components to the values inferred from the light curve solution, in addition to the previously fixed values of $\log\,g$. Our finally accepted spectroscopic parameters of both binary components are listed in Table~\ref{tab:gssp_params} (column designated as ``Run 2b''), and suggest that, within the quoted 1$\sigma$ uncertainties, both stars have equal effective temperatures and metallicities. On the other hand, the primary component is found to be more evolved and to have substantially lower projected rotational velocity than its companion star. The quality of the fit of the disentangled spectra corresponding to the finally accepted atmospheric parameters is illustrated in Figs~\ref{fig:gsspBalmerlinesfit} and \ref{fig:gsspfitlines}.

Adopting the light ratio determined in the TESS light curve analysis ($\ell_{\rm B}/\ell_{\rm A} = 0.212\pm0.014$) in the final iteration of the optimal fit of disentangled spectra needs justification since the TESS passband covers the red, and NIR part of the spectrum, from about $\lambda\lambda$ 6000 - 10\,000 {\AA}, with the effective wavelength about $\lambda = 7865$ {\AA} \citep[the TESS passband is centered at the standard Cousinns I passband as given in][]{Ricker2014}. The optimal fitting of disentangled spectra are performed on  500 {\AA} long spectral segment centered at about 5250 {\AA}, and 350 {\AA} long spectral segment centered at about 4570 {\AA}. We did calculation of the light ratio as a function of the wavelength using the spectral energy distribution for both stars. It was found that changes from spectral region centered at the 5200 {\AA}, to the spectral range covered by the TESS passband is within assigned uncertainties, and barely about 1\%. Therefore, we conclude that our assumption on the 'constant' light ratio for the entire spectral range analysed is justified.


\begin{figure}[h!]
\includegraphics[width=9.5cm]{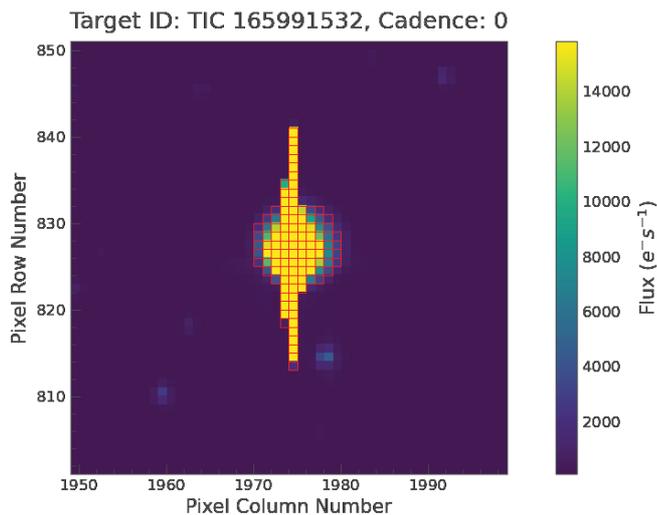}
\centering
\caption{
	The Target Pixel File (TPF) of $\alpha$ Dra in Sector 15. Our custom mask (red) and bright pixels of the target (yellow) are easily seen. The aperture to obtain the light curve that is represented with our custom mask (red) covers all the bright pixels of the target. To obtain the light curve, all pixels outside the mask that are background and nearby star light contamination are subtracted.
	}

\label{fig:tpf} 
\end{figure}

\begin{figure}[t!]
\centering
\includegraphics[width=9.cm]{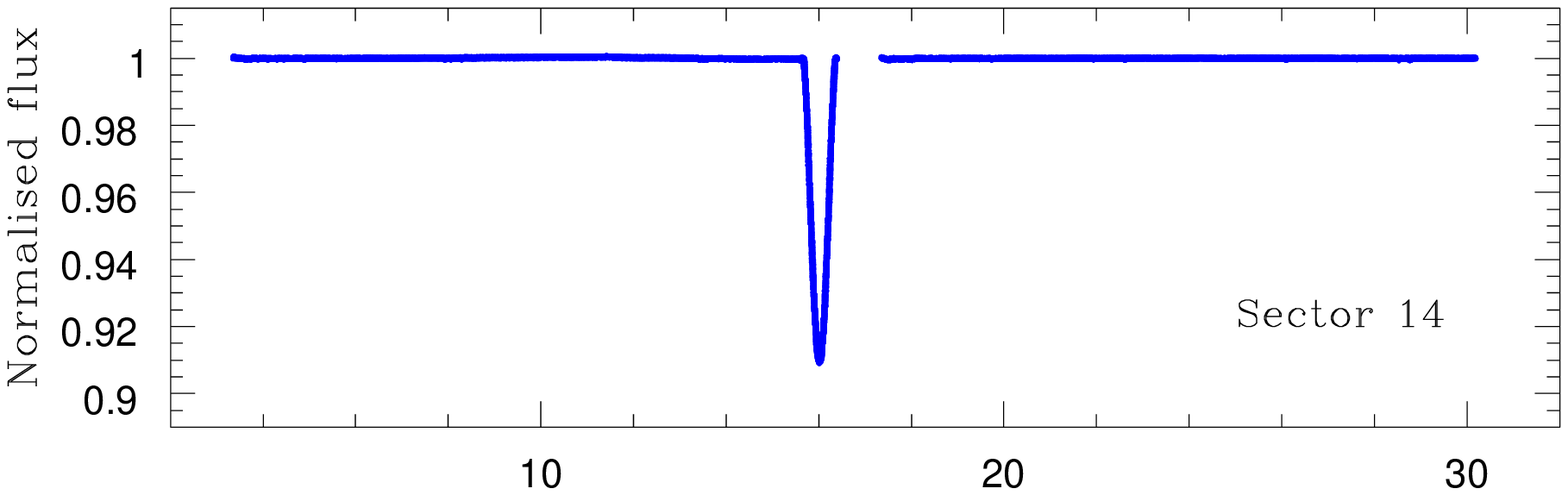} \\
\includegraphics[width=9.cm]{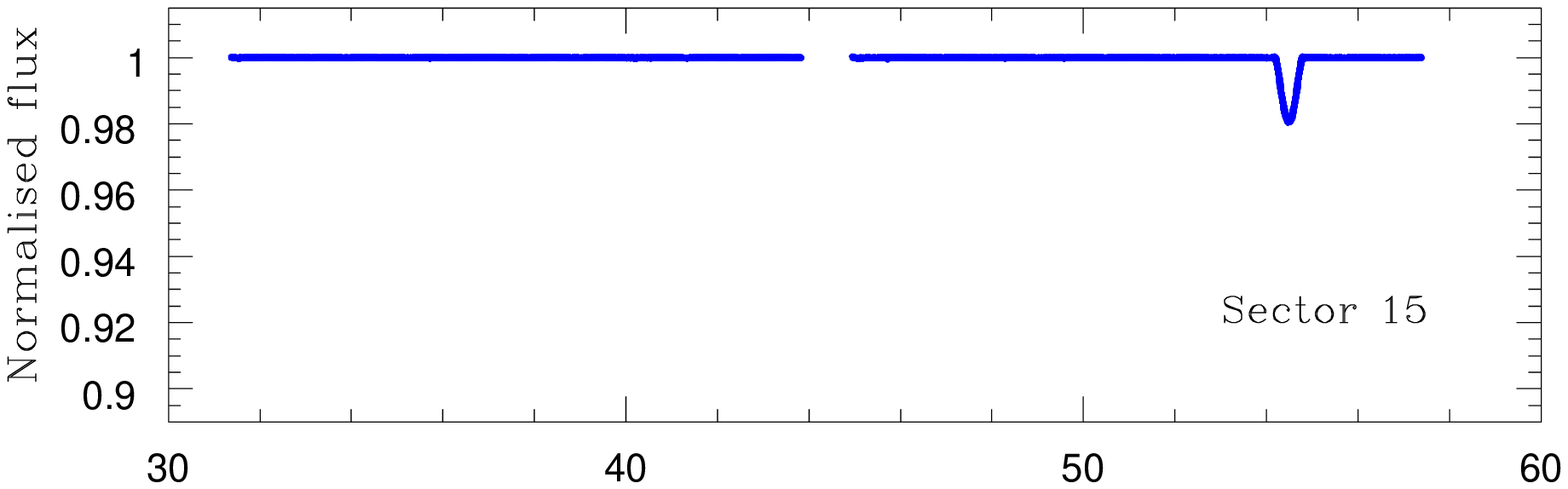} \\
\includegraphics[width=9.cm]{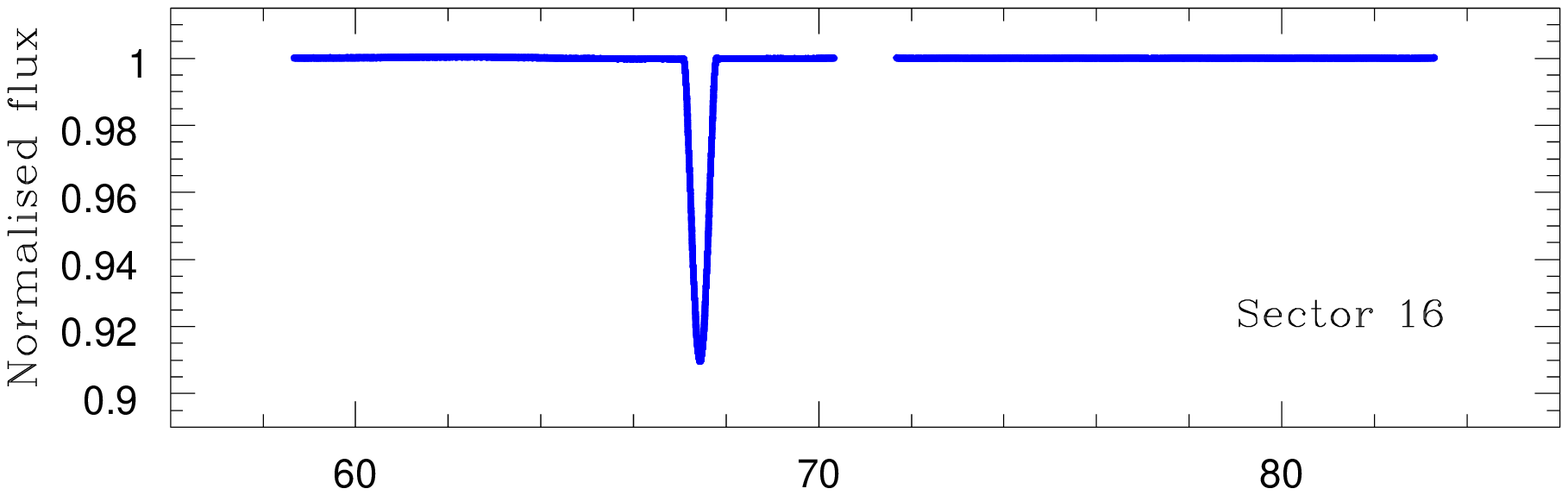} \\
\includegraphics[width=9.cm]{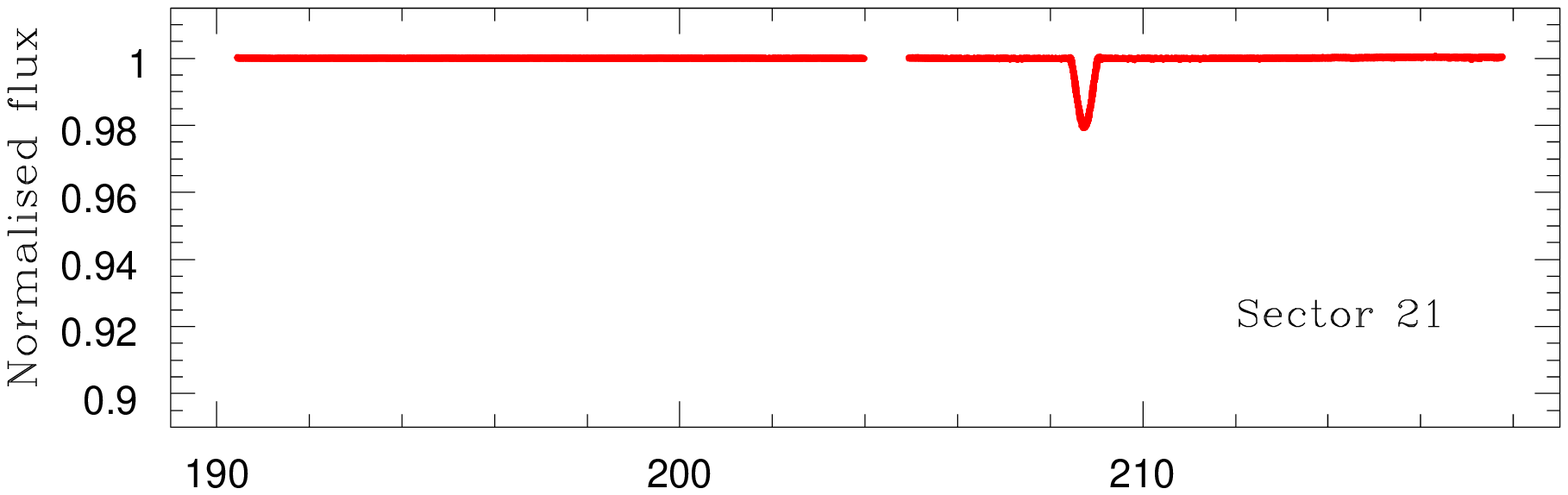} \\
\includegraphics[width=9.cm]{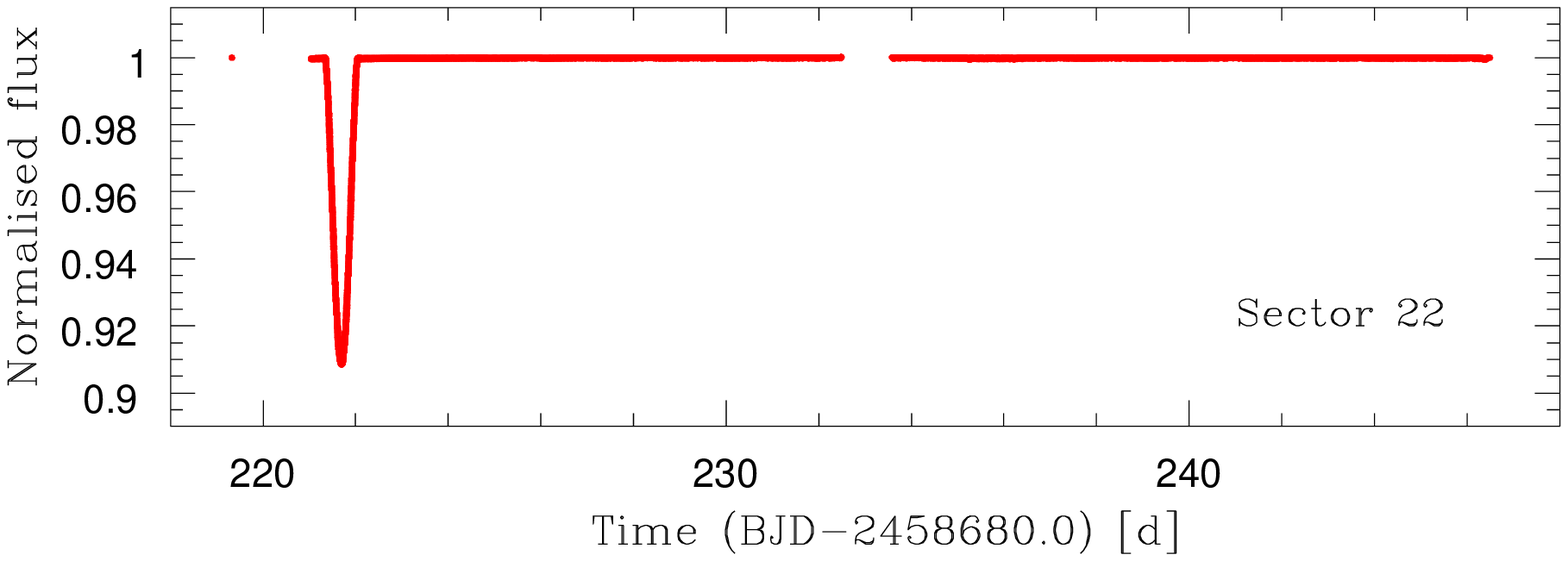}
\caption{TESS photometric measurements of $\alpha$\,Dra: normalised flux as a function of time, where $t=0$ is for BDJ 2\,458\,680.0 d. The light curve analysis presented in Sect.~\ref{sec:wd} was performed separately for the observations in Sectors 14, 15, and 16 (part 1, in blue), and Sectors 21, and 22 (part 2, in red).}
\label{fig:tess}
\end{figure}

\section{TESS space photometry}

\label{sect:tess}
\subsection{Observations}

$\alpha$~Dra (TIC 165991532) was observed by the Transiting Exoplanet Survey Satellite \citep[TESS,][]{Ricker2014,Ricker2015} at five different Sectors with a 2-min cadence.
At the time of this study, Sectors 14, 15, 16, 21 and 22 of the TESS data were available for $\alpha$~Dra. To extract the light curves, we used the Target Pixel Files (TPFs) from the Mikulski Archive for Space Telescopes (MAST) database. Because of the brightness of $\alpha$~Dra ($V = 3.68$ mag), the reduction methods utilised to construct the  Simple Aperture Photometry (SAP) or Pre-Search Data Conditioning Simple Aperture Photometry light curves from MAST \citep[PDCSAP,][]{Smithetal.2012} proved to be not ideal and led to anomalies in the extracted light curves. Instead, we employ the {\sc lightkurve} software package \citep{Lightkurve} to produce a target custom mask, and to estimate and subtract the background in each TESS sector. For that purpose, we utilise the median flux value computed across the TPF and the standard deviation estimated from the Median Absolute Deviation (MAD) of the data set. The MAD and the standard deviation are related through a scale factor $k$ that is taken to be 1.4826, assuming normally distributed data. All pixels whose flux values are found to exceed a certain threshold (computed as a number of standard deviations from the median flux value) are considered to belong to the target custom aperture while pixels with the flux values below the threshold are considered as background. Here, we find through a set of experiments the threshold value of nearly 30 percent of the maximum flux amplitude pixel to be the optimal one and consider all pixels with fluxes below (above) as background (target custom aperture). The background flux is estimated and subtracted per pixel over time which allows us to minimize the noise contribution and to remove light contamination from the nearby (background) stars. Figure~\ref{fig:tpf} shows an example of the TESS TPF and our custom aperture selection (red), used to obtain a light curve of $\alpha$~Dra from Sector~15 of the TESS data. The same method is used to extract light curves from other sectors of the available TESS data.

Altogether,  from the aforementioned five TESS Sectors, we obtained 91173 photometric measurements that cover three (two) primary (secondary) minima and out-of-eclipse phases of the $\alpha$\ Dra system.

\begin{figure*}
\includegraphics[width=8.6cm]{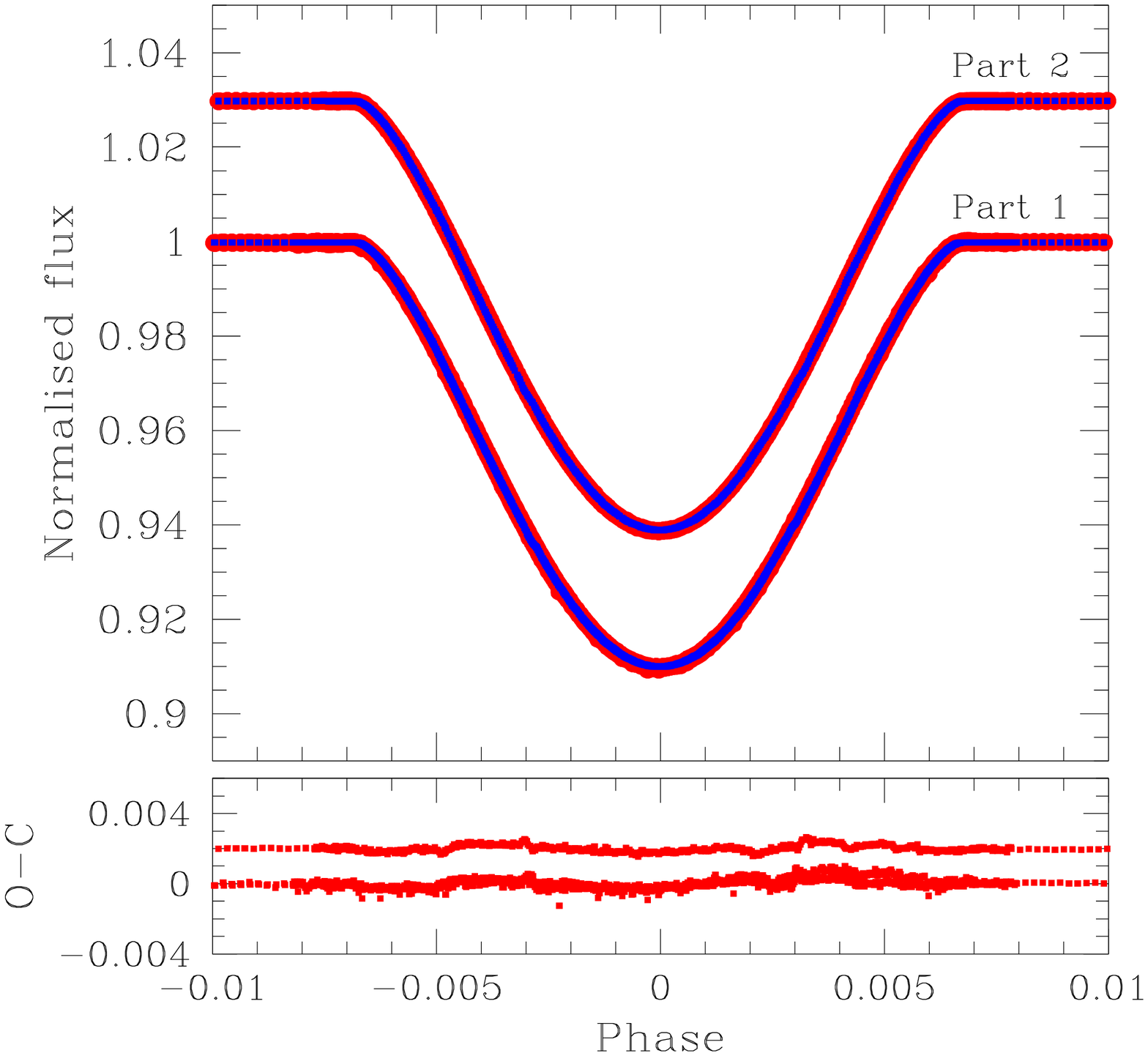}  \includegraphics[width=8.6cm]{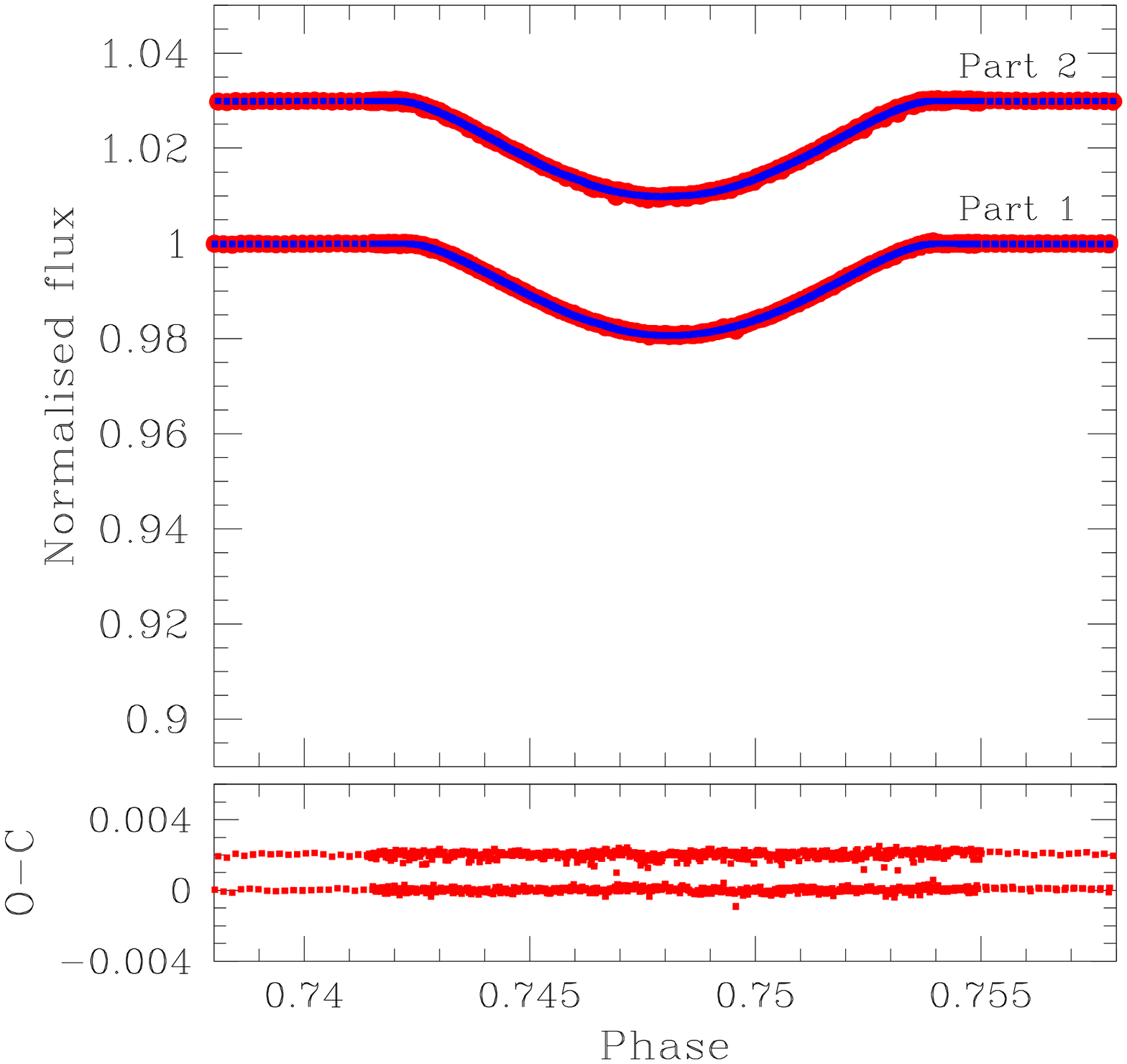}
\caption{Fits to the TESS light curve of $\alpha$~Dra for the primary (left) and secondary (right) minima. The normalised TESS measurements (upper panels) and O-C residuals (lower panels) are shown with red symbols while the best fit obtained with the {\sc wd} code is illustrated with blue solid line. See text for definition of the ``part 1'' and ``part 2'' light curves.    The phases are calculated from the ephemeris determined from the observed primary minima, $T_{\rm pr.~min} = (2458696.01948\pm0.00003)+(51.41891\pm0.00011)\,E$. The light curves and best fit, and the residuals, for the Part 2 are arbitrary shifted upward for better visibility.}
\label{fig:wdmodel} 
\end{figure*}

\begin{figure}[ht]
\includegraphics[width=8.6cm]{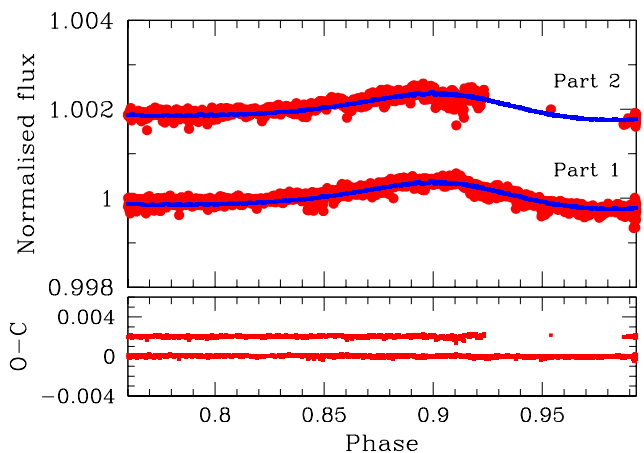} 
\caption{Same as Fig.~\ref{fig:wdmodel}, but in the portion of the light curve preceding the primary minimum. Brightening due to the proximity effect (reflection) is well below 1 millimag level, and is detected only thanks to the superb high-precision space photometry acquired by the TESS mission. The light curves and best fit for the Part 2 are arbitrary shifted upwards for better visibility.}
\label{fig:bump} 
\end{figure}

\subsection{The light curve analysis}
\label{sec:wd}
We performed the light curve analysis using the Wilson-Devinney ({\sc WD}) code, first formulated by \citet{Wilson_Devinney_1971} and subsequently generalised to eccentric binaries, also implementing complementary RVs analysis \citep{Wilson_1979}. The {\sc WD} code employs Roche equipotential surfaces to describe the shapes of stars forming a binary system and is commonly used for the analysis of various configurations of binary systems. In our analysis, we utilised the latest public version of the {\sc WD} code with designation {\sc WD2015} \citep{Wilson2014}, through a user-friendly designed {\sc gui} as well as a Python backend wrapper developed by \citet{Guzel2020}. 

We started to set up our initial binary model by taking into account the orbital and atmospheric parameters derived in previous sections and reported in Tables~\ref{tab:specorb} and \ref{tab:gssp_params}. According to the iterative process  described in Sec.~\ref{subsec:atmos}, we fixed the precisely determined temperature of the primary star ($T_{\rm eff,A}$), the mass ratio ($q$) and the projected separation of the orbit ($A \sin i$). Without prior knowledge of these parameters, it is impossible to break the degeneracy of the binary solutions which are only sensitive to the ratio of the effective temperatures and radii as well as the relative separation. An initial set of orbital parameters to be adjusted was prepared from the astrometric and spectroscopic orbital solutions (the period $P$, time of periastron passage $T_{\rm pp}$, eccentricity $e$, and an angle of periastron $\omega$), and the effective temperature of the secondary component  $T_{\rm eff,B}$. The surface gravities of both components $\log g_{\rm A,B}$ are also used for initial guesses of the surface  potentials of the components,  $\Omega_{\rm A, B}$. Since in light curve modelling only the ratio of the effective temperatures, and not temperatures themselves, can be determined, the effective temperature of the primary component $T_{\rm eff,A}$ was fixed. This component is dominant in the spectra of $\alpha$\,Dra, contributing the majority of the total flux of the binary system, and consequently its effective temperature is determined with high precision from the atmospheric analysis. In our initial runs, we optimised a total of nine free parameters: $P$, $T_{\rm pp}$, $e$, $\omega$, $T_{\rm eff,B}$, $\Omega_{\rm A}$, $\Omega_{\rm B}$, fractional light contribution of primary star $L_{\rm A}$, and inclination angle of the binary orbit $i$. Because {\sc WD2015} does not contain the TESS photometric bandpass, the {\sc Cousin} I$_c$ bandpass was used instead since its transmission is very close to that of TESS.  
 The linear limb-darkening law was used with the coefficients  interpolated from the tables of \citet{van_Hamme_1993} with a {\sc WD2015} built-in subroutine. The gravity-brightening coefficient 0.25 for radiative atmospheres was applied after seminal derivation by \citet{von_Zeipel_1924}.

\begin{table*}[t!]
\caption{The {\sc WD2015} best fit solutions for the TESS light curves of the $\alpha$~Dra system. The finally adopted solution is reported in the last column and represents the mean of the two individual solutions. See text for the definition of the ``Part 1'' and ``Part 2'' data sets.     }
\label{tab:wdsol}
\centering
\begin{tabular}{lcccc}
\hline\hline
Parameter & Unit &  Part 1 & Part 2 & Adopted  \\
    &      & Sectors 14,15,16  & Sectors 21,22 & solution \\
\hline
Orbital elements &  &  & & \\
\hline
Orbital period $P$ & [d]  &  $51.4198\pm0.0031$  & $51.4200\pm0.0032$ & $51.4199\pm0.0031$\\
Time of periastron $T_{\rm pp}$ & [d] & $2458753.107\pm0.019$  & $2458753.118\pm0.026$ & $2458753.112\pm0.023$ \\
Eccentricity $e$ &     & $0.4230\pm0.0013$ &  $ 0.4217\pm0.0026$  & $0.4223\pm0.0021$\\
Argument of periastron $\omega$ & [deg] &  $20.74\pm0.53$  & $20.27\pm0.78$ & $20.51\pm0.67$  \\
Orbital inclination $i$ & [deg]      &  $86.345\pm0.024$  & $ 86.358\pm0.026$ & $ 86.352\pm0.025 $  \\
\hline
Stars &  &  & \\
\hline
Surface potential of star A $\Omega_{\rm A}$ & & $22.22\pm0.14$ & $22.32\pm0.16 $ & $22.27\pm0.15$\\ 
Surface potential of star B $\Omega_{\rm B}$  & & $36.21\pm0.80$  &  $35.90\pm0.95$ &  $36.06\pm0.88 $\\ 
Fractional radius of star A $r_{\rm A}$  &     & $0.0479\pm0.0003$ & $0.0476\pm0.0004$ &  $0.0477\pm0.0003$     \\
Fractional radius of star B $r_{\rm B}$  &     & $0.0224\pm0.0005$  & $0.0226\pm0.0006$ & $0.0225\pm0.0005$ \\
Eff.\ temperature of star A $T_{\rm eff,A}$ & [K]  & $10\,225$ (fix)                &  $10\,225$ (fix) & 10\,225 (fix)\\ 
Eff.\ temperature of star B $T_{\rm eff,B}$  & [K] & $10\,031\pm248$   & $9\,837\pm375$ & $9930\pm328$   \\
Light of star A $L_{\rm A}$    &  & $101.61\pm1.44$    &  $100.61\pm1.78$  &  $101.11\pm1.62$  \\
Light of star B $L_{\rm B}$    &  & $21.57\pm1.24$    & $21.27\pm1.49$ & $21.42\pm1.37$  \\
Fractional light of star A $\ell_{\rm A}$  &   & $0.8249\pm0.0172$ & $0.8255\pm0.0214$  & $0.8252\pm0.0194$ \\
Fractional light of star B $\ell_{\rm B}$   &  & $0.1751\pm0.0104$  & $0.1745\pm0.0126$ & $0.1748\pm0.0115$  \\
Light ratio $\ell_{\rm B}/\ell_{\rm A}$  & & $0.2123\pm0.0125$   &  $0.2114\pm0.0152$  & $0.2118\pm0.0139$  \\
\hline
\end{tabular}
\end{table*}

$\alpha$\,Dra is a highly eccentric system ($e=0.423$) with a long orbital period resulting in very short eclipses which cover only about 2.5\,\% of the orbital cycle. In such cases, outside-of-eclipse measurements would have too large a weight in the light curve modelling. To suppress this unwanted effect, the TESS photometric observations were resampled; only each 10th measurement in the out-of-eclipses portions of the light curves is stored whilst all the observations in the eclipses are used. This reduces significantly the number of observations that are employed in the light curve modelling. In total, for all five sectors of the TESS observations of $\alpha$\,Dra (Fig.~\ref{fig:tess}), 15187 measurements survived our selection which constitute some 16.5\% of the initially extracted photometric measurements. We also increased the surface resolution to 60$\times$60 (i.e. N1 and N2 parameters of {\sc WD2015}) to make sure the required level of precision is achieved for modelling small-amplitude variations outside the eclipses (see below).

In the initial analysis, we detect sub-millimag level discrepancies in the descending branches of the secondary eclipses. Since the ascending and descending branches of the primary eclipses are in perfect agreement except for the depth in the last observed (third) primary eclipse, which is about 1 millimag deeper than in the previous two, we concluded that the orbital period is determined correctly. At present, it is not clear if the effect seen in the secondary eclipses is real or rather associated with some instrumental and/or data processing effects. To avoid any mix-up, we decided to perform the light curve analysis based on two ``observing parts'' separately, so that the results can ultimately be examined for consistency. This way, the nearly continuous observations in Sectors 14, 15, and 16 comprise the ``Part 1'' data set (9042 measurements) while the observations in Sectors 21 and 22 comprise the ``Part 2'' data set (6145 measurements).  This division is also indicated in Fig.~\ref{fig:tess} which shows raw TESS observations for all 5 sectors separately.

$\alpha$~Dra is a partially eclipsing system as is obvious from the characteristic V-shape of its light curve minima. Therefore, degeneracies and correlations of the parameters in a multidimensional parameter space are expected \citep{Conroy2020}. To explore the parameter space, as well as to obtain error estimates, many researchers utilize Markov-Chain Monte-Carlo  ({\sc MCMC}; \citealt{Goodman2010}) based optimisers (e.g. \citealt{Lehmann2020},  \citealt{Dervisoglu2018}). For the purpose of optimising the aforementioned nine free parameters, we developed a {\sc Python} framework ({\sc LC$_{\rm MCMC}$}) which uses PyWD2015 \citep{Guzel2020} as a {\sc WD} backend and the widely used {\sc emcee} package \citep{Emcee2013} as an ensemble sampler.

To obtain a good fit, we sampled parameter priors with a normal distribution around 10\,\% sigma. To guarantee the convergence, we used the integrated auto-correlation time module of {\sc emcee} to assure that enough number  iteration have been achieved.  Our final  {\sc MCMC} solution yielded approximately 200\,000 iterations per each part of the photometric data through 64 walkers.

In Table\,\ref{tab:wdsol}, we report the results of the posterior probability distributions of the optimised parameters. We calculated the relative radii $r_{\rm A,B}$, surface gravities $\log g_{\rm A,B}$, and the light ratio $\ell_{\rm B}/\ell_{\rm A}$, accordingly. The reported uncertainties and mean values were calculated from the 16th, 50th, and 84th percentiles of each distribution. For the finally adopted values, we assumed symmetric errors which is also the core assumption of the normal distributions (the last column in Table\,\ref{tab:wdsol}).

The best fit models are shown for the primary and secondary minima in both ``Part 1'' and ``Part 2'' data sets in Fig.\,\ref{fig:wdmodel}.  We also illustrate the accuracy of our models which successfully fitted the 1 millimag level of the out-of-eclipse brightening as shown in  Fig.\,\ref{fig:bump}. The corner plots of the samples in the marginalised posterior distributions of each part of the photometric data are also shown in Appendix (Figs.~\ref{fig:corner1} and \ref{fig:corner2}). The obvious correlation between the optimised parameters demonstrates the importance of adopting stochastic methods such as Bayesian inference for the multi-parameter model estimations.

It is evident from Figs.~\ref{fig:wdmodel} and \ref{fig:bump} that the decision to analyse the TESS light curves from the Sectors 14, 15, and 16 (``Part 1''), and Sectors 21 and 22 (``Part 2'') separately is fully justified. There are no  systematics in the residuals between the best fitting calculated light curves and observations in the secondary minima whilst the residuals are in sub-millimag level (right panel in Fig.~\ref{fig:wdmodel}). The same is true for the proximity (reflection) light excess at the orbital phases from 0.80 to 0.99 (Fig.~\ref{fig:bump}). The residuals from fitting 2 (1) primary minima in ``Part 1'' (``Part 2'') are somewhat larger, yet the amplitude does not exceed 1 millimag level (see left panel in Fig.~\ref{fig:wdmodel}). Comparison between the ``Part 1'' and ``Part 2'' light curve solutions reveals a high degree of consistency within the quoted 1$\sigma$  uncertainties for all inferred parameters (see Table~\ref{tab:wdsol}).

Orbital parameters inferred from interferometric, spectroscopic, and photometric data are listed in Tables~\ref{tab:astro_orbit}, \ref{tab:specorb}, and \ref{tab:wdsol}, respectively. Four parameters are in common between these determinations: the orbital period $P$, time of periastron passage $T_{\rm pp}$, orbital eccentricity $e$, and argument of periastron $\omega$. The former three parameters are found to be in good agreement within the quoted 1$\sigma$ uncertainties, while $\omega$ inferred from the interferometric data is 2$\sigma$ away from the respective values deduced from the spectroscopic and photometric data. Furthermore, the orbital inclination obtained in the astrometric analysis is also 2$\sigma$ away from the value inferred from photometry. However, we note that the astrometric value has an order of magnitude larger uncertainty. For that reason, we  use the orbital inclination value of $i=86^{\circ}.352\pm0^{\circ}.025$ as inferred from the light curve solution for the final determination of physical properties of the stars as well as for the dynamical parallax of the system.


\begin{figure*}[h!]
\includegraphics[width=9.0cm]{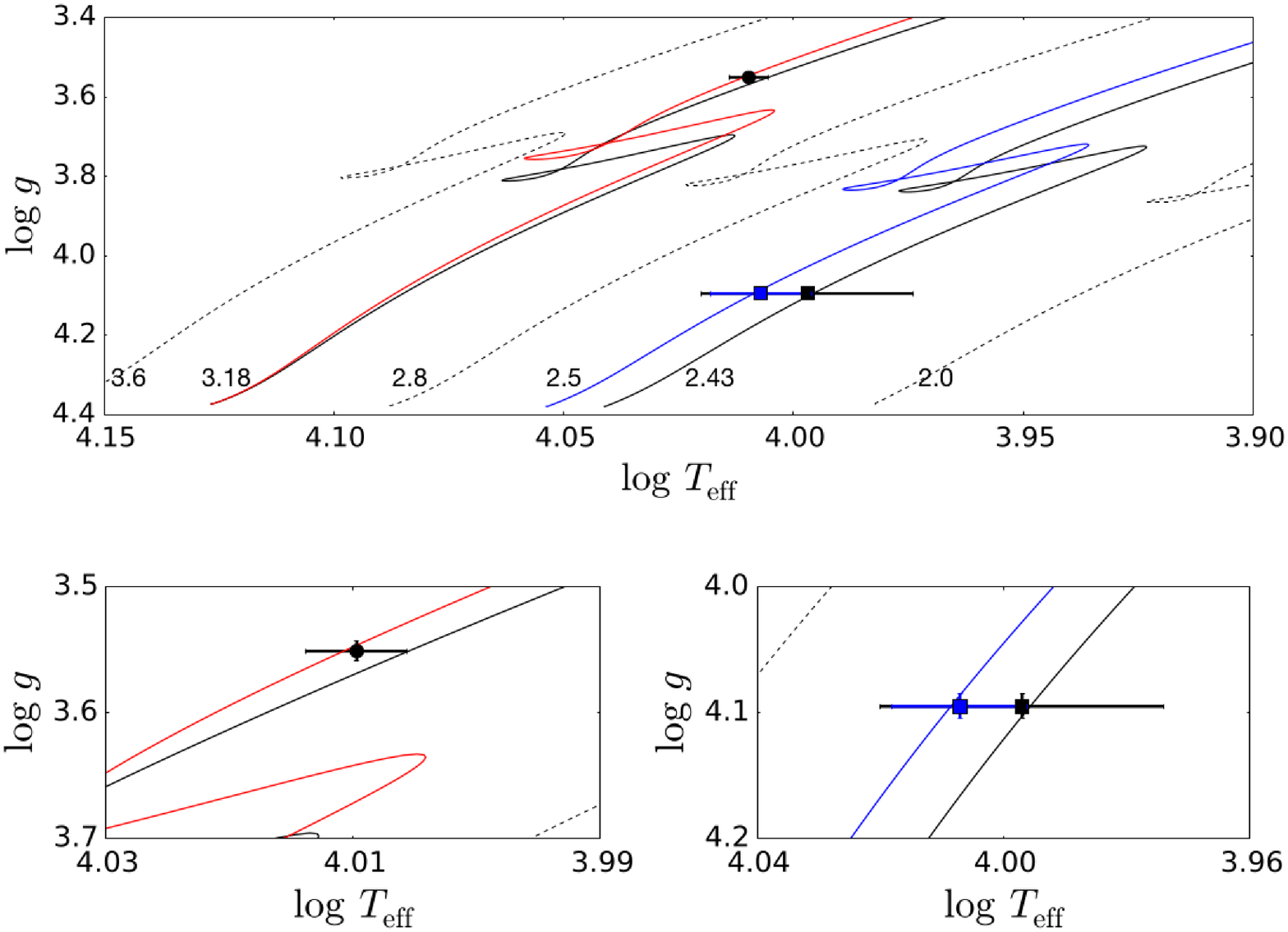}
\includegraphics[width=9.0cm]{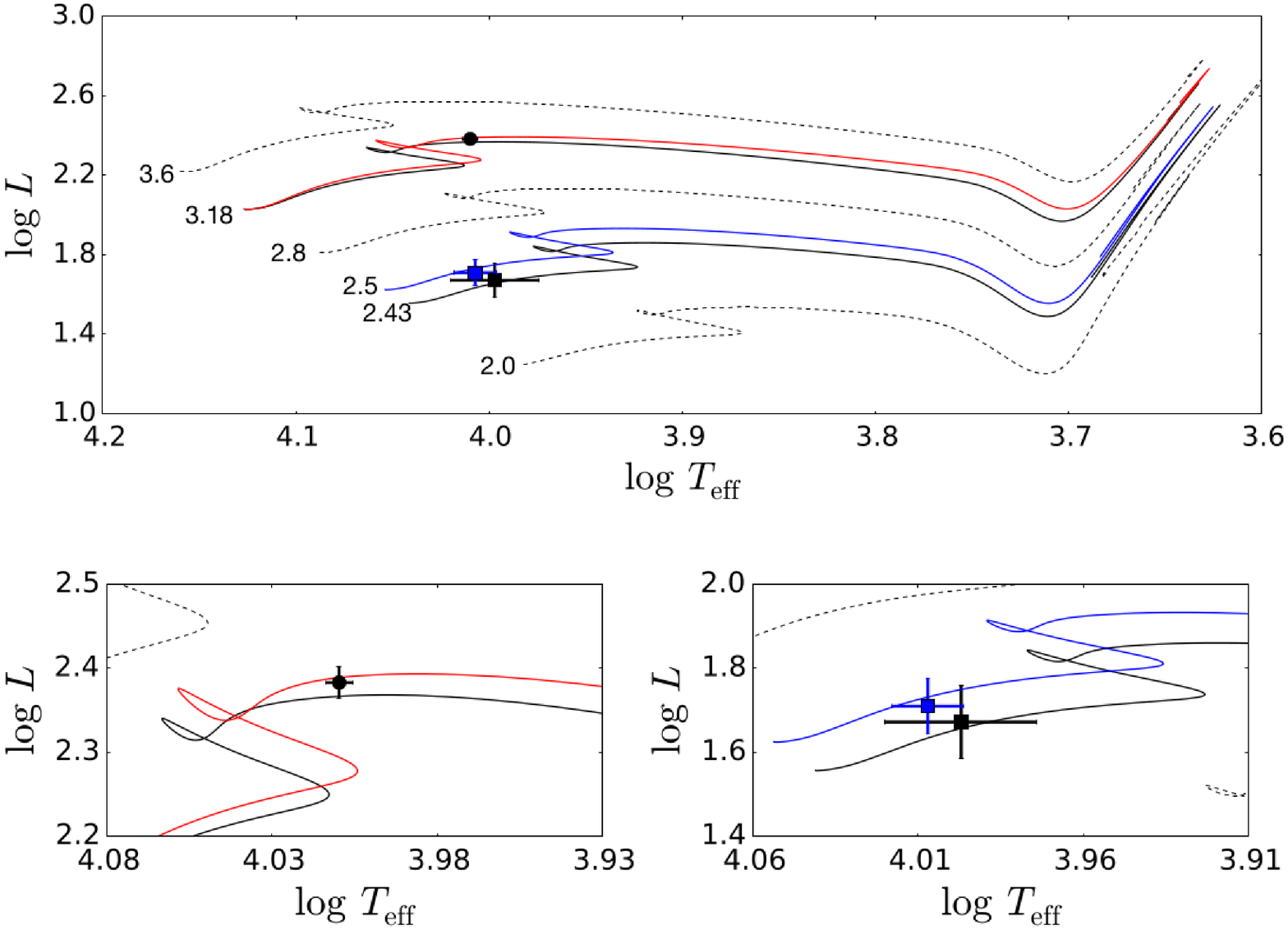}
\caption{{\bf Top:} positions of the primary (circle) and secondary (squares) components of $\alpha$~Dra in the Kiel (left column) and HRD (right column) diagrams. The black/blue square indicates the position of the secondary component corresponding to its photometric/spectroscopic value of $T_{\rm eff}$. {\sc MESA} evolutionary tracks are shown with lines, the exact values of the stellar mass (in M$_{\odot}$ units) are indicated in the plot. Black solid lines correspond to the dynamical masses of the stars and are computed for the overshooting parameter $f_{\rm ov}$=0.005 $H_{\rm p}$. The blue track is the best fit to the spectroscopic $T_{\rm eff}$ of the secondary; the red line indicates a {\sc MESA} evolutionary track corresponding to the dynamical mass of the primary component and computed with $f_{\rm ov}$=0.010 $H_{\rm p}$. {\bf Bottom:} close ups for the primary (left) and secondary (right) components of the binary systems.}
\label{fig:SSE} 
\end{figure*}

\section{Evolutionary models}\label{sec:SSE}
The combined spectroscopic, photometric, and astrometric analysis described in the previous sections allows us to put both components of the $\alpha$~Dra system in the $\log\,T_{\rm eff}$-$\log\,g$ Kiel diagram and assess its evolutionary status. We closely followed the procedure outlined in \citet{Tkachenko_2020} that makes use of a grid of {\sc mesa} stellar structure and evolution models \citep{Paxton2011,Paxton2013,Paxton2015,Paxton2018,Paxton2019} as presented in \citet{Johnston2019}. The grid employs the Ledoux criterion for convection and assumes a chemical mixture after \citet{Przybilla2008} and \citet{Nieva2012} (also known as the cosmic abundance standard). We set the initial helium and hydrogen fractions to Y = 0.276 and X = 0.71, respectively, and the mixing-length parameter $\alpha_{\rm mlt}$ to the solar calibrated value of 1.8. In addition, the grid employs a mixing profile in the radiative envelope as determined by \citet{Rogers2017} from 2D hydrodynamical simulations of internal gravity waves and implemented in {\sc mesa} by \citet{Pedersen2018}. Overall, the grid of \citet{Johnston2019} is optimised for intermediate-mass stars born with a convective core which makes it an excellent tool for the evolutionary analysis of the $\alpha$~Dra system.

\begin{table}
\caption{Stellar properties of the components in the binary system $\alpha$~Dra
determined from the combined spectroscopic and light curve analysis, and evolutionary calculations.}
\label{tab:abspar}
\centering
\begin{tabular}{lccc}
\hline\hline
Parameter & Unit & Primary & Secondary \\
\hline
$M$           & [M$_\odot$]     & $3.186\pm0.044$    &  $2.431\pm0.019$  \\
$R$           & [R$_\odot$]     & $4.932\pm0.036$    &  $2.326\pm0.052$  \\
$\log g$      & [dex]           & $3.555\pm0.006$    & $4.090\pm0.019$   \\
$T_{\rm eff}$ & [K]             & $10\,225\pm100$    & $9\,930\pm330$    \\
$\log L$      & [L$_\odot$]     & $2.380\pm0.018$    &  $1.677\pm0.059$  \\
$M_{\rm bol}$ & [mag]           & $-1.200\pm0.045$   & $0.557\pm0.147$   \\
$v\,\sin\, i$ &  [km\,s$^{-1}$] & $25.4\pm0.9$       & $168\pm11$        \\
Age           & [Myr]           & 280$\pm$10         & 345$\pm$ 25       \\
$M_{cc}$      & [M$_\odot$]     & 0.01$\pm$0.04      & 0.337$\pm$0.011   \\
X$_{\rm c}$ & & 0.00$\pm$0.005 & 0.41$\pm$0.03\\
$f_{\rm ov}$  &                 & 0.007$\pm$0.002    & 0.005$^*$         \\
\hline
\end{tabular}
\tablefoot{$^*$ secondary component is not sufficiently evolved to deduce its overshooting parameter from the Kiel or HRD diagram, hence $f_{\rm ov}$ is fixed to 0.005~$H_{\rm p}$.}
\end{table}

Figure~\ref{fig:SSE} shows the results of fitting {\sc mesa} evolutionary tracks to the positions of both binary components in the Kiel (left column) and HR (right column) diagrams. The positions of both components (black circle and square for the primary and secondary, respectively) are well reproduced within 1$\sigma$ uncertainties of the measured effective temperatures, surface gravities, and luminosities of the stars (see Table~\ref{tab:abspar} for numerical values) and assuming their dynamical masses (solid black tracks in Fig.~\ref{fig:SSE}). For the secondary component, we also considered a second scenario corresponding to its spectroscopic value of $T_{\rm eff}$ (cf. Table~\ref{tab:gssp_params}, column ``2b'' for numerical value and see blue square in Fig.~\ref{fig:SSE}). The best fit solution (solid blue track in Fig.~\ref{fig:SSE}) is obtained for the stellar mass that is some 3\% larger than the inferred dynamical mass of the star. Yet, owing to the large spectroscopic $\Delta T_{\rm eff}$ of 250~K, the corresponding position of the secondary in the Kiel/HR diagram is also consistent within 1$\sigma$ uncertainty of the dynamical mass track of the star. For the primary component, we find that a small amount of convective core overshooting cannot be excluded (solid red track in Fig.~\ref{fig:SSE}) although the $f_{\rm ov}$=0.005 track (corresponding to the lower boundary in the employed grid of {\sc mesa} models) is also consistent with the stellar position in the Kiel and HR diagrams.

In Table~\ref{tab:abspar}, we list the final accepted parameters (age, mass of the convective core, and overshooting parameter) inferred from our evolutionary analysis. We note that the parameters are based on the effective temperature and luminosity of the secondary as inferred from the TESS light curve. As discussed in detail in Sect.~\ref{subsec:atmos}, owing to a small contribution of the secondary component to the composite spectrum of the binary system, its disentangled spectrum is significantly more uncertain than that of the primary component. In particular, we found that the Balmer lines were considerably affected in the disentangling process and hence could not be used in the spectroscopic analysis. That being said, the quality of the spectroscopic information that we have at our disposal for the secondary component is inferior to the quality of the TESS light curve of the system, hence we put more trust in the photometrically inferred $\log T_{\rm eff}$\,/\,$\log\ L$ pair for the secondary component. According to our findings, the primary component is an evolved post-TAMS\footnote{Terminal Age Main Sequence} A-type star that has a tiny convective core ($M_{cc}=0.01\pm0.04$). On the other hand, the secondary component is a relatively unevolved star whose convective core mass constitutes some 14\% of the stellar mass. The age of the system is estimated by us to be about 300~Myr, with the individual ages being 280$\pm$10 and 345$\pm$25~Myr for the primary and secondary component, respectively. We note that the individual ages are in a good agreement with each other within 2$\sigma$ uncertainties, and the difference between them reduces to within 1$\sigma$ when the spectroscopic $T_{\rm eff}$ value of 10165$\pm$250~K is assumed for the secondary (age and the convective core mass of the star are respectively 315$\pm$22~Myr and $M_{cc}=0.354\pm0.010$~M$_{\odot}$ in that case).

Another test could be made to verify consistency between the light ratio determined from the observables ($\ell_{\rm B}/\ell_{\rm A} = 0.212\pm0.014$) in the TESS passband, and the light ratio inferred from the evolutionary tracks for determined dynamical masses of the components. The {\sc mist} interactive tool \citep{Dotter_2016, Choi_2016, Paxton2011, Paxton2013, Paxton2015, Paxton2018, Paxton2019} provided the synthetic photometry for the {\sc mesa} models for the various photometric passbands. First we compared the evolutionary tracks for the components from our grid, and the {\sc mist} calculations, and found an ecellent matches. Then for the TESS passband the calculations is giving the light ratio $({\ell_{\rm B}/\ell_{\rm A}})_{\rm MESA} = 0.221\pm0.015$ which is within 1-$\sigma$ uncertainty to the light ratio determined in the light curve analysis.

In this context, we should mention that the calculation of the light ratio in B, and V, passbands (the spectral range in which our spectroscopic analysis was performed) corroborate our previous assumption that changes with the wavelength is small, and within the uncertainties of our analysis:  $(\ell_{\rm B}/\ell_{\rm V})_{\rm B} = 0.213\pm0.018$, and
$(\ell_{\rm B}/\ell_{\rm V})_{\rm V} = 0.221\pm0.022$.

\section{Distance to $\alpha$~Dra}
\label{sec:parallax}

Prior to the launch of astrometric space missions, the trigonometric parallax for $\alpha$\,Dra was very uncertain. The entry for its parallax in the ``The General Catalogue of trigonometric [stellar] parallaxes'' \citep{van_Altena_1995} reads $\pi = 14.8\pm7.5$ mas. On the other hand, the Hipparcos parallax from the first 1997 reduction \citep{Perryman_1997} is $\pi_{\rm HIP} = 10.56\pm0.52$~mas and provides a considerable improvement over the previously published measurement. Moreover, the improved Hipparcos reduction from 2007 results in a slight increase of the absolute value of the parallax for $\alpha$\,Dra and in a significant increase in the precision of the measurement: $\pi_{\rm HIP} = 10.76\pm0.17$~mas \citep{van_Leeuwen_2007}. Finally, the {\sl Gaia} DR2 \citep{gaia_2018} reported trigonometric parallax of $\pi_{ Gaia} = 12.18\pm0.32$~mas  is somewhat larger than the 2007 Hipparcos value putting $\alpha$\,Dra at a shorter distance ($d_{Gaia} = 82.1\pm2.1$ pc).

\begin{figure}
    \centering
  \includegraphics[width=8.6cm]{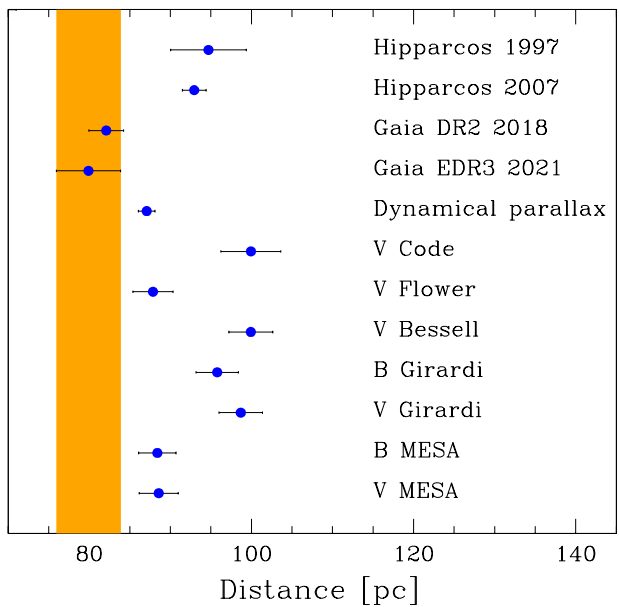}  
    \caption{Dynamical parallax of the binary system $\alpha$~Dra determined in this work compared to the trigonometric parallaxes measured by dedicated space astrometric missions, Hipparcos, and {\sl Gaia}, respectively. Distance determinations based on the geometric and radiative properties of the stars in $\alpha$~Dra using different bolometric corrections are shown, too. The latest {\sl Gaia} distance from the {\it Gaia} EDR3 is shown in coloured strip for easy comparison. 
    }
    \label{fig:gaia}
\end{figure}

The {\it Gaia} EDR3 parallax for $\alpha$ Dra is $\pi_{\rm EDR3} =12.516\pm0.203$ mas \citep{Gaia_Collaboration_2021}, slightly larger than quoted in the {\it Gaia} DR2. The parallax uncertainty may be underestimated by $\approx 30\%$ as described in \S6.3.1 of \citet{Gaia_Collaboration_2021}. The distance derived from the EDR3 parallax and published uncertainty is $d_{\rm EDR3} = 79.9 \pm 1.3$ pc, and considering the increased uncertainty it is $d_{GAIA} = 79.9^{+4.1}_{-3.7}$ pc. As seen in Fig.~\ref{fig:gaia} this distance places $\alpha$ Dra closer than other distance determinations. The solution in EDR3 is a 5-parameter solution, and the {\it Gaia} does not yet provide solutions for binary stars. There are hints within the statistics that $\alpha$ Dra might be resolved. The renormalized unit weight error {\tt ruwe} is 2.4 and this is greater than the 1.4 level that may indicate a resolved double. However, the related statistic 
{\tt ipd\_gof\_harmonic\_amplitude} is 0.056 which is less than the 0.1 that is associated with resolved doubles \citep{Gaia_Collaboration_2021}. If $\alpha$ Dra were resolved by {\it Gaia} the PSF should be elongated along the position angle of the binary at epoch 2016.0, the epoch of EDR3 \citep{Gaia_Collaboration_2021, Fabricius_2021}. Statistic {\tt ipd\_gof\_harmonic\_phase} is 108$^{\circ}$.091. For the orbital elements in Table 2 the predicted ephemeris gives a position angle of 252$^{\circ}$.76 and a separation of 7.4 mas. As the phase of the PSF elongation is modulo 180$^{\circ}$ 
the rotated ephemeris position angle to consider is 72$^{\circ}$.76. This is offset from the {\tt ipd\_gof\_harmonic\_amplitude} by 31$^{\circ}$. Until a binary solution from {\it Gaia} is provided it is speculative to associate this 31$^{\circ}$ offset for a 7 mas binary with the flux ratio determined in this work. In summary, there are two statistics suggestive of a resolved binary: 
{\tt ruwe} and {\tt ipd\_gof\_harmonic\_phase}. Although, it is not clear that the orientation of the phase supports agreement with the position angle of the binary at this time. The {\tt ipd\_gof\_harmonic\_amplitude}, which should be considered with the {\tt ruwe}, does not suggest a resolved binary. We eagerly await future releases from {\it Gaia} that include results for resolved doubles. 

Our complementary observations of the $\alpha$\,Dra system enable direct determination of its distance through the measurements of the angular semi-major axis of the orbit $a''$ (see Table~\ref{tab:astro_orbit}) and the semi-major axis of the spectroscopic orbit (see Table~\ref{tab:specorb}). With $A\,\sin i=103.37\pm0.49$~R$_\odot$, $a'' = 5.52\pm0.06$~mas, and $i = 86.35\pm0.03$ deg inferred from interferometry, spectroscopy, and combined spectroscopic and photometric analysis, respectively, we obtain a distance to the $\alpha$\,Dra binary system of $d = 87.07\pm1.03$~pc, or expressed as the dynamical parallax $\pi_{\rm dyn} = 11.48\pm0.13$~mas. Figure~\ref{fig:gaia} provides a graphical representation of the comparison of the distance derived in this work with the distances inferred from the Hipparcos and {\sl Gaia} parallaxes.

The distance to a binary system can also be derived from stellar luminosities and the measured apparent brightness, provided the fractional light contribution of each component to the total light of the binary system is known. Since the individual radii, effective temperatures and fractional light contributions of the components are determined in the present analysis, we can also provide alternative distance estimates to the system. \citet{Oja1993} measured the $BV$ brightness of $\alpha$~Dra to be $B = 3.640\pm0.007$~mag and $V = 3.680\pm0.009$~mag. Absolute bolometric magnitudes, $M_{\rm bol}$, also given in Table\,\ref{tab:abspar}, are transformed to absolute visual magnitudes, $M_{\rm V}$, hence the bolometric correction BC$_{V}$ are applied. Here, we use the bolometric corrections from four different sources: empirical tabulation from \citet{Code_1976}, and \citet{Flower_1996}\footnote{See corrections given in \citet{Torres_corr_2010}}, and  theoretical estimates from model atmospheres for given photometric bandpasses by \citet{Bessell_1998} and \citet{Girardi_2002}. We use an estimate of the colour excess for $\alpha$\,Dra $E(B-V) = 0.025$ from \citet{Zorec_2009} who compute it as an average from several independent determinations with different methods. In \citet{Kervella_2019}, a somewhat lower colour excess of $E(B-V) = 0.007$ is reported using an interactive code {\sc stilsm} which implements 3D maps of the local ISM \citep{Lallement_2014, Lallement_2018}. Our calculations are performed with the code {\sc jktabsdim} \citep{Southworth_2005} where in addition to the parameters listed above we also assume interstellar absorption. Taking interstellar attenuation of light into account  resulted in shorter distances compared to the case when the effect is ignored. 

Calculations of the distance to $\alpha$\,Dra using the stellar properties of its components determined in the present work eventually give two groups of the results, clustered at a ``short'' distance of $d \sim 90$ pc, and at a ``long'' distance of $d \sim 100$ pc. None of our estimates are in agreement with the {\sc gaia} parallax-based distance measurement, while some overlap within 1-$\sigma$ uncertainties is obtained with the Hipparcos parallax-based distance determinations (see Fig.~\ref{fig:gaia}). We note that among the different calibrations of the bolometric corrections, the empirical tabulation given by \citet{Flower_1996} gives the distance $d_{\rm FLO,V} = 87.8\pm2.5$~pc which is in a good agreement within the quoted 1-$\sigma$ uncertainty with the distance measured from the dynamical parallax in this work. We also note that the {\sl Gaia} parallax measurement did not take into account the fact that $\alpha$~Dra is a binary system, hence the corresponding distance estimate is expected to be uncertain.

A group of the ``long'' distance determinations comprises distances determined from the theoretical bolometric corrections: those of \citet{Girardi_2002} being close to distance of 100~pc,  $d_{\rm GIR,B} = 95.8\pm2.6$~pc, and  $d_{\rm GIR,V} = 98.7\pm2.7$~pc, while application of the theoretical bolometric corrections from \citet{Bessell_1998} gives the largest distance of all, $d_{\rm BES,V} = 99.9\pm2.7$~pc. The largest distance came for the empirical calibration given by \citet{Code_1976}, $d_{\rm COD, V} = 99.8\pm3.7$ pc, also with the largest uncertainty.  Distances quoted are for the calculations of the binary system as a whole. With a prevailing flux coming from the primary component, less precisely determined physical properties of the secondary component have no decisive role in the error budget in distance determinations.

The {\sc mist} interactive tool using the {\sc mesa} models we use in this work also provide synthetic photometry with built-in bolometric corrections. These enables calculations of the absolute magnitudes
in a given photometric passband. The absolute magnitudes in B, and V, passbands calculated for the MESA models with the physical quantities given in Table~\ref{tab:abspar} are $M_{\rm B} = -1.195$ mag, and $M_{\rm V} = -1.134$ mag. Using the photometry for $\alpha$~Dra from \citet{Oja1993}, and $E(B-V) = 0.075$ from \citet{Zorec_2009}, the inferred distance to $\alpha$~Dra is $d_{\rm MESA, B} = 88.4\pm2.3$ pc, and $d_{\rm MESA, V} = 88.6\pm2.4$ pc. This is rather close to the distance we determined from dynamical parallax, as well to the distance calculated with the bolometric correction from \citet{Flower_1996}, as represented in Fig.~\ref{fig:gaia}.

\section{Summary and Discussion}
\label{sec:dis}

The detailed analysis of the newly obtained high-resolution, high S/N {\sc HERMES} optical spectra unambiguously reveals the spectral contribution of an A-type secondary component in the observed composite spectra of the $\alpha$~Dra system. The companion star is found to contribute some  17.5\% of the total flux of the system, however, its rapid rotation implied by $v \sin i$ of $168\pm11$~km\,s$^{-1}$ results in the apparent spectral contribution of only 1\% of the continuum level in the regions of metal lines. Therefore, it does not come as a surprise that $\alpha$~Dra has for long been classified in the literature as a SB1 system. Our newly obtained data set also allows us to achieve the precision of $\sim$0.1\% in the measurement of the primary's RV semi-amplitude. A similar precision has been achieved by \citet{Kallinger_2004} and Hey et al. (2021, in review) while other measurements available in the literature are significantly inferior \citep[e.g.][]{Elst_Nelles_1983, Bischoff_2017}.

The combined analysis of the {\sc HERMES} high-resolution spectroscopic and the high-quality TESS space-based photometric data enabled us to determine the fundamental and atmospheric properties of both components of the $\alpha$ Dra system with high confidence. The precision achieved for the mass, radius, and effective temperature of the primary (secondary) component are some 1.4\% (0.8\%), 1.1\% (3.1\%), and 1.0\% (3.0\%). We find the mass and radius of the primary component to be respectively some 14\% and 15\% lower than the values of 3.7$\pm$0.1~M$_{\odot}$ and 5.8$\pm$0.1~R$_{\odot}$ reported by Hey et al. (2021, in review). We note that the $\alpha$~Dra system will be re-observed by the TESS mission during Sectors 48 and 49\footnote{$\alpha$\,Dra was also observed by the TESS mission during Sector 41, however, only out-of-eclipse phases were covered during the respective 27.4 days of observations.} of the cycle 4 observations, offering a prospect for further improvement of the aforementioned precision, and in particular for the radius of the secondary component.

The primary component of the $\alpha$~Dra system is an evolved A-type star, which is an extremely rare case when it comes to eclipsing double-lined binaries. Querying DEBCat\footnote{\url{https://www.astro.keele.ac.uk/jkt/debcat/}}, an on-line catalogue of detached eclipsing binaries with the masses and radii determined with a precision better than 2\% \citep[maintained and regularly updated by][]{Southworth_2015}, results in only one binary system with an evolved A-type star, namely $\psi$\,Cen. This binary system comprises a pair of A0\,IV and A1\,V stars that reside in a 38.8~d orbit. Fundamental and atmospheric properties of both components of the $\psi$\,Cen system have been determined with high precision, thanks to the available WIRE space-based photometric \citep{Bruntt_2006} and ground-based high-resolution spectroscopic data \citep{Mantegazza_2010, Gallenne_2019}. For example, \citet{Gallenne_2019} report the mass and radius of the primary component of $\psi$\,Cen to be $M = 3.187\pm0.031$ M$_\odot$ and $R = 3.814\pm0.007$ R$_\odot$. Therefore, the primary components of the $\psi$\,Cen and $\alpha$~Dra are strikingly similar in terms of their masses, while the difference of about 1 R$_\odot$ in their radii (see Table~\ref{tab:abspar} for the parameters of $\alpha$~Dra) indicates the primary component of $\alpha$~Dra is evolutionary more advanced. This in turn suggests a slight difference in ages of the two systems, and indeed, we find $\alpha$~Dra to be $310\pm25$~Myr old, which is to be compared to the age determinations of $\psi$\,Cen of $\tau\,({\rm \psi\,Cen\,A}) = 280\pm10$~Myr \citep{Gallenne_2019} and $\tau\,({\rm \psi\,Cen}) \sim 290$~Myr \citep{Bruntt_2006}. Looking for ``a replica'' of the secondary component of $\alpha$~Dra in terms of mass and age, the closest matches are the TZ\,Men\,A \citep{Andersen_1987}  and V541\,Cyg\,A \citep{Torres_2017} systems. Their masses and surface gravities (used as a proxy for age) are matching the respective quantities of $\alpha$~Dra~B within 0.1 M$_\odot$ and 0.1~dex

\citet{Royer_2007} report a bi-modal distribution of the projected rotational velocity $v\, \sin i$ in early A-type stars, with ``slowly'' and ``rapidly'' rotating stars having distributions centered at 45~km\,s$^{-1}$ and 200~km\,s$^{-1}$, respectively. It is thus not unexpected to find a binary system composed of two early A-type stars with significantly different projected rotational velocities, especially when one of the components is at an advanced evolutionary stage. Our measurement of $v\,\sin i$ for the primary component is in excellent agreement with the value of $v\,\sin i = 25\pm2$~km\,s$^{-1}$ inferred by \citet{Royer_2002b} with the Fourier method. The projected rotational velocity of the secondary component of 168$\pm$11~km\,s$^{-1}$ determined by us corroborates well the speculation by Hey et al. (2021, in review) that a companion star must be a rapidly rotating A-type star with $v\,\sin i \sim 200$~km\,s$^{-1}$. A-type stars with discordant projected rotational velocities measured from different spectral lines were identified in observational studies by \citet{Zverko_2011,Zverko_2018} and \citet{Vanko_2020}, and references therein. Systematic spectroscopic observations revealed that this disparity in $v\,\sin\,i$ is a consequence of binarity itself, and the $\alpha$~Dra system adds extra value to that argument.

The angular diameters of both components of the $\alpha$\,Dra binary system were resolved in our high-angular resolution interferometric observations with the NPOI.  The measured angular diameters for the primary and secondary components are $\theta_{\rm A} = 0.62\pm 0.05$~mas and  $\theta_{\rm B} = 0.28\pm0.05$~mas, respectively. Combining angular diameters with the dynamical parallax of $\pi = 11.48\pm0.13$~mas ($d = 87.07\pm1.03$~pc; see Sect.~\ref{sec:parallax}), we obtain linear radii of $R_{\rm lin,A} = 5.80\pm0.47$ R$_\odot$ and  $R_{\rm lin,B} = 2.62\pm0.47$ R$_\odot$ for the primary and secondary component, respectively. We note that uncertainties in the interferometric measurements are the main contributor to the total error budget of the linear radii. The aforementioned values are within 2-$\sigma$ in agreement with the radius for the primary, and 1-$\sigma$ for the radius of the secondary component inferred from the combined spectroscopic and light curve analysis.

The high precision achieved in the measurement of the dynamical parallax of the system in turn leads to the 1.1\% uncertainty on its distance. The distance from {\sl Gaia} EDR3 trigonometric parallax is about 7~pc shorter than the value derived from our dynamical parallax. Due to the brightness of $\alpha$~Dra, the uncertainties of {\it Gaia} parallax when corrected are large and within 2$\sigma$ in agreement with our dynamical parallax. At the same time, inference of the distance to $\alpha$\,Dra from physical properties of the components derived in this work (radii, effective temperatures, and accounting for the light ratio between the two components) and bolometric corrections gave a broad range of results. The closest distance to our measurement of the dynamical parallax is obtained from the empirical bolometric correction by \citet{Flower_1996}, $d_{\rm FLO} = 87.9\pm2.5$ pc. On the other hand, using theoretically calculated bolometric corrections from \citet{Bessell_1998}, and \citet{Girardi_2002} both yield a distance that is about 10~pc further away than the distance obtained in this work (see Fig.~\ref{fig:gaia}). We attribute the difference between the {\sl Gaia} and dynamical parallaxes to the (current) neglect of binarity in the {\sl Gaia} solution, where the orbital movement produces changes in the  photocenter of the star thus affects the determination of the trigonometric parallax, as well produces anomalous proper motion.

Inspired by the results of Hey et al. (2021, in review) who reported a complex abundance pattern for the primary component of $\alpha$~Dra, we inferred abundances of helium, silicon, and scandium to check whether they are consistent with the solar metallicity of the star inferred from its disentangled spectrum. We focus here on He, Si, and Sc as these three elements are reported by Hey et al. (2021, in review) to show some of the largest under-abundances, while a detailed chemical composition analysis of both components of $\alpha$~Dra is the subject of a forthcoming paper. We confirm a slight under-abundance of He (though consistent with the solar composition within the quoted errors) at the level of 0.1$\pm$0.1~dex, whereas abundances of Si and Sc are found to be in excellent agreement with the deduced solar metallicity of the star. Hence, we speculate that the complex abundance pattern reported by Hey et al. (2021, in review) for the primary component from the analysis of a co-added SOPHIE spectrum is a consequence of unaccounted spectral contribution of the secondary component to the observed composite spectrum of the system. Analysis of the spectrum of $\alpha$~Dra as if it was a single star is also likely the reason why \citet{Andersen_1987} and \citet{Adelman_2001} classified the system as a $\lambda$~Boo chemically peculiar star. The latter are known for their metal deficiency, an effect that is easily mimicked by light dilution from an unseen companion in spectroscopic data.

Finally, from an evolutionary analysis of the system, we find that the primary component of $\alpha$~Dra is an evolved post-TAMS A-type star. The secondary component is shown to be relatively unevolved, whose convective core mass constitutes some 14\% of the apparent mass of the star. Fundamental and atmospheric properties of both binary components are consistent with low values of the convective core overshooting, with $f_{\rm ov}$ just below 0.010~$H_{\rm p}$.

\section{Conclusion and future work}\label{sec:conclusions}

The analysis of $\alpha$~Dra presented in this work builds upon two fundamental findings in the literature: (i) spatially resolving the binary components based on the high-resolution NPOI interferometric data \citep{Hutter_2016}, and (ii) detection of eclipses in TESS space-based photometric data \citep{Bedding_2019}. Both these data sets suggest that $\alpha$~Dra is unlikely to be a SB1 system as previously reported in the literature, where \citet{Hutter_2016} give an estimate of the light ratio of the two stars of about 0.19. Intrigued by the fact that such a rare system with an evolved A-type primary component has  been spectroscopically misclassified as a SB1 so far, due to a low apparent contribution of the secondary, we collected a new data set of high-resolution, high S/N optical spectroscopic data to attempt for spectroscopic detection of a companion star, and for detailed analysis of the system based on combined spectroscopic, interferometric, and photometric data sets. We report the first unambiguous detection of the secondary component in the $\alpha$~Dra system which makes it an eclipsing spectroscopic double-lined (SB2) binary. 

The main results and conclusions of our work are: \begin{itemize}
    \item the companion star is found  to contribute some 17\% of the total light of the system, with the apparent spectroscopic line depths of about 1\% in metal line regions, owing to the high rotational velocity of the star;
    \item dynamical masses and radii of the primary and secondary components are found to be $M_{\rm A} = 3.186\pm0.044$~M$_{\odot}$ and $R_{\rm A} = 4.932\pm0.036$~R$_{\odot}$, and $M_{\rm B} = 2.431\pm0.019$~M$_{\odot}$ and $R_{\rm B} = 2.326\pm0.052$~R$_{\odot}$, respectively. High precision of the above-mentioned parameters,  coupled with precise measurements of the effective temperature for both stars, allows us to constrain the evolutionary status of both components and the age of the system as a whole;
    \item we confirm that the primary component is an evolved post-TAMS early A-type star. The companion is, on the other hand, a fairly unevolved main-sequence star whose convective core mass constitutes some 14\% of its apparent total mass. The age of the system is estimated to be 310$\pm$25~Myr;
    \item the high quality of our interferometric observations with an optimal distribution over the orbital cycle,  allow us to infer the angular semi-major axis with a precision of about 1\%. This high precision in turn allows for the inference of a precise dynamical parallax of $\pi = 11.48\pm0.13$~mas, ultimately leading to the distance estimate of $d = 87.07\pm1.03$~pc.
\end{itemize} 

Both components of the $\alpha$~Dra system are interesting targets to consider in the context of the mass discrepancy and near-core mixing levels in detached eclipsing double-lined binaries. The term mass discrepancy refers to the difference between the measured dynamical mass of the star and its mass that is inferred from fitting evolutionary models to the position of the star in the Kiel- or HR-diagram. Until recently, it was common to associate the mass discrepancy with large amounts of the near-core mixing in the form of overshooting \citep[e.g.][]{Guinan2000,Pavlovski_2009}, hence the two effects are not totally disconnected. \citet{Claret2019} report a sharp increase in the $f_{\rm ov}$ overshooting parameter up to a mass of some 2~M$_{\odot}$ and a subsequent levelling off at the value of some 0.017~$H_{\rm p}$ beyond 2~M$_{\odot}$ and at least up to 4.4~M$_{\odot}$. On the other hand, \citet{Tkachenko_2020} find that (i) the mass discrepancy is a strong function of the surface gravity of the star while it does not show any dependence on the stellar mass itself; (ii) the mass discrepancy can be only partially explained by insufficient near-core mixing and convective core mass predicted by evolutionary models; (iii) there is evidence that the mass discrepancy is strongly connected with incorrect spectroscopic inference of the effective temperature of the star due to ignoring the contribution of the turbulent and radiative pressure in the stellar atmosphere. We note, however, that the study by \citet{Tkachenko_2020} concerned generally higher mass stars, in the mass range between some 4.5~M$_{\odot}$ and 17~M$_{\odot}$. Our evolutionary analysis of the $\alpha$~Dra system does not support the conclusion of \citet{Claret2019} as to the dependence of the convective core overshooting parameter on stellar mass. We find low overshooting values well below $f_{\rm ov}=0.010\,H_{\rm p}$ for both binary components, while according to the dynamical masses and empirical findings by \citet{Claret_Torres_2016} the stars should show significantly higher levels of core overshooting. Perfect agreement between the dynamical and evolutionary mass for the evolved primary component of $\alpha$~Dra is also seemingly in contradiction with the findings of \citet{Tkachenko_2020} that evolved stars tend to show large values of the mass discrepancy. However, we stress that both components of $\alpha$~Dra have substantially lower masses than the bulk sample studied by \citet{Tkachenko_2020}, hence we do not expect their $T_{\rm eff}$ inference be significantly influenced by the turbulent and/or radiative pressure in the stellar atmosphere. Therefore, $\alpha$~Dra provides an extra argument for the mass discrepancy in eclipsing binaries to be strongly connected with inferior modelling of stellar atmospheres in certain mass regimes, and probably less so with the predictive power of evolutionary models as to the convective core mass and amount of near-core mixing.

Finally, by looking at abundances of He, Si, and Sc in the atmosphere of the primary component of $\alpha$~Dra, we find the former to be slightly under-abundant, while the other two elements show abundances consistent with the solar metallicity of the star. This result suggests the primary component might be a chemically ordinary A-type star, however a more in-depth chemical composition analysis is required before any firm conclusions can be drawn. Detailed chemical composition analysis of both components of $\alpha$ Dra and subsequent comparison of the inferred  abundance patterns with predictions from stellar structure and evolution models are a subject of a forthcoming paper (Sahin et al., in prep.).

\begin{acknowledgements} 
We would like to thank the referee for a thorough review of the paper and for very useful comments that helped us to improve the manuscript. 
We acknowledge Daniel Hey for kindly communicating his manuscript before publication. Based on observations obtained with the HERMES spectrograph, which is supported by the Research Foundation - Flanders (FWO), Belgium, the Research Council of KU Leuven, Belgium, the Fonds National de la Recherche Scientifique (F.R.S.-FNRS), Belgium, the Royal Observatory of Belgium, the Observatoire de Genève, Switzerland and the Thüringer Landessternwarte Tautenburg, Germany. The Navy Precision Optical Interferometer is a joint project of the Naval Research Laboratory and the US Naval Observatory, in cooperation with Lowell Observatory and is funded by the Office of Naval Research and the Oceanographer of the Navy. The authors thank Dr. J.A.\ Benson and the NPOI observational support staff whose efforts made this project possible. This article made use of the Smithsonian/NASA Astrophysics Data System (ADS), of the Centre de Données astronomiques de Strasbourg (CDS) and of the Jean-Marie Mariotti Center (JMMC). The research leading to these results has (partially) received funding from the European Research Council (ERC) under the European Union's Horizon 2020 research and innovation programme (grant agreement N$^\circ$670519: MAMSIE), from the KU~Leuven Research Council (grant C16/18/005: PARADISE), from the Research Foundation Flanders (FWO) under grant agreement G0H5416N (ERC Runner Up Project), and a PhD Fellowship to R.B., and S.G. under contract No.\ 11E5620N, as well as from the BELgian federal Science Policy Office (BELSPO) through PRODEX grant PLATO. DMB gratefully acknowledges a senior postdoctoral fellowship from the Research Foundation Flanders (FWO) with grant agreement no. 1286521N. J.B. acknowledges support from the FWO Odysseus program under project G0F8H6N. 
A.D. and C.K. acknowledge support by Erciyes University Scientific Research Projects Coordination Unit under grant number MAP-2020-9749.
This work has made use of data from the European Space Agency (ESA) mission {\it Gaia} (\url{https://www.cosmos.esa.int/gaia}), processed by the {\it Gaia} Data Processing and Analysis Consortium (DPAC, \url{https://www.cosmos.esa.int/web/gaia/dpac/consortium}). Funding for the DPAC has been provided by national institutions, in particular the institutions participating in the {\it Gaia} Multilateral Agreement.

\end{acknowledgements}
 
%
\bibliographystyle{aa} 
\bibliography{draft_alpha_dra} 

\begin{thebibliography}{127}
\expandafter\ifx\csname natexlab\endcsname\relax\def\natexlab#1{#1}\fi

\bibitem[{{Adelman} {et~al.}(2001){Adelman}, {Caliskan}, {Kocer}, {Kablan},
  {Y{\"u}ce}, \& {Engin}}]{Adelman_2001}
{Adelman}, S.~J., {Caliskan}, H., {Kocer}, D., {et~al.} 2001, \aap, 371, 1078

\bibitem[{{Aller} {et~al.}(1982){Aller}, {Appenzeller}, {Baschek}, {Duerbeck},
  {Herczeg}, {Lamla}, {Meyer-Hofmeister}, {Schmidt-Kaler}, {Scholz},
  {Seggewiss}, {Seitter}, \& {Weidemann}}]{1982lbg6.conf.....A}
{Aller}, L.~H., {Appenzeller}, I., {Baschek}, B., {et~al.}, eds. 1982,
  {Landolt-B{\"o}rnstein: Numerical Data and Functional Relationships in
  Science and Technology - New Series '' Gruppe/Group 6 Astronomy and
  Astrophysics '' Volume 2 Schaifers/Voigt: Astronomy and Astrophysics /
  Astronomie und Astrophysik '' Stars and Star Clusters / Sterne und
  Sternhaufen}

\bibitem[{{Almeida} {et~al.}(2017){Almeida}, {Sana}, {Taylor}, {Barb{\'a}},
  {Bonanos}, {Crowther}, {Damineli}, {de Koter}, {de Mink}, {Evans}, {Gieles},
  {Grin}, {H{\'e}nault-Brunet}, {Langer}, {Lennon}, {Lockwood}, {Ma{\'\i}z
  Apell{\'a}niz}, {Moffat}, {Neijssel}, {Norman}, {Ram{\'\i}rez-Agudelo},
  {Richardson}, {Schootemeijer}, {Shenar}, {Soszy{\'n}ski}, {Tramper}, \&
  {Vink}}]{Almeida2017}
{Almeida}, L.~A., {Sana}, H., {Taylor}, W., {et~al.} 2017, \aap, 598, A84

\bibitem[{{Andersen} {et~al.}(1987){Andersen}, {Clausen}, \&
  {Nordstrom}}]{Andersen_1987}
{Andersen}, J., {Clausen}, J.~V., \& {Nordstrom}, B. 1987, \aap, 175, 60

\bibitem[{{Armstrong} {et~al.}(1998){Armstrong}, {Mozurkewich}, {Rickard},
  {Hutter}, {Benson}, {Bowers}, {Elias}, {Hummel}, {Johnston}, {Buscher},
  {Clark}, {Ha}, {Ling}, {White}, \& {Simon}}]{Armstrong_1998}
{Armstrong}, J.~T., {Mozurkewich}, D., {Rickard}, L.~J., {et~al.} 1998, \apj,
  496, 550

\bibitem[{{Balona} {et~al.}(2015){Balona}, {Baran},
  {Daszy{\'n}ska-Daszkiewicz}, \& {De Cat}}]{Balona_2015}
{Balona}, L.~A., {Baran}, A.~S., {Daszy{\'n}ska-Daszkiewicz}, J., \& {De Cat},
  P. 2015, \mnras, 451, 1445

\bibitem[{{Banyard} {et~al.}(2021){Banyard}, {Sana}, {Mahy}, {Bodensteiner},
  {Villase{\~n}or}, \& {Evans}}]{Banyard2021}
{Banyard}, G., {Sana}, H., {Mahy}, L., {et~al.} 2021, arXiv e-prints,
  arXiv:2108.07814

\bibitem[{{Bedding} {et~al.}(2019){Bedding}, {Hey}, \& {Murphy}}]{Bedding_2019}
{Bedding}, T.~R., {Hey}, D.~R., \& {Murphy}, S.~J. 2019, Research Notes of the
  American Astronomical Society, 3, 163

\bibitem[{{Bessell} {et~al.}(1998){Bessell}, {Castelli}, \&
  {Plez}}]{Bessell_1998}
{Bessell}, M.~S., {Castelli}, F., \& {Plez}, B. 1998, \aap, 333, 231

\bibitem[{{Bischoff} {et~al.}(2017){Bischoff}, {Mugrauer}, {Zehe},
  {W{\"o}ckel}, {Pannicke}, {Lux}, {Wagner}, {Heyne}, {Adam}, \&
  {Neuh{\"a}user}}]{Bischoff_2017}
{Bischoff}, R., {Mugrauer}, M., {Zehe}, T., {et~al.} 2017, Astronomische
  Nachrichten, 338, 671

\bibitem[{{Bodensteiner} {et~al.}(2021){Bodensteiner}, {Sana}, {Wang},
  {Langer}, {Mahy}, {Banyard}, {de Koter}, {de Mink}, {Evans}, {G{\"o}tberg},
  {Patrick}, {Schneider}, \& {Tramper}}]{Bodensteiner2021}
{Bodensteiner}, J., {Sana}, H., {Wang}, C., {et~al.} 2021, \aap, 652, A70

\bibitem[{{Bruntt} {et~al.}(2006){Bruntt}, {Southworth}, {Torres}, {Penny},
  {Clausen}, \& {Buzasi}}]{Bruntt_2006}
{Bruntt}, H., {Southworth}, J., {Torres}, G., {et~al.} 2006, \aap, 456, 651

\bibitem[{{Choi} {et~al.}(2016){Choi}, {Dotter}, {Conroy}, {Cantiello},
  {Paxton}, \& {Johnson}}]{Choi_2016}
{Choi}, J., {Dotter}, A., {Conroy}, C., {et~al.} 2016, \apj, 823, 102

\bibitem[{{Claret} \& {Torres}(2016)}]{Claret_Torres_2016}
{Claret}, A. \& {Torres}, G. 2016, \aap, 592, A15

\bibitem[{{Claret} \& {Torres}(2017)}]{Claret2017}
{Claret}, A. \& {Torres}, G. 2017, \apj, 849, 18

\bibitem[{{Claret} \& {Torres}(2018)}]{Claret2018}
{Claret}, A. \& {Torres}, G. 2018, \apj, 859, 100

\bibitem[{{Claret} \& {Torres}(2019)}]{Claret2019}
{Claret}, A. \& {Torres}, G. 2019, \apj, 876, 134

\bibitem[{{Code} {et~al.}(1976){Code}, {Bless}, {Davis}, \&
  {Brown}}]{Code_1976}
{Code}, A.~D., {Bless}, R.~C., {Davis}, J., \& {Brown}, R.~H. 1976, \apj, 203,
  417

\bibitem[{{Cohen} {et~al.}(2020){Cohen}, {Geller}, \& {von Hippel}}]{Cohen2020}
{Cohen}, R.~E., {Geller}, A.~M., \& {von Hippel}, T. 2020, \aj, 159, 11

\bibitem[{{Conroy} {et~al.}(2020){Conroy}, {Kochoska}, {Hey}, {Pablo},
  {Hambleton}, {Jones}, {Giammarco}, {Abdul-Masih}, \&
  {Pr{\v{s}}a}}]{Conroy2020}
{Conroy}, K.~E., {Kochoska}, A., {Hey}, D., {et~al.} 2020, \apjs, 250, 34

\bibitem[{{Dervi{\c{s}}o{\v{g}}lu} {et~al.}(2018){Dervi{\c{s}}o{\v{g}}lu},
  {Pavlovski}, {Lehmann}, {Southworth}, \& {Bewsher}}]{Dervisoglu2018}
{Dervi{\c{s}}o{\v{g}}lu}, A., {Pavlovski}, K., {Lehmann}, H., {Southworth}, J.,
  \& {Bewsher}, D. 2018, \mnras, 481, 5660

\bibitem[{{Dotter}(2016)}]{Dotter_2016}
{Dotter}, A. 2016, \apjs, 222, 8

\bibitem[{{Drimmel} {et~al.}(2003){Drimmel}, {Cabrera-Lavers}, \&
  {L{\'o}pez-Corredoira}}]{Drimmel_2003}
{Drimmel}, R., {Cabrera-Lavers}, A., \& {L{\'o}pez-Corredoira}, M. 2003, \aap,
  409, 205

\bibitem[{{Duch{\^e}ne} \& {Kraus}(2013)}]{Duchene_Kraus_2013}
{Duch{\^e}ne}, G. \& {Kraus}, A. 2013, \araa, 51, 269

\bibitem[{{Elson} {et~al.}(1998){Elson}, {Sigurdsson}, {Davies}, {Hurley}, \&
  {Gilmore}}]{Elson_1998}
{Elson}, R. A.~W., {Sigurdsson}, S., {Davies}, M., {Hurley}, J., \& {Gilmore},
  G. 1998, \mnras, 300, 857

\bibitem[{{Elst} \& {Nelles}(1983)}]{Elst_Nelles_1983}
{Elst}, E.~W. \& {Nelles}, B. 1983, \aaps, 53, 215

\bibitem[{{Fabricius} {et~al.}(2021){Fabricius}, {Luri}, {Arenou}, {Babusiaux},
  {Helmi}, {Muraveva}, {Reyl{\'e}}, {Spoto}, {Vallenari}, {Antoja}, {Balbinot},
  {Barache}, {Bauchet}, {Bragaglia}, {Busonero}, {Cantat-Gaudin}, {Carrasco},
  {Diakit{\'e}}, {Fabrizio}, {Figueras}, {Garcia-Gutierrez}, {Garofalo},
  {Jordi}, {Kervella}, {Khanna}, {Leclerc}, {Licata}, {Lambert}, {Marrese},
  {Masip}, {Ramos}, {Robichon}, {Robin}, {Romero-G{\'o}mez}, {Rubele}, \&
  {Weiler}}]{Fabricius_2021}
{Fabricius}, C., {Luri}, X., {Arenou}, F., {et~al.} 2021, \aap, 649, A5

\bibitem[{{Feiden} \& {Chaboyer}(2012)}]{Feiden2012}
{Feiden}, G.~A. \& {Chaboyer}, B. 2012, \apj, 761, 30

\bibitem[{{Feiden} \& {Chaboyer}(2013)}]{Feiden2013}
{Feiden}, G.~A. \& {Chaboyer}, B. 2013, \apj, 779, 183

\bibitem[{{Feiden} \& {Chaboyer}(2014)}]{Feiden2014}
{Feiden}, G.~A. \& {Chaboyer}, B. 2014, \apj, 789, 53

\bibitem[{{Flower}(1996)}]{Flower_1996}
{Flower}, P.~J. 1996, \apj, 469, 355

\bibitem[{{Foreman-Mackey} {et~al.}(2013){Foreman-Mackey}, {Hogg}, {Lang}, \&
  {Goodman}}]{Emcee2013}
{Foreman-Mackey}, D., {Hogg}, D.~W., {Lang}, D., \& {Goodman}, J. 2013, \pasp,
  125, 306

\bibitem[{{Frankowski} {et~al.}(2007){Frankowski}, {Jancart}, \&
  {Jorissen}}]{Frankowski_2007}
{Frankowski}, A., {Jancart}, S., \& {Jorissen}, A. 2007, \aap, 464, 377

\bibitem[{{Gaia Collaboration} {et~al.}(2018){Gaia Collaboration}, {Brown},
  {Vallenari}, {Prusti}, {de Bruijne}, {Babusiaux}, {Bailer-Jones}, {Biermann},
  {Evans}, {Eyer}, {Jansen}, {Jordi}, {Klioner}, {Lammers}, {Lindegren},
  {Luri}, {Mignard}, {Panem}, {Pourbaix}, {Randich}, {Sartoretti}, {Siddiqui},
  {Soubiran}, {van Leeuwen}, {Walton}, {Arenou}, {Bastian}, {Cropper},
  {Drimmel}, {Katz}, {Lattanzi}, {Bakker}, {Cacciari}, {Casta{\~n}eda},
  {Chaoul}, {Cheek}, {De Angeli}, {Fabricius}, {Guerra}, {Holl}, {Masana},
  {Messineo}, {Mowlavi}, {Nienartowicz}, {Panuzzo}, {Portell}, {Riello},
  {Seabroke}, {Tanga}, {Th{\'e}venin}, {Gracia-Abril}, {Comoretto},
  {Garcia-Reinaldos}, {Teyssier}, {Altmann}, {Andrae}, {Audard},
  {Bellas-Velidis}, {Benson}, {Berthier}, {Blomme}, {Burgess}, {Busso},
  {Carry}, {Cellino}, {Clementini}, {Clotet}, {Creevey}, {Davidson}, {De
  Ridder}, {Delchambre}, {Dell'Oro}, {Ducourant},
  {Fern{\'a}ndez-Hern{\'a}ndez}, {Fouesneau}, {Fr{\'e}mat}, {Galluccio},
  {Garc{\'\i}a-Torres}, {Gonz{\'a}lez-N{\'u}{\~n}ez}, {Gonz{\'a}lez-Vidal},
  {Gosset}, {Guy}, {Halbwachs}, {Hambly}, {Harrison}, {Hern{\'a}ndez},
  {Hestroffer}, {Hodgkin}, {Hutton}, {Jasniewicz}, {Jean-Antoine-Piccolo},
  {Jordan}, {Korn}, {Krone-Martins}, {Lanzafame}, {Lebzelter}, {L{\"o}ffler},
  {Manteiga}, {Marrese}, {Mart{\'\i}n-Fleitas}, {Moitinho}, {Mora}, {Muinonen},
  {Osinde}, {Pancino}, {Pauwels}, {Petit}, {Recio-Blanco}, {Richards},
  {Rimoldini}, {Robin}, {Sarro}, {Siopis}, {Smith}, {Sozzetti}, {S{\"u}veges},
  {Torra}, {van Reeven}, {Abbas}, {Abreu Aramburu}, {Accart}, {Aerts},
  {Altavilla}, {{\'A}lvarez}, {Alvarez}, {Alves}, {Anderson}, {Andrei},
  {Anglada Varela}, {Antiche}, {Antoja}, {Arcay}, {Astraatmadja}, {Bach},
  {Baker}, {Balaguer-N{\'u}{\~n}ez}, {Balm}, {Barache}, {Barata}, {Barbato},
  {Barblan}, {Barklem}, {Barrado}, {Barros}, {Barstow}, {Bartholom{\'e}
  Mu{\~n}oz}, {Bassilana}, {Becciani}, {Bellazzini}, {Berihuete}, {Bertone},
  {Bianchi}, {Bienaym{\'e}}, {Blanco-Cuaresma}, {Boch}, {Boeche}, {Bombrun},
  {Borrachero}, {Bossini}, {Bouquillon}, {Bourda}, {Bragaglia}, {Bramante},
  {Breddels}, {Bressan}, {Brouillet}, {Br{\"u}semeister}, {Brugaletta},
  {Bucciarelli}, {Burlacu}, {Busonero}, {Butkevich}, {Buzzi}, {Caffau},
  {Cancelliere}, {Cannizzaro}, {Cantat-Gaudin}, {Carballo}, {Carlucci},
  {Carrasco}, {Casamiquela}, {Castellani}, {Castro-Ginard}, {Charlot},
  {Chemin}, {Chiavassa}, {Cocozza}, {Costigan}, {Cowell}, {Crifo}, {Crosta},
  {Crowley}, {Cuypers}, {Dafonte}, {Damerdji}, {Dapergolas}, {David}, {David},
  {de Laverny}, {De Luise}, {De March}, {de Martino}, {de Souza}, {de Torres},
  {Debosscher}, {del Pozo}, {Delbo}, {Delgado}, {Delgado}, {Di Matteo},
  {Diakite}, {Diener}, {Distefano}, {Dolding}, {Drazinos}, {Dur{\'a}n},
  {Edvardsson}, {Enke}, {Eriksson}, {Esquej}, {Eynard Bontemps}, {Fabre},
  {Fabrizio}, {Faigler}, {Falc{\~a}o}, {Farr{\`a}s Casas}, {Federici},
  {Fedorets}, {Fernique}, {Figueras}, {Filippi}, {Findeisen}, {Fonti},
  {Fraile}, {Fraser}, {Fr{\'e}zouls}, {Gai}, {Galleti}, {Garabato},
  {Garc{\'\i}a-Sedano}, {Garofalo}, {Garralda}, {Gavel}, {Gavras}, {Gerssen},
  {Geyer}, {Giacobbe}, {Gilmore}, {Girona}, {Giuffrida}, {Glass}, {Gomes},
  {Granvik}, {Gueguen}, {Guerrier}, {Guiraud}, {Guti{\'e}rrez-S{\'a}nchez},
  {Haigron}, {Hatzidimitriou}, {Hauser}, {Haywood}, {Heiter}, {Helmi}, {Heu},
  {Hilger}, {Hobbs}, {Hofmann}, {Holland}, {Huckle}, {Hypki}, {Icardi},
  {Jan{\ss}en}, {Jevardat de Fombelle}, {Jonker}, {Juh{\'a}sz}, {Julbe},
  {Karampelas}, {Kewley}, {Klar}, {Kochoska}, {Kohley}, {Kolenberg},
  {Kontizas}, {Kontizas}, {Koposov}, {Kordopatis}, {Kostrzewa-Rutkowska},
  {Koubsky}, {Lambert}, {Lanza}, {Lasne}, {Lavigne}, {Le Fustec}, {Le
  Poncin-Lafitte}, {Lebreton}, {Leccia}, {Leclerc}, {Lecoeur-Taibi},
  {Lenhardt}, {Leroux}, {Liao}, {Licata}, {Lindstr{\o}m}, {Lister}, {Livanou},
  {Lobel}, {L{\'o}pez}, {Managau}, {Mann}, {Mantelet}, {Marchal}, {Marchant},
  {Marconi}, {Marinoni}, {Marschalk{\'o}}, {Marshall}, {Martino}, {Marton},
  {Mary}, {Massari}, {Matijevi{\v{c}}}, {Mazeh}, {McMillan}, {Messina},
  {Michalik}, {Millar}, {Molina}, {Molinaro}, {Moln{\'a}r}, {Montegriffo},
  {Mor}, {Morbidelli}, {Morel}, {Morris}, {Mulone}, {Muraveva}, {Musella},
  {Nelemans}, {Nicastro}, {Noval}, {O'Mullane}, {Ord{\'e}novic},
  {Ord{\'o}{\~n}ez-Blanco}, {Osborne}, {Pagani}, {Pagano}, {Pailler},
  {Palacin}, {Palaversa}, {Panahi}, {Pawlak}, {Piersimoni}, {Pineau}, {Plachy},
  {Plum}, {Poggio}, {Poujoulet}, {Pr{\v{s}}a}, {Pulone}, {Racero}, {Ragaini},
  {Rambaux}, {Ramos-Lerate}, {Regibo}, {Reyl{\'e}}, {Riclet}, {Ripepi}, {Riva},
  {Rivard}, {Rixon}, {Roegiers}, {Roelens}, {Romero-G{\'o}mez}, {Rowell},
  {Royer}, {Ruiz-Dern}, {Sadowski}, {Sagrist{\`a} Sell{\'e}s}, {Sahlmann},
  {Salgado}, {Salguero}, {Sanna}, {Santana-Ros}, {Sarasso}, {Savietto},
  {Schultheis}, {Sciacca}, {Segol}, {Segovia}, {S{\'e}gransan}, {Shih},
  {Siltala}, {Silva}, {Smart}, {Smith}, {Solano}, {Solitro}, {Sordo}, {Soria
  Nieto}, {Souchay}, {Spagna}, {Spoto}, {Stampa}, {Steele},
  {Steidelm{\"u}ller}, {Stephenson}, {Stoev}, {Suess}, {Surdej}, {Szabados},
  {Szegedi-Elek}, {Tapiador}, {Taris}, {Tauran}, {Taylor}, {Teixeira},
  {Terrett}, {Teyssandier}, {Thuillot}, {Titarenko}, {Torra Clotet}, {Turon},
  {Ulla}, {Utrilla}, {Uzzi}, {Vaillant}, {Valentini}, {Valette}, {van Elteren},
  {Van Hemelryck}, {van Leeuwen}, {Vaschetto}, {Vecchiato}, {Veljanoski},
  {Viala}, {Vicente}, {Vogt}, {von Essen}, {Voss}, {Votruba}, {Voutsinas},
  {Walmsley}, {Weiler}, {Wertz}, {Wevers}, {Wyrzykowski}, {Yoldas},
  {{\v{Z}}erjal}, {Ziaeepour}, {Zorec}, {Zschocke}, {Zucker}, {Zurbach}, \&
  {Zwitter}}]{gaia_2018}
{Gaia Collaboration}, {Brown}, A.~G.~A., {Vallenari}, A., {et~al.} 2018, \aap,
  616, A1

\bibitem[{{Gaia Collaboration} {et~al.}(2021){Gaia Collaboration}, {Brown},
  {Vallenari}, {Prusti}, {de Bruijne}, {Babusiaux}, {Biermann}, {Creevey},
  {Evans}, {Eyer}, {Hutton}, {Jansen}, {Jordi}, {Klioner}, {Lammers},
  {Lindegren}, {Luri}, {Mignard}, {Panem}, {Pourbaix}, {Randich}, {Sartoretti},
  {Soubiran}, {Walton}, {Arenou}, {Bailer-Jones}, {Bastian}, {Cropper},
  {Drimmel}, {Katz}, {Lattanzi}, {van Leeuwen}, {Bakker}, {Cacciari},
  {Casta{\~n}eda}, {De Angeli}, {Ducourant}, {Fabricius}, {Fouesneau},
  {Fr{\'e}mat}, {Guerra}, {Guerrier}, {Guiraud}, {Jean-Antoine Piccolo},
  {Masana}, {Messineo}, {Mowlavi}, {Nicolas}, {Nienartowicz}, {Pailler},
  {Panuzzo}, {Riclet}, {Roux}, {Seabroke}, {Sordo}, {Tanga}, {Th{\'e}venin},
  {Gracia-Abril}, {Portell}, {Teyssier}, {Altmann}, {Andrae}, {Bellas-Velidis},
  {Benson}, {Berthier}, {Blomme}, {Brugaletta}, {Burgess}, {Busso}, {Carry},
  {Cellino}, {Cheek}, {Clementini}, {Damerdji}, {Davidson}, {Delchambre},
  {Dell'Oro}, {Fern{\'a}ndez-Hern{\'a}ndez}, {Galluccio}, {Garc{\'\i}a-Lario},
  {Garcia-Reinaldos}, {Gonz{\'a}lez-N{\'u}{\~n}ez}, {Gosset}, {Haigron},
  {Halbwachs}, {Hambly}, {Harrison}, {Hatzidimitriou}, {Heiter},
  {Hern{\'a}ndez}, {Hestroffer}, {Hodgkin}, {Holl}, {Jan{\ss}en}, {Jevardat de
  Fombelle}, {Jordan}, {Krone-Martins}, {Lanzafame}, {L{\"o}ffler}, {Lorca},
  {Manteiga}, {Marchal}, {Marrese}, {Moitinho}, {Mora}, {Muinonen}, {Osborne},
  {Pancino}, {Pauwels}, {Petit}, {Recio-Blanco}, {Richards}, {Riello},
  {Rimoldini}, {Robin}, {Roegiers}, {Rybizki}, {Sarro}, {Siopis}, {Smith},
  {Sozzetti}, {Ulla}, {Utrilla}, {van Leeuwen}, {van Reeven}, {Abbas}, {Abreu
  Aramburu}, {Accart}, {Aerts}, {Aguado}, {Ajaj}, {Altavilla}, {{\'A}lvarez},
  {{\'A}lvarez Cid-Fuentes}, {Alves}, {Anderson}, {Anglada Varela}, {Antoja},
  {Audard}, {Baines}, {Baker}, {Balaguer-N{\'u}{\~n}ez}, {Balbinot}, {Balog},
  {Barache}, {Barbato}, {Barros}, {Barstow}, {Bartolom{\'e}}, {Bassilana},
  {Bauchet}, {Baudesson-Stella}, {Becciani}, {Bellazzini}, {Bernet}, {Bertone},
  {Bianchi}, {Blanco-Cuaresma}, {Boch}, {Bombrun}, {Bossini}, {Bouquillon},
  {Bragaglia}, {Bramante}, {Breedt}, {Bressan}, {Brouillet}, {Bucciarelli},
  {Burlacu}, {Busonero}, {Butkevich}, {Buzzi}, {Caffau}, {Cancelliere},
  {C{\'a}novas}, {Cantat-Gaudin}, {Carballo}, {Carlucci}, {Carnerero},
  {Carrasco}, {Casamiquela}, {Castellani}, {Castro-Ginard}, {Castro Sampol},
  {Chaoul}, {Charlot}, {Chemin}, {Chiavassa}, {Cioni}, {Comoretto}, {Cooper},
  {Cornez}, {Cowell}, {Crifo}, {Crosta}, {Crowley}, {Dafonte}, {Dapergolas},
  {David}, {David}, {de Laverny}, {De Luise}, {De March}, {De Ridder}, {de
  Souza}, {de Teodoro}, {de Torres}, {del Peloso}, {del Pozo}, {Delbo},
  {Delgado}, {Delgado}, {Delisle}, {Di Matteo}, {Diakite}, {Diener},
  {Distefano}, {Dolding}, {Eappachen}, {Edvardsson}, {Enke}, {Esquej}, {Fabre},
  {Fabrizio}, {Faigler}, {Fedorets}, {Fernique}, {Fienga}, {Figueras},
  {Fouron}, {Fragkoudi}, {Fraile}, {Franke}, {Gai}, {Garabato},
  {Garcia-Gutierrez}, {Garc{\'\i}a-Torres}, {Garofalo}, {Gavras}, {Gerlach},
  {Geyer}, {Giacobbe}, {Gilmore}, {Girona}, {Giuffrida}, {Gomel}, {Gomez},
  {Gonzalez-Santamaria}, {Gonz{\'a}lez-Vidal}, {Granvik},
  {Guti{\'e}rrez-S{\'a}nchez}, {Guy}, {Hauser}, {Haywood}, {Helmi}, {Hidalgo},
  {Hilger}, {H{\l}adczuk}, {Hobbs}, {Holland}, {Huckle}, {Jasniewicz},
  {Jonker}, {Juaristi Campillo}, {Julbe}, {Karbevska}, {Kervella}, {Khanna},
  {Kochoska}, {Kontizas}, {Kordopatis}, {Korn}, {Kostrzewa-Rutkowska},
  {Kruszy{\'n}ska}, {Lambert}, {Lanza}, {Lasne}, {Le Campion}, {Le Fustec},
  {Lebreton}, {Lebzelter}, {Leccia}, {Leclerc}, {Lecoeur-Taibi}, {Liao},
  {Licata}, {Lindstr{\o}m}, {Lister}, {Livanou}, {Lobel}, {Madrero Pardo},
  {Managau}, {Mann}, {Marchant}, {Marconi}, {Marcos Santos}, {Marinoni},
  {Marocco}, {Marshall}, {Martin Polo}, {Mart{\'\i}n-Fleitas}, {Masip},
  {Massari}, {Mastrobuono-Battisti}, {Mazeh}, {McMillan}, {Messina},
  {Michalik}, {Millar}, {Mints}, {Molina}, {Molinaro}, {Moln{\'a}r},
  {Montegriffo}, {Mor}, {Morbidelli}, {Morel}, {Morris}, {Mulone}, {Munoz},
  {Muraveva}, {Murphy}, {Musella}, {Noval}, {Ord{\'e}novic}, {Orr{\`u}},
  {Osinde}, {Pagani}, {Pagano}, {Palaversa}, {Palicio}, {Panahi}, {Pawlak},
  {Pe{\~n}alosa Esteller}, {Penttil{\"a}}, {Piersimoni}, {Pineau}, {Plachy},
  {Plum}, {Poggio}, {Poretti}, {Poujoulet}, {Pr{\v{s}}a}, {Pulone}, {Racero},
  {Ragaini}, {Rainer}, {Raiteri}, {Rambaux}, {Ramos}, {Ramos-Lerate}, {Re
  Fiorentin}, {Regibo}, {Reyl{\'e}}, {Ripepi}, {Riva}, {Rixon}, {Robichon},
  {Robin}, {Roelens}, {Rohrbasser}, {Romero-G{\'o}mez}, {Rowell}, {Royer},
  {Rybicki}, {Sadowski}, {Sagrist{\`a} Sell{\'e}s}, {Sahlmann}, {Salgado},
  {Salguero}, {Samaras}, {Sanchez Gimenez}, {Sanna}, {Santove{\~n}a},
  {Sarasso}, {Schultheis}, {Sciacca}, {Segol}, {Segovia}, {S{\'e}gransan},
  {Semeux}, {Shahaf}, {Siddiqui}, {Siebert}, {Siltala}, {Slezak}, {Smart},
  {Solano}, {Solitro}, {Souami}, {Souchay}, {Spagna}, {Spoto}, {Steele},
  {Steidelm{\"u}ller}, {Stephenson}, {S{\"u}veges}, {Szabados}, {Szegedi-Elek},
  {Taris}, {Tauran}, {Taylor}, {Teixeira}, {Thuillot}, {Tonello}, {Torra},
  {Torra}, {Turon}, {Unger}, {Vaillant}, {van Dillen}, {Vanel}, {Vecchiato},
  {Viala}, {Vicente}, {Voutsinas}, {Weiler}, {Wevers}, {Wyrzykowski}, {Yoldas},
  {Yvard}, {Zhao}, {Zorec}, {Zucker}, {Zurbach}, \&
  {Zwitter}}]{Gaia_Collaboration_2021}
{Gaia Collaboration}, {Brown}, A.~G.~A., {Vallenari}, A., {et~al.} 2021, \aap,
  649, A1

\bibitem[{{Gallenne} {et~al.}(2019){Gallenne}, {Pietrzy{\'n}ski}, {Graczyk},
  {Pilecki}, {Storm}, {Nardetto}, {Taormina}, {Gieren}, {Tkachenko},
  {Kervella}, {M{\'e}rand}, \& {Weber}}]{Gallenne_2019}
{Gallenne}, A., {Pietrzy{\'n}ski}, G., {Graczyk}, D., {et~al.} 2019, \aap, 632,
  A31

\bibitem[{{Girardi} {et~al.}(2002){Girardi}, {Bertelli}, {Bressan}, {Chiosi},
  {Groenewegen}, {Marigo}, {Salasnich}, \& {Weiss}}]{Girardi_2002}
{Girardi}, L., {Bertelli}, G., {Bressan}, A., {et~al.} 2002, \aap, 391, 195

\bibitem[{{Goodman} \& {Weare}(2010)}]{Goodman2010}
{Goodman}, J. \& {Weare}, J. 2010, Communications in Applied Mathematics and
  Computational Science, 5, 65

\bibitem[{{Gray}(2014)}]{Gray_2014}
{Gray}, D.~F. 2014, \aj, 147, 81

\bibitem[{{Guinan} {et~al.}(2000){Guinan}, {Ribas}, {Fitzpatrick},
  {Gim{\'e}nez}, {Jordi}, {McCook}, \& {Popper}}]{Guinan2000}
{Guinan}, E.~F., {Ribas}, I., {Fitzpatrick}, E.~L., {et~al.} 2000, \apj, 544,
  409

\bibitem[{{G{\"u}zel} \& {{\"O}zdarcan}(2020)}]{Guzel2020}
{G{\"u}zel}, O. \& {{\"O}zdarcan}, O. 2020, Contributions of the Astronomical
  Observatory Skalnate Pleso, 50, 535

\bibitem[{{Hadrava}(1995)}]{Hadrava_1995}
{Hadrava}, P. 1995, \aaps, 114, 393

\bibitem[{{Harper}(1907)}]{Harper_1907}
{Harper}, W.~E. 1907, \jrasc, 1, 237

\bibitem[{{Hensberge} {et~al.}(2000){Hensberge}, {Pavlovski}, \&
  {Verschueren}}]{Hensberge_2000}
{Hensberge}, H., {Pavlovski}, K., \& {Verschueren}, W. 2000, \aap, 358, 553

\bibitem[{{Higl} \& {Weiss}(2017)}]{Higl_Weiss_2017}
{Higl}, J. \& {Weiss}, A. 2017, \aap, 608, A62

\bibitem[{{Hummel} {et~al.}(2003){Hummel}, {Benson}, {Hutter}, {Johnston},
  {Mozurkewich}, {Armstrong}, {Hindsley}, {Gilbreath}, {Rickard}, \&
  {White}}]{Hummel_2003}
{Hummel}, C.~A., {Benson}, J.~A., {Hutter}, D.~J., {et~al.} 2003, \aj, 125,
  2630

\bibitem[{{Hummel} {et~al.}(1998){Hummel}, {Mozurkewich}, {Armstrong},
  {Hajian}, {Elias}, \& {Hutter}}]{Hummel_1998}
{Hummel}, C.~A., {Mozurkewich}, D., {Armstrong}, J.~T., {et~al.} 1998, \aj,
  116, 2536

\bibitem[{{Hutter} {et~al.}(2016){Hutter}, {Zavala}, {Tycner}, {Benson},
  {Hummel}, {Sanborn}, {Franz}, \& {Johnston}}]{Hutter_2016}
{Hutter}, D.~J., {Zavala}, R.~T., {Tycner}, C., {et~al.} 2016, \apjs, 227, 4

\bibitem[{{Ilijic} {et~al.}(2004){Ilijic}, {Hensberge}, {Pavlovski}, \&
  {Freyhammer}}]{Ilijic_2004}
{Ilijic}, S., {Hensberge}, H., {Pavlovski}, K., \& {Freyhammer}, L.~M. 2004, in
  Astronomical Society of the Pacific Conference Series, Vol. 318,
  Spectroscopically and Spatially Resolving the Components of the Close Binary
  Stars, ed. R.~W. {Hilditch}, H.~{Hensberge}, \& K.~{Pavlovski}, 111--113

\bibitem[{{Johnston} {et~al.}(2019{\natexlab{a}}){Johnston}, {Pavlovski}, \&
  {Tkachenko}}]{Johnston2019b}
{Johnston}, C., {Pavlovski}, K., \& {Tkachenko}, A. 2019{\natexlab{a}}, \aap,
  628, A25

\bibitem[{{Johnston} {et~al.}(2019{\natexlab{b}}){Johnston}, {Tkachenko},
  {Aerts}, {Molenberghs}, {Bowman}, {Pedersen}, {Buysschaert}, \&
  {P{\'a}pics}}]{Johnston2019}
{Johnston}, C., {Tkachenko}, A., {Aerts}, C., {et~al.} 2019{\natexlab{b}},
  \mnras, 482, 1231

\bibitem[{{Kallinger} {et~al.}(2004){Kallinger}, {Iliev}, {Lehmann}, \&
  {Weiss}}]{Kallinger_2004}
{Kallinger}, T., {Iliev}, I., {Lehmann}, H., \& {Weiss}, W.~W. 2004, in The
  A-Star Puzzle, ed. J.~{Zverko}, J.~{Ziznovsky}, S.~J. {Adelman}, \& W.~W.
  {Weiss}, Vol. 224, 848--852

\bibitem[{{Kervella} {et~al.}(2019){Kervella}, {Arenou}, {Mignard}, \&
  {Th{\'e}venin}}]{Kervella_2019}
{Kervella}, P., {Arenou}, F., {Mignard}, F., \& {Th{\'e}venin}, F. 2019, \aap,
  623, A72

\bibitem[{{Kolbas} {et~al.}(2015){Kolbas}, {Pavlovski}, {Southworth}, {Lee},
  {Lee}, {Lee}, {Kim}, {Kim}, {Smalley}, \& {Tkachenko}}]{Kolbas_2015}
{Kolbas}, V., {Pavlovski}, K., {Southworth}, J., {et~al.} 2015, \mnras, 451,
  4150

\bibitem[{{Kroupa} {et~al.}(1993){Kroupa}, {Tout}, \& {Gilmore}}]{Kroupa_1993}
{Kroupa}, P., {Tout}, C.~A., \& {Gilmore}, G. 1993, \mnras, 262, 545

\bibitem[{{Lallement} {et~al.}(2018){Lallement}, {Capitanio}, {Ruiz-Dern},
  {Danielski}, {Babusiaux}, {Vergely}, {Elyajouri}, {Arenou}, \&
  {Leclerc}}]{Lallement_2018}
{Lallement}, R., {Capitanio}, L., {Ruiz-Dern}, L., {et~al.} 2018, \aap, 616,
  A132

\bibitem[{{Lallement} {et~al.}(2014){Lallement}, {Vergely}, {Valette},
  {Puspitarini}, {Eyer}, \& {Casagrande}}]{Lallement_2014}
{Lallement}, R., {Vergely}, J.~L., {Valette}, B., {et~al.} 2014, \aap, 561, A91

\bibitem[{{Lehmann} {et~al.}(2020){Lehmann}, {Dervi{\c{s}}o{\u{g}}lu},
  {Mkrtichian}, {Pertermann}, {Tkachenko}, \& {Tsymbal}}]{Lehmann2020}
{Lehmann}, H., {Dervi{\c{s}}o{\u{g}}lu}, A., {Mkrtichian}, D.~E., {et~al.}
  2020, \aap, 644, A121

\bibitem[{{Lightkurve Collaboration} {et~al.}(2018){Lightkurve Collaboration},
  {Cardoso}, {Hedges}, {Gully-Santiago}, {Saunders}, {Cody}, {Barclay}, {Hall},
  {Sagear}, {Turtelboom}, {Zhang}, {Tzanidakis}, {Mighell}, {Coughlin}, {Bell},
  {Berta-Thompson}, {Williams}, {Dotson}, \& {Barentsen}}]{Lightkurve}
{Lightkurve Collaboration}, {Cardoso}, J.~V.~d.~M., {Hedges}, C., {et~al.}
  2018, {Lightkurve: Kepler and TESS time series analysis in Python},
  Astrophysics Source Code Library

\bibitem[{{Luo} {et~al.}(2021){Luo}, {Zhao}, {Li}, {Guo}, \& {Liu}}]{Luo2021}
{Luo}, F., {Zhao}, Y.-H., {Li}, J., {Guo}, Y.-J., \& {Liu}, C. 2021, arXiv
  e-prints, arXiv:2108.11120

\bibitem[{{Makarov} \& {Kaplan}(2005)}]{Makarov_Kaplan_2005}
{Makarov}, V.~V. \& {Kaplan}, G.~H. 2005, \aj, 129, 2420

\bibitem[{{Mantegazza} {et~al.}(2010){Mantegazza}, {Rainer}, \&
  {Antonello}}]{Mantegazza_2010}
{Mantegazza}, L., {Rainer}, M., \& {Antonello}, E. 2010, \aap, 512, A42

\bibitem[{{Martinet} {et~al.}(2021){Martinet}, {Meynet}, {Ekstr{\"o}m},
  {Sim{\'o}n-D{\'\i}az}, {Holgado}, {Castro}, {Georgy}, {Eggenberger},
  {Buldgen}, {Salmon}, {Hirschi}, {Groh}, {Farrell}, \&
  {Murphy}}]{Martinet_2021}
{Martinet}, S., {Meynet}, G., {Ekstr{\"o}m}, S., {et~al.} 2021, \aap, 648, A126

\bibitem[{{Mayer} {et~al.}(2013){Mayer}, {Harmanec}, \&
  {Pavlovski}}]{Mayer_2013}
{Mayer}, P., {Harmanec}, P., \& {Pavlovski}, K. 2013, \aap, 550, A2

\bibitem[{{Monier}(2021)}]{Monier_2021}
{Monier}, R. 2021, Research Notes of the American Astronomical Society, 5, 163

\bibitem[{{Mowlavi} {et~al.}(2013){Mowlavi}, {Barblan}, {Saesen}, \&
  {Eyer}}]{Mowlavi2013}
{Mowlavi}, N., {Barblan}, F., {Saesen}, S., \& {Eyer}, L. 2013, \aap, 554, A108

\bibitem[{{Mozurkewich} {et~al.}(2003){Mozurkewich}, {Armstrong}, {Hindsley},
  {Quirrenbach}, {Hummel}, {Hutter}, {Johnston}, {Hajian}, {Elias}, {Buscher},
  \& {Simon}}]{Mozurkewich_2003}
{Mozurkewich}, D., {Armstrong}, J.~T., {Hindsley}, R.~B., {et~al.} 2003, \aj,
  126, 2502

\bibitem[{{Murphy} {et~al.}(2015){Murphy}, {Corbally}, {Gray}, {Cheng}, {Neff},
  {Koen}, {Kuehn}, {Newsome}, \& {Riggs}}]{Murphy_2015}
{Murphy}, S.~J., {Corbally}, C.~J., {Gray}, R.~O., {et~al.} 2015, \pasa, 32,
  e036

\bibitem[{{Nieva} \& {Przybilla}(2012)}]{Nieva2012}
{Nieva}, M.~F. \& {Przybilla}, N. 2012, \aap, 539, A143

\bibitem[{{Oja}(1993)}]{Oja1993}
{Oja}, T. 1993, \aaps, 100, 591

\bibitem[{{Pavlovski} \& {Hensberge}(2005)}]{Pavlovski_Hensberge_2005}
{Pavlovski}, K. \& {Hensberge}, H. 2005, \aap, 439, 309

\bibitem[{{Pavlovski} \& {Hensberge}(2010)}]{Pavlovski_Hensberge_2010}
{Pavlovski}, K. \& {Hensberge}, H. 2010, in Astronomical Society of the Pacific
  Conference Series, Vol. 435, Binaries - Key to Comprehension of the Universe,
  ed. A.~{Pr{\v{s}}a} \& M.~{Zejda}, 207

\bibitem[{{Pavlovski} {et~al.}(2018){Pavlovski}, {Southworth}, \&
  {Tamajo}}]{Pavlovski_2018}
{Pavlovski}, K., {Southworth}, J., \& {Tamajo}, E. 2018, \mnras, 481, 3129

\bibitem[{{Pavlovski} {et~al.}(2009){Pavlovski}, {Tamajo}, {Koubsk{\'y}},
  {Southworth}, {Yang}, \& {Kolbas}}]{Pavlovski_2009}
{Pavlovski}, K., {Tamajo}, E., {Koubsk{\'y}}, P., {et~al.} 2009, \mnras, 400,
  791

\bibitem[{{Paxton} {et~al.}(2011){Paxton}, {Bildsten}, {Dotter}, {Herwig},
  {Lesaffre}, \& {Timmes}}]{Paxton2011}
{Paxton}, B., {Bildsten}, L., {Dotter}, A., {et~al.} 2011, \apjs, 192, 3

\bibitem[{{Paxton} {et~al.}(2013){Paxton}, {Cantiello}, {Arras}, {Bildsten},
  {Brown}, {Dotter}, {Mankovich}, {Montgomery}, {Stello}, {Timmes}, \&
  {Townsend}}]{Paxton2013}
{Paxton}, B., {Cantiello}, M., {Arras}, P., {et~al.} 2013, \apjs, 208, 4

\bibitem[{{Paxton} {et~al.}(2015){Paxton}, {Marchant}, {Schwab}, {Bauer},
  {Bildsten}, {Cantiello}, {Dessart}, {Farmer}, {Hu}, {Langer}, {Townsend},
  {Townsley}, \& {Timmes}}]{Paxton2015}
{Paxton}, B., {Marchant}, P., {Schwab}, J., {et~al.} 2015, \apjs, 220, 15

\bibitem[{{Paxton} {et~al.}(2018){Paxton}, {Schwab}, {Bauer}, {Bildsten},
  {Blinnikov}, {Duffell}, {Farmer}, {Goldberg}, {Marchant}, {Sorokina},
  {Thoul}, {Townsend}, \& {Timmes}}]{Paxton2018}
{Paxton}, B., {Schwab}, J., {Bauer}, E.~B., {et~al.} 2018, \apjs, 234, 34

\bibitem[{{Paxton} {et~al.}(2019){Paxton}, {Smolec}, {Schwab}, {Gautschy},
  {Bildsten}, {Cantiello}, {Dotter}, {Farmer}, {Goldberg}, {Jermyn}, {Kanbur},
  {Marchant}, {Thoul}, {Townsend}, {Wolf}, {Zhang}, \& {Timmes}}]{Paxton2019}
{Paxton}, B., {Smolec}, R., {Schwab}, J., {et~al.} 2019, \apjs, 243, 10

\bibitem[{{Pearce}(1957)}]{Pearce_1957}
{Pearce}, J.~A. 1957, Publications of the Dominion Astrophysical Observatory
  Victoria, 10, 331

\bibitem[{{Pedersen} {et~al.}(2018){Pedersen}, {Aerts}, {P{\'a}pics}, \&
  {Rogers}}]{Pedersen2018}
{Pedersen}, M.~G., {Aerts}, C., {P{\'a}pics}, P.~I., \& {Rogers}, T.~M. 2018,
  \aap, 614, A128

\bibitem[{{Perryman} {et~al.}(1997){Perryman}, {Lindegren}, {Kovalevsky},
  {Hog}, {Bastian}, {Bernacca}, {Creze}, {Donati}, {Grenon}, {Grewing}, {van
  Leeuwen}, {van der Marel}, {Mignard}, {Murray}, {Le Poole}, {Schrijver},
  {Turon}, {Arenou}, {Froeschle}, \& {Petersen}}]{Perryman_1997}
{Perryman}, M.~A.~C., {Lindegren}, L., {Kovalevsky}, J., {et~al.} 1997, \aap,
  500, 501

\bibitem[{{Pols} {et~al.}(1997){Pols}, {Tout}, {Schroder}, {Eggleton}, \&
  {Manners}}]{Pols_1997}
{Pols}, O.~R., {Tout}, C.~A., {Schroder}, K.-P., {Eggleton}, P.~P., \&
  {Manners}, J. 1997, \mnras, 289, 869

\bibitem[{{Przybilla} {et~al.}(2008){Przybilla}, {Nieva}, \&
  {Butler}}]{Przybilla2008}
{Przybilla}, N., {Nieva}, M.-F., \& {Butler}, K. 2008, \apjl, 688, L103

\bibitem[{{Raskin} {et~al.}(2011){Raskin}, {van Winckel}, {Hensberge},
  {Jorissen}, {Lehmann}, {Waelkens}, {Avila}, {de Cuyper}, {Degroote},
  {Dubosson}, {Dumortier}, {Fr{\'e}mat}, {Laux}, {Michaud}, {Morren}, {Perez
  Padilla}, {Pessemier}, {Prins}, {Smolders}, {van Eck}, \&
  {Winkler}}]{Raskin_2011}
{Raskin}, G., {van Winckel}, H., {Hensberge}, H., {et~al.} 2011, \aap, 526, A69

\bibitem[{{Ricker} {et~al.}(2014){Ricker}, {Winn}, {Vanderspek}, {Latham},
  {Bakos}, {Bean}, {Berta-Thompson}, {Brown}, {Buchhave}, {Butler}, {Butler},
  {Chaplin}, {Charbonneau}, {Christensen-Dalsgaard}, {Clampin}, {Deming},
  {Doty}, {De Lee}, {Dressing}, {Dunham}, {Endl}, {Fressin}, {Ge}, {Henning},
  {Holman}, {Howard}, {Ida}, {Jenkins}, {Jernigan}, {Johnson}, {Kaltenegger},
  {Kawai}, {Kjeldsen}, {Laughlin}, {Levine}, {Lin}, {Lissauer}, {MacQueen},
  {Marcy}, {McCullough}, {Morton}, {Narita}, {Paegert}, {Palle}, {Pepe},
  {Pepper}, {Quirrenbach}, {Rinehart}, {Sasselov}, {Sato}, {Seager},
  {Sozzetti}, {Stassun}, {Sullivan}, {Szentgyorgyi}, {Torres}, {Udry}, \&
  {Villasenor}}]{Ricker2014}
{Ricker}, G.~R., {Winn}, J.~N., {Vanderspek}, R., {et~al.} 2014, in Society of
  Photo-Optical Instrumentation Engineers (SPIE) Conference Series, Vol. 9143,
  Space Telescopes and Instrumentation 2014: Optical, Infrared, and Millimeter
  Wave, ed. J.~{Oschmann}, Jacobus~M., M.~{Clampin}, G.~G. {Fazio}, \& H.~A.
  {MacEwen}, 914320

\bibitem[{{Ricker} {et~al.}(2015){Ricker}, {Winn}, {Vanderspek}, {Latham},
  {Bakos}, {Bean}, {Berta-Thompson}, {Brown}, {Buchhave}, {Butler}, {Butler},
  {Chaplin}, {Charbonneau}, {Christensen-Dalsgaard}, {Clampin}, {Deming},
  {Doty}, {De Lee}, {Dressing}, {Dunham}, {Endl}, {Fressin}, {Ge}, {Henning},
  {Holman}, {Howard}, {Ida}, {Jenkins}, {Jernigan}, {Johnson}, {Kaltenegger},
  {Kawai}, {Kjeldsen}, {Laughlin}, {Levine}, {Lin}, {Lissauer}, {MacQueen},
  {Marcy}, {McCullough}, {Morton}, {Narita}, {Paegert}, {Palle}, {Pepe},
  {Pepper}, {Quirrenbach}, {Rinehart}, {Sasselov}, {Sato}, {Seager},
  {Sozzetti}, {Stassun}, {Sullivan}, {Szentgyorgyi}, {Torres}, {Udry}, \&
  {Villasenor}}]{Ricker2015}
{Ricker}, G.~R., {Winn}, J.~N., {Vanderspek}, R., {et~al.} 2015, Journal of
  Astronomical Telescopes, Instruments, and Systems, 1, 014003

\bibitem[{{Rogers} \& {McElwaine}(2017)}]{Rogers2017}
{Rogers}, T.~M. \& {McElwaine}, J.~N. 2017, \apjl, 848, L1

\bibitem[{{Royer} {et~al.}(2002){Royer}, {Grenier}, {Baylac}, {G{\'o}mez}, \&
  {Zorec}}]{Royer_2002b}
{Royer}, F., {Grenier}, S., {Baylac}, M.~O., {G{\'o}mez}, A.~E., \& {Zorec}, J.
  2002, \aap, 393, 897

\bibitem[{{Royer} {et~al.}(2007){Royer}, {Zorec}, \& {G{\'o}mez}}]{Royer_2007}
{Royer}, F., {Zorec}, J., \& {G{\'o}mez}, A.~E. 2007, \aap, 463, 671

\bibitem[{{Sana} {et~al.}(2013){Sana}, {de Koter}, {de Mink}, {Dunstall},
  {Evans}, {H{\'e}nault-Brunet}, {Ma{\'\i}z Apell{\'a}niz},
  {Ram{\'\i}rez-Agudelo}, {Taylor}, {Walborn}, {Clark}, {Crowther}, {Herrero},
  {Gieles}, {Langer}, {Lennon}, \& {Vink}}]{Sana2013}
{Sana}, H., {de Koter}, A., {de Mink}, S.~E., {et~al.} 2013, \aap, 550, A107

\bibitem[{{Sana} {et~al.}(2012){Sana}, {de Mink}, {de Koter}, {Langer},
  {Evans}, {Gieles}, {Gosset}, {Izzard}, {Le Bouquin}, \&
  {Schneider}}]{Sana2012}
{Sana}, H., {de Mink}, S.~E., {de Koter}, A., {et~al.} 2012, Science, 337, 444

\bibitem[{{Sana} {et~al.}(2014){Sana}, {Le Bouquin}, {Lacour}, {Berger},
  {Duvert}, {Gauchet}, {Norris}, {Olofsson}, {Pickel}, {Zins}, {Absil}, {de
  Koter}, {Kratter}, {Schnurr}, \& {Zinnecker}}]{Sana_2014}
{Sana}, H., {Le Bouquin}, J.~B., {Lacour}, S., {et~al.} 2014, \apjs, 215, 15

\bibitem[{{Sch{\"o}ller} {et~al.}(2010){Sch{\"o}ller}, {Correia}, {Hubrig}, \&
  {Ageorges}}]{Scholler_2010}
{Sch{\"o}ller}, M., {Correia}, S., {Hubrig}, S., \& {Ageorges}, N. 2010, \aap,
  522, A85

\bibitem[{{Schr{\"o}der} \& {Schmitt}(2007)}]{Schroder_Schmitt_2007}
{Schr{\"o}der}, C. \& {Schmitt}, J.~H.~M.~M. 2007, \aap, 475, 677

\bibitem[{{Serenelli} {et~al.}(2021){Serenelli}, {Weiss}, {Aerts}, {Angelou},
  {Baroch}, {Bastian}, {Beck}, {Bergemann}, {Bestenlehner}, {Czekala},
  {Elias-Rosa}, {Escorza}, {Van Eylen}, {Feuillet}, {Gandolfi}, {Gieles},
  {Girardi}, {Lebreton}, {Lodieu}, {Martig}, {Miller Bertolami}, {Mombarg},
  {Morales}, {Moya}, {Nsamba}, {Pavlovski}, {Pedersen}, {Ribas}, {Schneider},
  {Silva Aguirre}, {Stassun}, {Tolstoy}, {Tremblay}, \&
  {Zwintz}}]{Serenelli_2021}
{Serenelli}, A., {Weiss}, A., {Aerts}, C., {et~al.} 2021, \aapr, 29, 4

\bibitem[{{Shao} {et~al.}(1988){Shao}, {Colavita}, {Hines}, {Staelin},
  {Hutter}, {Johnston}, {Mozurkewich}, {Simon}, {Hershey}, {Hughes}, \&
  {Kaplan}}]{Shao_1988}
{Shao}, M., {Colavita}, M.~M., {Hines}, B.~E., {et~al.} 1988, \aap, 193, 357

\bibitem[{{Shorlin} {et~al.}(2002){Shorlin}, {Wade}, {Donati}, {Landstreet},
  {Petit}, {Sigut}, \& {Strasser}}]{Shorlin_2002}
{Shorlin}, S.~L.~S., {Wade}, G.~A., {Donati}, J.~F., {et~al.} 2002, \aap, 392,
  637

\bibitem[{{Shulyak} {et~al.}(2004){Shulyak}, {Tsymbal}, {Ryabchikova},
  {St{\"u}tz}, \& {Weiss}}]{Shulyak2004}
{Shulyak}, D., {Tsymbal}, V., {Ryabchikova}, T., {St{\"u}tz}, C., \& {Weiss},
  W.~W. 2004, \aap, 428, 993

\bibitem[{{Simon} \& {Sturm}(1994)}]{Simon_Sturm_1994}
{Simon}, K.~P. \& {Sturm}, E. 1994, \aap, 281, 286

\bibitem[{{Smith} {et~al.}(2012){Smith}, {Stumpe}, {Van Cleve}, {Jenkins},
  {Barclay}, {Fanelli}, {Girouard}, {Kolodziejczak}, {McCauliff}, {Morris}, \&
  {Twicken}}]{Smithetal.2012}
{Smith}, J.~C., {Stumpe}, M.~C., {Van Cleve}, J.~E., {et~al.} 2012, \pasp, 124,
  1000

\bibitem[{{Southworth}(2015)}]{Southworth_2015}
{Southworth}, J. 2015, in Astronomical Society of the Pacific Conference
  Series, Vol. 496, Living Together: Planets, Host Stars and Binaries, ed.
  S.~M. {Rucinski}, G.~{Torres}, \& M.~{Zejda}, 164

\bibitem[{{Southworth} {et~al.}(2005){Southworth}, {Maxted}, \&
  {Smalley}}]{Southworth_2005}
{Southworth}, J., {Maxted}, P.~F.~L., \& {Smalley}, B. 2005, \aap, 429, 645

\bibitem[{{Stelzer} {et~al.}(2011){Stelzer}, {Hummel}, {Sch{\"o}ller},
  {Hubrig}, \& {Cowley}}]{Stelzer_2011}
{Stelzer}, B., {Hummel}, C.~A., {Sch{\"o}ller}, M., {Hubrig}, S., \& {Cowley},
  C. 2011, \aap, 529, A29

\bibitem[{{Struve}(1955)}]{Struve_1955}
{Struve}, O. 1955, \skytel, 14, 461

\bibitem[{{Struve} {et~al.}(1957){Struve}, {Sahade}, {Lynds}, \&
  {Huang}}]{Struve_1957}
{Struve}, O., {Sahade}, J., {Lynds}, C.~R., \& {Huang}, S.~S. 1957, \apj, 125,
  115

\bibitem[{{Theme{\ss}l} {et~al.}(2018){Theme{\ss}l}, {Hekker}, {Southworth},
  {Beck}, {Pavlovski}, {Tkachenko}, {Angelou}, {Ball}, {Barban}, {Corsaro},
  {Elsworth}, {Handberg}, \& {Kallinger}}]{Themessl_2018}
{Theme{\ss}l}, N., {Hekker}, S., {Southworth}, J., {et~al.} 2018, \mnras, 478,
  4669

\bibitem[{{Tkachenko}(2015)}]{Tkachenko2015}
{Tkachenko}, A. 2015, \aap, 581, A129

\bibitem[{{Tkachenko} {et~al.}(2020){Tkachenko}, {Pavlovski}, {Johnston},
  {Pedersen}, {Michielsen}, {Bowman}, {Southworth}, {Tsymbal}, \&
  {Aerts}}]{Tkachenko_2020}
{Tkachenko}, A., {Pavlovski}, K., {Johnston}, C., {et~al.} 2020, \aap, 637, A60

\bibitem[{{Torres}(2010)}]{Torres_corr_2010}
{Torres}, G. 2010, \aj, 140, 1158

\bibitem[{{Torres} {et~al.}(2010){Torres}, {Andersen}, \&
  {Gim{\'e}nez}}]{Torres_2010}
{Torres}, G., {Andersen}, J., \& {Gim{\'e}nez}, A. 2010, \aapr, 18, 67

\bibitem[{{Torres} {et~al.}(2017){Torres}, {McGruder}, {Siverd}, {Rodriguez},
  {Pepper}, {Stevens}, {Stassun}, {Lund}, \& {James}}]{Torres_2017}
{Torres}, G., {McGruder}, C.~D., {Siverd}, R.~J., {et~al.} 2017, \apj, 836, 177

\bibitem[{{Torres} {et~al.}(2014){Torres}, {Sandberg Lacy}, {Pavlovski},
  {Feiden}, {Sabby}, {Bruntt}, \& {Viggo Clausen}}]{Torres_2014}
{Torres}, G., {Sandberg Lacy}, C.~H., {Pavlovski}, K., {et~al.} 2014, \apj,
  797, 31

\bibitem[{{Tsymbal}(1996)}]{Tsymbal1996}
{Tsymbal}, V. 1996, in Astronomical Society of the Pacific Conference Series,
  Vol. 108, M.A.S.S., Model Atmospheres and Spectrum Synthesis, ed. S.~J.
  {Adelman}, F.~{Kupka}, \& W.~W. {Weiss}, 198

\bibitem[{{van Altena} {et~al.}(1995){van Altena}, {Lee}, \&
  {Hoffleit}}]{van_Altena_1995}
{van Altena}, W.~F., {Lee}, J.~T., \& {Hoffleit}, E.~D. 1995, {The general
  catalogue of trigonometric [stellar] parallaxes}

\bibitem[{{van Belle} {et~al.}(2009){van Belle}, {Creech-Eakman}, \&
  {Hart}}]{van_Belle_2009}
{van Belle}, G.~T., {Creech-Eakman}, M.~J., \& {Hart}, A. 2009, \mnras, 394,
  1925

\bibitem[{{van Hamme}(1993)}]{van_Hamme_1993}
{van Hamme}, W. 1993, \aj, 106, 2096

\bibitem[{{van Leeuwen}(2007)}]{van_Leeuwen_2007}
{van Leeuwen}, F. 2007, \aap, 474, 653

\bibitem[{{Va{\v{n}}ko} {et~al.}(2020){Va{\v{n}}ko}, {Pribulla},
  {Hamb{\'a}lek}, {Kundra}, {Kom{\v{z}}{\'\i}k}, {Garai}, {Budaj}, {Paunzen},
  {Zieli{\'n}ski}, \& {Zverko}}]{Vanko_2020}
{Va{\v{n}}ko}, M., {Pribulla}, T., {Hamb{\'a}lek}, {\v{L}}., {et~al.} 2020,
  Contributions of the Astronomical Observatory Skalnate Pleso, 50, 632

\bibitem[{{von Zeipel}(1924)}]{von_Zeipel_1924}
{von Zeipel}, H. 1924, \mnras, 84, 665

\bibitem[{{White} {et~al.}(2017){White}, {Pope}, {Antoci}, {P{\'a}pics},
  {Aerts}, {Gies}, {Gordon}, {Huber}, {Schaefer}, {Aigrain}, {Albrecht},
  {Barclay}, {Barentsen}, {Beck}, {Bedding}, {Fredslund Andersen}, {Grundahl},
  {Howell}, {Ireland}, {Murphy}, {Nielsen}, {Silva Aguirre}, \&
  {Tuthill}}]{White_2017}
{White}, T.~R., {Pope}, B.~J.~S., {Antoci}, V., {et~al.} 2017, \mnras, 471,
  2882

\bibitem[{{Wilson}(1979)}]{Wilson_1979}
{Wilson}, R.~E. 1979, \apj, 234, 1054

\bibitem[{{Wilson} \& {Devinney}(1971)}]{Wilson_Devinney_1971}
{Wilson}, R.~E. \& {Devinney}, E.~J. 1971, \apj, 166, 605

\bibitem[{{Wilson} \& {Van Hamme}(2014)}]{Wilson2014}
{Wilson}, R.~E. \& {Van Hamme}, W. 2014, \apj, 780, 151

\bibitem[{{Zorec} {et~al.}(2009){Zorec}, {Cidale}, {Arias}, {Fr{\'e}mat},
  {Muratore}, {Torres}, \& {Martayan}}]{Zorec_2009}
{Zorec}, J., {Cidale}, L., {Arias}, M.~L., {et~al.} 2009, \aap, 501, 297

\bibitem[{{Zverko} {et~al.}(2018){Zverko}, {Iliev}, {Romanyuk}, {Stateva},
  {Kudryavtsev}, \& {Semenko}}]{Zverko_2018}
{Zverko}, J., {Iliev}, I., {Romanyuk}, I.~I., {et~al.} 2018, Astrophysical
  Bulletin, 73, 351

\bibitem[{{Zverko} {et~al.}(2011){Zverko}, {{\v{Z}}i{\v{z}}{\v{n}}ovsk{\'y}},
  {Iliev}, {Barzova}, {Stateva}, {Romanyuk}, {Kudryavtsev}, \&
  {Semenko}}]{Zverko_2011}
{Zverko}, J., {{\v{Z}}i{\v{z}}{\v{n}}ovsk{\'y}}, J., {Iliev}, I., {et~al.}
  2011, Astrophysical Bulletin, 66, 325

\end{thebibliography}
%

\begin{appendix}

\section{Interferometric observations of $\alpha$~Dra}

\begin{table*}[h!]
\caption{Mark III (1989 - 1992) and NPOI (1997 - 2015) observations}
\label{tab:log}
\centering
\begin{tabular}{llllll}
\hline
\hline
UT Date&Julian Year&Triangles and baselines&$B_{\rm min} [m]$&$B_{\rm max} [m]$&Calibrators \\
 (1) & (2) & (3) & (4) & (5) & (6) \\
\hline
1989 Apr 18&1989.2945&NF-SC& 16& 17&FK5\,509\\
1991 Jun 19&1991.4627&NF-SC& 14& 17&FK5\,913\\
1992 Feb 28&1992.1584&NF-SC& 13& 17&FK5\,447 FK5\,456\\
1992 Apr 28&1992.3225&NF-SD& 19& 20&HR\,5062 FK5\,447 FK5\,509\\
1992 May 02&1992.3337&NC-SC&  8& 10&HR\,5062 HR\,5329 FK5\,456\\
1992 Jun 08&1992.4349&NF-SC& 12& 17&HR\,5062 FK5\,456\\
1992 Jul 06&1992.5113&NA-SA&  9& 10&HR\,5062\\
1997 Mar 19&1997.2125&AC-AE-AW& 18& 37&FK5\,472\\
1997 Mar 26&1997.2318&AC-AE-AW& 16& 37&FK5\,447 FK5\,472\\
1997 Apr 15&1997.2864&AC-AE-AW& 17& 37&FK5\,423 FK5\,472\\
1997 Apr 18&1997.2946&AC-AE-AW& 17& 37&FK5\,472\\
1997 May 08&1997.3493&AC-AE-AW& 17& 37&FK5\,423 FK5\,472 FK5\,668\\
1997 May 29&1997.4067&AC-AE-AW& 16& 37&FK5\,423 FK5\,668\\
1997 Jun 19&1997.4640&AC-AE-AW& 16& 37&FK5\,447 FK5\,456 FK5\,472 FK5\,509 FK5\,668\\
1997 Jun 21&1997.4695&AC-AE-AW& 17& 37&FK5\,447 FK5\,472 FK5\,509\\
1997 Jun 22&1997.4723&AC-AE-AW& 14& 37&FK5\,447 FK5\,472 FK5\,509 FK5\,668\\
1997 Jun 26&1997.4834&AC-AE-AW& 15& 36&FK5\,472 FK5\,509 FK5\,668\\
1997 Jun 27&1997.4859&AC-AE-AW& 17& 37&FK5\,472 FK5\,509\\
1997 Jul 04&1997.5051&AC-AE-AW& 14& 37&FK5\,447 FK5\,509\\
1997 Jul 08&1997.5159&AC-AE-AW& 17& 37&FK5\,509 FK5\,668\\
1997 Jul 09&1997.5188&AC-AE-AW& 15& 36&FK5\,509\\
1997 Jul 13&1997.5298&AC-AE-AW& 16& 37&FK5\,509\\
1997 Jul 14&1997.5325&AC-AE-AW& 15& 37&FK5\,509\\
1997 Jul 18&1997.5435&AC-AE-AW& 14& 35&FK5\,509\\
2013 Mar 05&2013.1743&AW-AC& 19& 53&FK5\,0472\\
2013 Mar 06&2013.1769&AW-AC& 19& 53&FK5\,0472\\
2013 Mar 07&2013.1797&AW-AC& 19& 53&FK5\,472\\
2013 May 24&2013.3931&AW-AC& 15& 50&FK5\,472\\
2013 May 25&2013.3956&AC-AW-E6& 16& 53&FK5\,0472\\
2013 May 26&2013.3985&AW-AC& 16& 50&FK5\,472\\
2013 May 27&2013.4012&AC-AW-E6& 16& 53&FK5\,472\\
2013 May 30&2013.4094&AC-AW-E6& 15& 53&FK5\,472\\
2013 May 31&2013.4121&AC-AW-E6& 16& 53&FK5\,472\\
2013 Jun 01&2013.4148&AC-AW-E6& 16& 53&FK5\,472\\
2013 Jun 03&2013.4203&AC-AW-E6& 16& 53&FK5\,472\\
2013 Jun 04&2013.4230&AC-AW-E6& 16& 53&FK5\,472\\
2013 Jun 05&2013.4258&AC-AW-E6& 15& 53&FK5\,472\\
2013 Jun 07&2013.4315&AW-AC& 15& 50&FK5\,472\\
2013 Jun 08&2013.4341&AW-AC& 16& 50&FK5\,423 FK5\,472\\
2013 Jun 10&2013.4395&AW-AC& 15& 50&FK5\,423 FK5\,472\\
2013 Jun 11&2013.4423&AW-AC& 15& 50&FK5\,423 FK5\,472\\
2013 Jun 13&2013.4477&AW-AC& 17& 51&FK5\,423 FK5\,472\\
2013 Jun 15&2013.4532&AW-AC& 16& 51&FK5\,423 FK5\,472\\
2013 Jun 16&2013.4559&AW-AC& 16& 51&FK5\,423 FK5\,472\\
2013 Jun 17&2013.4586&AW-AC& 16& 50&FK5\,423 FK5\,472\\
2014 Jun 24&2014.4772&AE-AW-E6 AW-E6-W7& 15& 79&HR\,5329\\
2014 Jun 25&2014.4799&AE-AW-E6 AW-E6-W7& 15& 79&HR\,5329\\
2014 Jun 26&2014.4826&AE-AW-E6 AW-E6-W7& 15& 79&HR\,5329\\
2015 Mar 28&2015.2360&AC-AW-E03& 10& 32&HR\,5329 FK5\,423 FK5\,622\\
2015 Mar 29&2015.2387&AC-AW-E03& 10& 32&HR\,5329 FK5\,423 FK5\,622\\
2015 Apr 21&2015.3015&AC-AW-E03& 10& 32&HR\,5329 FK5\,423 FK5\,622\\
2015 Apr 28&2015.3206&AC-AW-E03& 10& 32&HR\,5329 FK5\,622\\
2015 Apr 30&2015.3261&AC-AW-E03& 10& 32&HR\,5329 FK5\,423 FK5\,622\\
2015 May 20&2015.3808&AC-AW-E03& 10& 32&HR\,5329 FK5\,677\\
2015 May 28&2015.4026&AC-AW-E03& 10& 32&HR\,5329 FK5\,423 FK5\,677\\
2015 Jun 01&2015.4136&AC-AW-E03& 10& 32&HR\,5329 FK5\,423 FK5\,677\\
\hline
\end{tabular}
\flushleft
Notes: column 3 lists station names used in baselines or triangles, col.\ 4 and 5 list the minimum and maximum projected baseline lengths during the observations, and col.\ 6 the calibrator stars used.
\end{table*}

\begin{table*}[h!]
\caption{Astrometric results}
\label{tab:astro_results}
\begin{tabular}{lccrrrrrrr}
\hline
\hline
UT Date&Julian year&Number of vis.&
$\rho$&$\theta$&$\sigma_{\rm maj}$&$\sigma_{\rm min}$&$\phi$&$O-C_\rho$&$O-C_\theta$\\
     &     &     & mas & deg.& mas & mas & deg. & mas & deg. \\
 (1) & (2) & (3) & (4) & (5) & (6) & (7) & (8) & (9) & (10) \\
\hline
Apr 18&1989.2946&  44&  3.47&  70.28&  0.371&  0.079&102.4&  0.26&  -0.3\\
Jun 19&1991.4629&  59&  7.22& 251.89&  0.208&  0.098& 65.3& -0.19&  -0.6\\
Feb 28&1992.1583&  39&  6.98& 253.18&  0.198&  0.075&112.3&  0.23&   1.8\\
Apr 28&1992.3226&  75&  8.27& 254.54&  0.287&  0.119& 84.1&  0.74&   0.4\\
May 02&1992.3336&  61&  7.47& 256.13&  0.327&  0.166& 83.5&  0.60&   0.7\\
Jun 08&1992.4349&  87&  6.39& 249.32&  0.222&  0.062& 62.6&  0.26&  -1.3\\
Mar 19&1997.2123& 549&  4.62& 251.34&  0.264&  0.192& 10.1&  0.02&   2.6\\
Mar 26&1997.2315& 549&  7.25& 252.08&  0.226&  0.131&135.0&  0.04&  -0.0\\
May 08&1997.3492& 337&  3.73& 247.60&  0.409&  0.156&121.7& -0.00&   0.3\\
Jun 19&1997.4642& 234&  3.12&  77.49&  0.345&  0.136&135.0& -0.01&   4.0\\
Jun 21&1997.4697& 316&  2.08&  77.18&  0.414&  0.183&154.8& -0.10&  -1.1\\
Jun 22&1997.4724& 450&  1.47&  76.79&  0.345&  0.136&135.0&  0.02&  -6.9\\
Jun 26&1997.4833& 237&  2.27& 250.11&  0.478&  0.181&116.5&  0.28&   8.4\\
Jun 27&1997.4861& 328&  2.82& 236.87&  0.264&  0.192& 10.1&  0.07&  -8.0\\
Jul 04&1997.5052& 656&  6.26& 248.55&  0.226&  0.131&135.0& -0.21&  -2.5\\
Jul 08&1997.5162& 246&  7.42& 254.26&  0.453&  0.117&135.0&  0.01&   1.8\\
Jul 09&1997.5189& 246&  7.39& 252.07&  0.740&  0.186&122.1& -0.14&  -0.7\\
Jul 13&1997.5299& 137&  7.85& 257.55&  0.789&  0.268&135.0&  0.28&   3.6\\
Jul 14&1997.5326& 410&  7.32& 252.20&  0.414&  0.183&115.2& -0.15&  -2.1\\
Mar 05&2013.1740& 420&  6.25& 256.07&  0.226&  0.131&135.0& -0.01&  -0.3\\
Mar 06&2013.1767& 360&  5.62& 256.83&  0.292&  0.143&157.7& -0.31&   0.0\\
Mar 07&2013.1794& 330&  5.39& 256.09&  0.582&  0.226& 22.4& -0.18&  -1.2\\
May 24&2013.3930& 445&  2.62& 242.84&  0.226&  0.131&135.0& -0.01&  -1.7\\
May 25&2013.3957&1208&  3.43& 244.31&  0.264&  0.192& 10.1&  0.09&  -2.2\\
May 26&2013.3985& 480&  4.11& 245.67&  0.374&  0.221&121.5&  0.11&  -2.1\\
May 27&2013.4012&1530&  4.53& 250.81&  0.292&  0.143&157.7& -0.07&   2.1\\
May 30&2013.4094&1649&  6.04& 251.26&  0.226&  0.131&135.0& -0.01&   0.7\\
May 31&2013.4122&1829&  6.30& 253.41&  0.264&  0.192& 10.1& -0.12&   2.4\\
Jun 01&2013.4149&1620&  6.66& 253.59&  0.226&  0.131&135.0& -0.07&   2.2\\
Jun 03&2013.4204&1304&  7.27& 252.46&  0.292&  0.143&157.7&  0.05&   0.4\\
Jun 04&2013.4231&1920&  7.40& 253.14&  0.264&  0.192& 10.1&  0.01&   0.7\\
Jun 05&2013.4259&1608&  7.60& 253.22&  0.226&  0.131&135.0&  0.09&   0.5\\
Jun 07&2013.4313&  60&  7.56& 255.34&  1.058&  0.490&115.7& -0.07&   2.0\\
Jun 08&2013.4341& 339&  7.63& 257.33&  0.576&  0.317&143.8&  0.00&   3.7\\
Jun 10&2013.4395& 450&  7.63& 255.59&  0.264&  0.192& 10.1&  0.14&   1.4\\
Jun 11&2013.4423& 457&  7.72& 256.67&  0.414&  0.183&154.8&  0.35&   2.1\\
Jun 13&2013.4478& 171&  7.15& 258.39&  0.385&  0.243&162.3&  0.13&   3.2\\
Jun 15&2013.4532& 300&  6.77& 258.07&  0.688&  0.171&161.5&  0.26&   2.1\\
Jun 16&2013.4560& 300&  6.38& 256.03&  0.345&  0.136&135.0&  0.17&  -0.4\\
Jun 17&2013.4587& 359&  5.95& 256.65&  0.483&  0.197&136.6&  0.08&  -0.2\\
Jun 24&2014.4772&4248&  0.80&  12.46&  0.226&  0.131&135.0&  0.14& -15.8\\
Jun 25&2014.4799&4186&  1.23&  41.58&  0.226&  0.131&135.0&  0.07&  -9.7\\
Jun 26&2014.4827&4334&  1.81&  57.29&  0.226&  0.131&135.0&  0.09&  -2.6\\
Mar 28&2015.2356& 875&  5.39& 247.62&  0.226&  0.131&135.0& -0.04&  -2.2\\
Mar 29&2015.2383& 620&  5.84& 246.62&  0.483&  0.197&133.4& -0.04&  -3.7\\
Apr 21&2015.3013& 895&  4.11& 256.71&  0.344&  0.126&135.0&  0.18&  -3.5\\
May 20&2015.3807& 476&  5.93& 249.06&  0.514&  0.261&177.8& -0.18&  -1.6\\
May 28&2015.4026&1065&  7.46& 249.77&  0.345&  0.136&135.0& -0.17&  -3.6\\
Jun 01&2015.4135& 630&  7.23& 251.26&  0.483&  0.197&136.6& -0.12&  -3.3\\
\hline
\end{tabular}
\flushleft
Notes: column 3 lists the number of visibility measurements obtained, col.\ 4 and 5 separation and position angle of the binary components (at local midnight of the date of observation), col.\ 6, 7 and 8 the major and minor axis, and the position angle of the astrometric error ellipse, respectively. Columns 9 and 10 list the offsets in separation and angle between the measurements and the predicted orbit positions.
\end{table*}

\section{EMCEE calculations of the uncertainties in the light curve analysis}

\begin{figure*}[ht!]
\includegraphics[width=17cm]{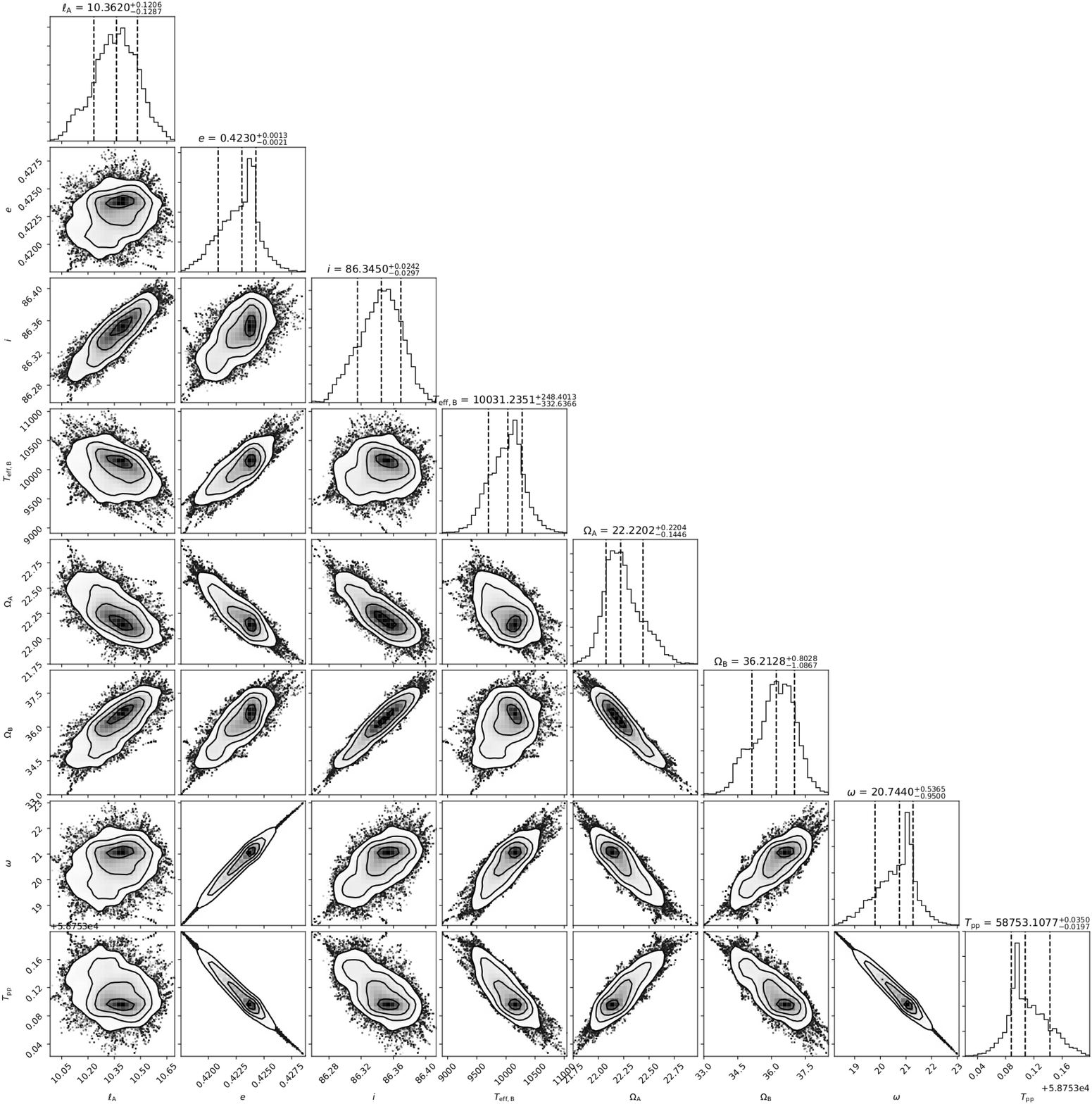} 
\caption{
The MCMC determination of the light curve parameters and accompanying uncertainties for TESS observations of $\alpha$~Dra in Part 1 (Sectors 14, 15, and 16). A strong correlation between parameters is present, as expected for partial (almost grazing) eclipses. The posterior distribution densities  are plotted (solid contours). The histogram distributions (solid lines) are plotted across the associated quantity with 16th, 50th, and 84th percentiles levels (dashed lines).
}
\label{fig:corner1} 
\end{figure*}

\begin{figure*}[ht!]
\includegraphics[width=17cm]{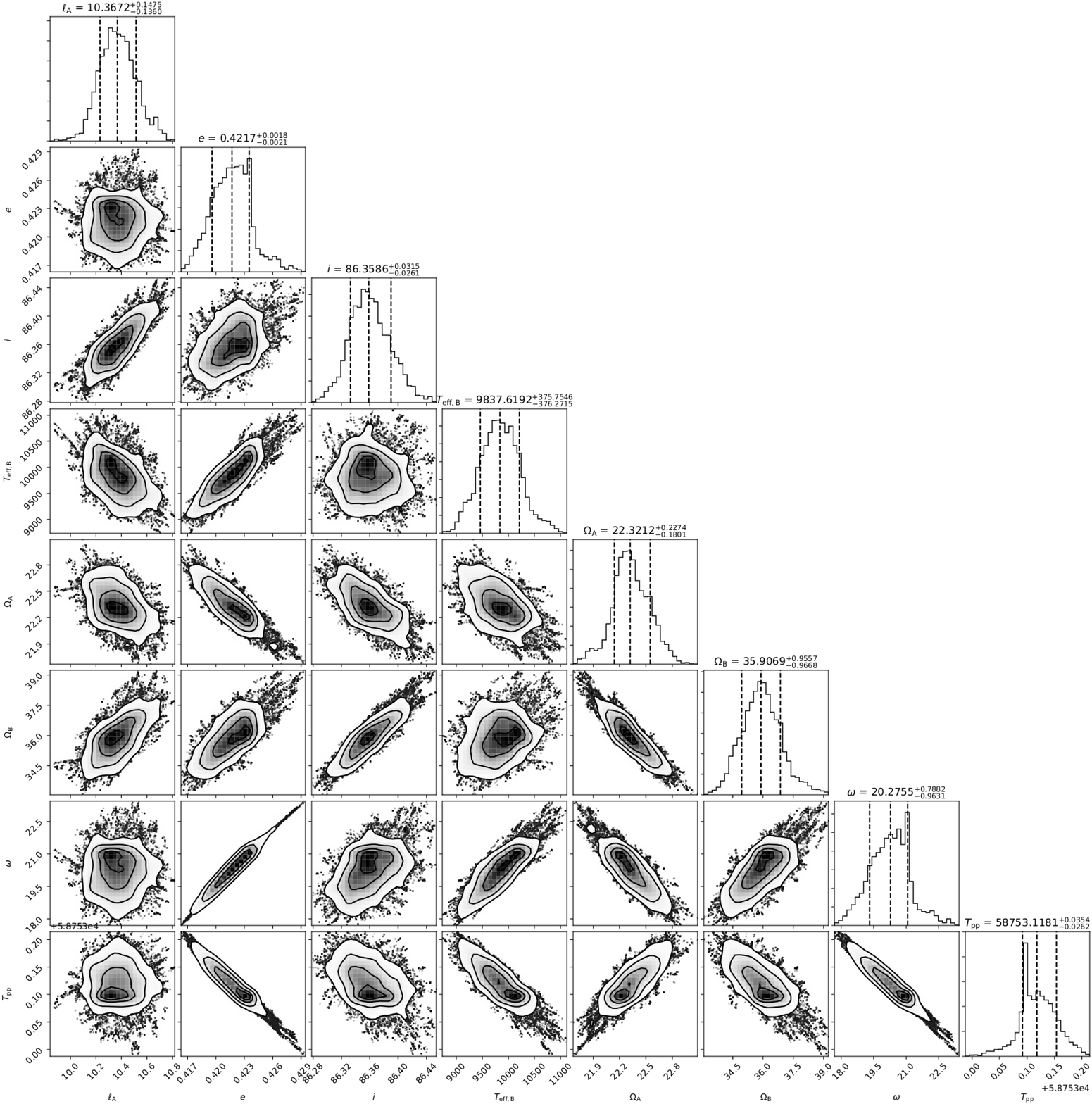} 
\caption{
Same as Fig.~\ref{fig:corner1} but for TESS observations of $\alpha$~Dra in Part 2 (Sectors 21, and 22).
}
\label{fig:corner2} 
\end{figure*}

\end{appendix}

\end{document}